\documentclass[pra,twocolumn,preprintnumbers,amsmath,amssymb,floatfix]{revtex4}
\usepackage{graphicx}
\usepackage{graphics}
\usepackage{psfrag}
\usepackage{hyperref}
\usepackage{multirow}
\usepackage[sort&compress]{natbib}
\usepackage{amssymb}
\usepackage{amsmath}
\usepackage{amsthm}
\usepackage{bbold}

\newcommand{\vare}{\varepsilon}

\newcommand{\rmi}{{\rm i}}
\newcommand{\mathJ}{\mathcal{J}}

\begin{document}

\hypersetup{pdftitle={Anisotropy induces non-Fermi-liquid behavior and nematic magnetic order in three-dimensional Luttinger semimetals}}
\title{Anisotropy induces non-Fermi-liquid behavior and nematic magnetic order in three-dimensional Luttinger semimetals}

\author{Igor Boettcher}
\affiliation{Department of Physics, Simon Fraser University, Burnaby, British Columbia, Canada}
\author{Igor F. Herbut}
\affiliation{Department of Physics, Simon Fraser University, Burnaby, British Columbia, Canada}

\begin{abstract}
We illuminate the intriguing role played by spatial anisotropy in three-dimensional Luttinger semimetals featuring quadratic band touching and long-range Coulomb interactions. We observe the anisotropy to be subject to an exceptionally slow renormalization group (RG) evolution so that it can be considered approximately constant when computing the impact of quantum fluctuations on the remaining couplings of the system. Using perturbative RG we then study the competition of all local short-range interactions that are generated from the long-range interactions for fixed anisotropy. Two main effects come to light for sufficiently strong anisotropy. First, the three-dimensional system features an Abrikosov non-Fermi liquid ground state. Second, there appear qualitatively new fixed points which describe quantum phase transitions into phases with nemagnetic orders -- higher-rank tensor orders that break time-reversal symmetry, and thus have both nematic and magnetic character. In real materials these phases may be realized through sufficiently strong microscopic short-range interactions. On the pyrochlore lattice, the anisotropy-induced fixed points determine the onset of all-in-all-out or spin ice ordering of local magnetic moments of the electrons.
\end{abstract}

\maketitle

\section{Introduction}

Understanding Fermi points at high-symmetry band crossings in semimetallic materials constitutes a promising portal towards designing exotic states of matter \cite{doi:10.1146/annurev-conmatphys-031113-133841}. The intriguing properties of systems with linear band touching described by a low-energy effective Dirac Hamiltonian have been studied extensively in graphene both theoretically and experimentally \cite{herbut2006,RevModPhys.81.109,RevModPhys.83.407}. The recent advance in realizing three-dimensional Weyl semimetals and the subsequent experimental verification of their peculiar properties marks another milestone on this road \cite{Xu613,PhysRevX.5.031013}. A quadratic band touching (QBT) point is realized in bilayer graphene and three-dimensional Luttinger semimetals, the latter being described by the Luttinger Hamiltonian \cite{Luttinger}, which includes GaAs, HgTe, $\alpha$-Sn, or the recently actively studied class of Pyrochlore Iridates \cite{murakami,moon,doi:10.1146/annurev-conmatphys-020911-125138}.

Systems with QBT become particularly interesting when strong spin-orbit coupling leads to a band inversion such that both positive and negative energy states meet at the Fermi point. Furthermore, whereas the short-range part of Coulomb interactions is screened in three dimensions, its long-range part substantially influences the many-body electron correlations. It has been pointed out by Abrikosov \cite{abrben,abrikosov} and reinvestigated more recently \cite{moon} that the interplay between an inverted QBT point and long-range Coulomb repulsion leads to a non-Fermi liquid (NFL) ground state of the system. Both requirements for Abrikosov's scenario are realized in the Pyrochlore Iridates \cite{PhysRevLett.96.087204,MachidaNature,doi:10.1143/JPSJ.80.044708,PhysRevLett.106.217204,PhysRevB.87.155101,PhysRevB.91.115124,kondo,PhysRevLett.117.056403,2016arXiv160308022L}, which therefore constitutes an ideal experimental platform to test its predictions. It is found that for most members of the material class the ground state is a magnet with spins ordered in all-in-all-out (AIAO) states on the tetrahedra of the pyrochlore lattice. An exception is given by Pr$_2$Ir$_2$O$_7$, which remains a bad metal for the lowest temperatures explored in experiment \cite{doi:10.1146/annurev-conmatphys-020911-125138}.

Although the density of states is proportional to the square root of energy at a QBT point in three dimensions so that the short-range part of the Coulomb repulsion is technically irrelevant, short-range interactions can still influence the phase structure of the model in two ways. First, if the corresponding coupling constant exceeds a certain critical value, they can lead to qualitatively different ordered states, represented in the renormalization group (RG) by a runaway flow. Second, even if initially absent, short-range interactions are generated during the RG flow. The feedback of those short-range couplings onto each other leads to the appearance of further fixed points, which may collide with Abrikosov's infrared fixed point and annihilate it. In fact, within an $\vare$-expansion around four dimensions extrapolated to three dimensions, this does indeed happen for the isotropic system, and the Abrikosov fixed point is removed through precisely this mechanism \cite{PhysRevLett.113.106401}. The system is left with a runaway flow towards an ordered nematic state \cite{PhysRevB.92.045117,PhysRevB.93.165109}.

In this work we extend the analysis of the influence of short-range interactions on Abrikosov's scenario by incorporating the effect of spatial anisotropy of the band structure at the QBT point. In fact, most real materials are not fully rotationally symmetric, but rather feature only cubic rotation invariance -- even in the low-energy description encoded in the Luttinger Hamiltonian. Whereas Abrikosov points out that a stable fixed point can only be isotropic \cite{abrikosov}, Savary, Moon, and Balents \cite{savary} find a stable quantum critical point towards AIAO order at (maximally) strong anisotropy within a controlled $1/N$-expansion. Here we investigate within the $\vare$-expansion whether (i) anisotropy can lead to a stable NFL fixed point in three dimension, and (ii) whether it can yield further unstable directions, such as towards the AIAO state. A key finding is that the RG flow of the anisotropy parameter is negligibly slow so that the anisotropy can simply be considered constant for all practical purposes; i. e. an approximately marginal coupling. Within this approximation both (i) and (ii) will be found to be answered in the affirmative.

This work is organized as follows. In Sec. \ref{SecField} we introduce the field theoretic framework describing the anisotropic Luttinger semimetal together with the RG flow of the anisotropy parameter and Abrikosov's NFL scenario. In Sec. \ref{SecShort} we study the RG fixed point structure and influence of short-range interactions in the anisotropic system. We close with a discussion of our results in Sec. \ref{SecConcl}. Extensive appendices are devoted to deriving the full set of RG equations (\ref{AppRG}), constructing irreducible spin tensors (\ref{AppTens}), Fierz identities (\ref{AppFierz}), and computing the functions $f_i(\delta)$ used throughout the text (\ref{AppCub}).

\section{Field theoretic framework}\label{SecField}

\subsection{Lagrangian}

The physics of three-dimensional Luttinger fermions with long-range Coulomb repulsion is captured by the Lagrangian
\begin{align}
 \label{field2} L = \psi^\dagger (\partial_\tau + H+\rmi a) \psi +\frac{1}{2e^2} (\nabla a)^2,
\end{align}
where $\psi$ is a four-component Grassmann field, $a$ is the real electrostatic photon field, $\tau$ denotes imaginary time, $e$ is electric charge, and $H$ is the Luttinger Hamiltonian \cite{Luttinger}. The chemical potential is tuned to be at $\mu=0$. We assume time-reversal invariance at the single-particle level, no external magnetic field shall be applied. Under these conditions, each of the two bands touching quadratically is doubly degenerate, and Luttinger showed that the most general single-particle Hamiltonian is given by
\begin{align}
 \nonumber H =\ &\frac{\hbar^2}{2m^*}\Bigl[\Bigl(\alpha_1+\frac{5}{2}\alpha_2\Bigr)p^2\mathbb{1}_4-2\alpha_3(\vec{p}\cdot\vec{J})^2\\
 \label{field3} &+2(\alpha_3-\alpha_2)\sum_{i=1}^3p_i^2J_i^2\Bigr],
\end{align}
with Luttinger parameters $\alpha_{1,2,3}$, effective electron mass $m^*$, and momentum operator $\vec{p}=-\rmi \nabla$. The $4\times 4$ matrices $\vec{J}=(J_x,J_y,J_z)^{\rm t}$ represent the spin-3/2 angular momentum operators. We set $\hbar=2m^*=1$ in the following. We have written the Lagrangian (\ref{field2}) in a manner such that the dynamic critical exponent $z=2$ at the noninteracting Gaussian fixed point. Further, $\mathbb{1}_L$ denotes the $L \times L$ unit matrix.

The Luttinger Hamiltonian can be written in the computationally more advantageous form \cite{murakami,moon,PhysRevLett.113.106401,PhysRevB.93.205138}
\begin{align}
 \nonumber H=\ & \alpha_1 p^2 \mathbb{1}_4-(\alpha_2+\alpha_3)\sum_{a=1}^5d_a(\vec{p})\gamma_a\\
 \label{field4} &+(\alpha_2-\alpha_3)\sum_{a=1}^5s_ad_a(\vec{p})\gamma_a,
\end{align}
with five $4\times 4$ Hermitean matrices $\{\gamma_a\}_{a=1,\dots,5}$ satisfying the Clifford algebra,
\begin{align}
 \label{field5} \{\gamma_a,\gamma_b\}= 2\delta_{ab} \mathbb{1}_4,
\end{align}
the functions $d_a$ being given by $d_1 = \frac{\sqrt{3}}{2}(p_x^2-p_y^2)$, $d_2 = \frac{1}{2}(2p_z^2-p_x^2-p_y^2)$, $d_3 = \sqrt{3}p_zp_x$, $d_4 = \sqrt{3}p_yp_z$, $d_5 = \sqrt{3}p_xp_y$, and we have $s_{1,2}=-1$ and $s_{3,4,5}=+1$. We define the matrices $\{\gamma_a\}$ below in a more general context. The functions $d_a$ constitute the real $\ell=2$ spherical harmonics on a sphere of radius $p$.

For the field theoretic treatment we rescale the Hamiltonian by $A_\psi=-(\alpha_2+\alpha_3)$, normalize the field $\psi$ such that $A_\psi=1$, and introduce the particle-hole asymmetry and anisotropy parameters, $x$ and $\delta$, via
\begin{align}
 \label{field6} x = -\frac{\alpha_1}{\alpha_2+\alpha_3},\ \delta = -\frac{\alpha_2-\alpha_3}{\alpha_2+\alpha_3}.
\end{align}
In the following we set $x=0$, corresponding to particle-hole symmetry, which will be shown to emerge dynamically during the renormalization group flow.
With the normalization in Eq. (\ref{field6}), the parameter $\delta$ lies within the real interval $[-1,1]$. We eventually arrive at
\begin{align}
 \label{field7} H =\sum_{a=1}^5 (1+\delta s_a)d_a(\vec{p})\gamma_a.
\end{align}
Squaring the Hamiltonian yields
\begin{align}
 \label{field8} H^2 = \Bigl( (1-\delta)^2 p^4  +12\delta \sum_{i<j} p_i^2p_j^2\Bigr)\mathbb{1}_4
\end{align}
due to $\sum_{a=1}^5 d_a^2 = p^4$ and $\sum_{a=3,4,5} d_a^2 = 3\sum_{i<j}p_i^2p_j^2$. The roots of this expression determine the doubly degenerate spectrum of the Hamiltonian. We observe the spectrum to be rotation symmetric (a function of $p^2=p_x^2+p_y^2+p_z^2$ alone) only for $\delta =0$.

The latter observation is a key element of this work. Under a spatial rotation of coordinates $x_i \mapsto \mathcal{R}_{ij}x_j$ with $\mathcal{R}\in\text{SO}(3)$, $\mathcal{R}^{\rm t}\mathcal{R}=\mathbb{1}_3$, the operators $\vec{p}$ and $\vec{J}$ transform as vectors, i.e., in the same manner as $\vec{x}$. Obviously, the first line in Eq. (\ref{field3}) is rotation invariant. The same is true for the first line in the representation of Eq. (\ref{field4}) -- although less obviously so at this point. It will become apparent once we define the $\gamma$-matrices as components of the second rank tensor $S_{ij}$ below: the term $\sum_a d_a\gamma_a$ is then seen to be proportional to $S_{ij}p_ip_j$, which is clearly rotation invariant.

Rotation invariance of the Hamiltonian is broken for $\delta \propto (\alpha_2-\alpha_3)\neq 0$. However, rotations with certain fixed angles still leave the expression invariant, namely those which rotate the individual coordinate axes onto each other. Roughly, those transformations permute the coordinate labels $x,y,z$. If this symmetry is present we say that the system has cubic symmetry. The Luttinger Hamiltonian exhausts all cubic invariant terms to order $p^2$, so that three Luttinger parameters suffice to parametrize a quadratic band touching point. The full rotation and cubic rotation groups are $\text{SO}(3)$ and $\text{O}_{\rm h}$, respectively.

\subsection{RG flow of the anisotropy}

Performing the usual Wilson's integration of the fermionic modes within the momentum shell $[\Lambda/b,\Lambda]$ and with all frequencies, we derive a flow equation for the anisotropy parameter $\delta$ from the renormalization of the fermion self-energy. The computation is presented in detail in App. \ref{AppRG}. For this, angular integrations are performed in three dimensions, but the qualitative results remain invariant when performing the angular integration in four dimensions, which constitutes the upper critical dimension.

The RG flow equations presented in this work are valid for arbitrary values of $\delta$. Except for some special values, such as $\delta=0$, the $\beta$-function can only be determined numerically. To make them more accessible, however, we introduce functions $f_i(\delta)$ with the following properties: We have $f_i(0)=1$ for $\delta=0$, and for general $\delta\in[-1,1]$, $f_i(\delta)$ is nonzero, positive, and of order unity. In this way, the qualitative and mostly quantitative aspects of the RG flow can be understood by setting
\begin{align}
 \label{field10} f_i(\delta) \approx 1
\end{align}
in the $\beta$-functions. The functions $f_i(\delta)$ are computed in App. \ref{AppCub} and shown in Fig. \ref{Figfi}.

\begin{figure}[t]
\centering
\includegraphics[width=8.5cm]{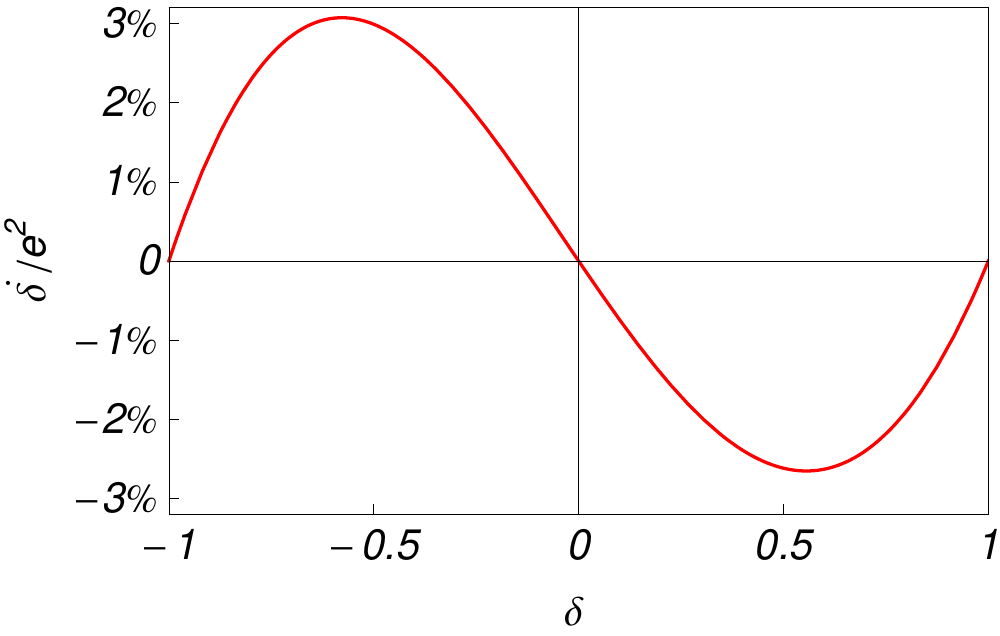}
\caption{RG flow of the anisotropy $\delta$. We observe an attractive fixed point at $\delta_\star=0$ (negative slope), and two repulsive ones at $\pm1$. This, however, is hardly of practical relevance since the prefactor multiplying $e^2$ in Eq. (\ref{field11}) is exceptionally small so that $\delta$ can effectively be treated as a constant parameter with $\dot{\delta}\approx 0$ during the running of couplings. Put differently, the tiny value of the beta function also implies very small gradients $\mbox{d}\dot{\delta}/\mbox{d}\delta$ and thus an anomalously slow RG flow. As a consequence, $\delta(b)$ is approximately constant for those momentum rescale factors  $b$ where renormalizations of the charge and of the generated short-range interactions are important.}
\label{FigAniso}
\end{figure}

The RG flow for the anisotropy parameter $\delta$ to leading order in $e^2$ reads
\begin{align}
 \label{field11} \dot{\delta} = \frac{\mbox{d}\delta}{\mbox{d}\log b} = -\frac{2}{15} (1-\delta^2)\Bigl[f_{1\rm e}(\delta)-f_{1\rm t}(\delta)\Bigr] e^2.
\end{align}
Since $f_i(0)=1$, we immediately discern three fixed points at $\delta_\star=0,\pm1$. The coefficient multiplying $e^2$, however, is exceptionally small. Close to the attractive fixed point $\delta_\star =0$, for example, the linearized flow reads
\begin{align}
 \label{field12} \dot{\delta} \simeq -\frac{8}{105} e^2\delta .
\end{align}
This signals an extremely slow flow towards the fixed point. In fact, the entire prefactor multiplying $e^2$ in Eq. (\ref{field11}) remains below $3\%$ in magnitude for all values of $\delta$, see Fig. \ref{FigAniso}. Consequently, the parameter $\delta$ can approximately be considered to be marginal - in contrast to a running coupling.

\subsection{Abrikosov's NFL fixed point}

\begin{figure}[t]
\centering
\includegraphics[width=8.5cm]{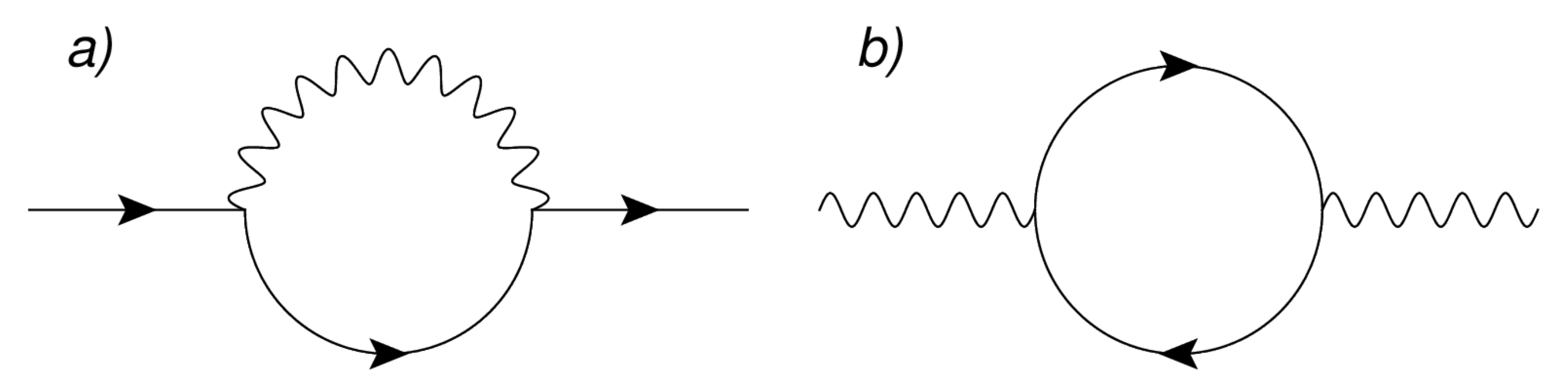}
\caption{One-loop self-energy contributions to the RG flow. Diagrams a) and b) show the fermion and photon self-energy, respectively. A straight line represents a fermion propagator, a wiggly line a photon propagator. Contribution a) generates the anomalous fermion scaling, expressed by $\eta$ and $z$, as well as the exceptionally weak running of the anisotropy $\delta$. Diagram b) results in the charge renormalization. Both diagrams taken together yield Abrikosov's NFL fixed point scenario.}
\label{FigLoops}
\end{figure}

The loop corrections to the fermion self-energy determine the RG flow of the couplings $x$ and $\delta$, the fermion anomalous dimension $\eta$, and the dynamic critical exponent $z$. The photon self-energy leads to a renormalization of the charge $e^2$. The diagrammatic one-loop contributions to the fermion and photon self-energies are displayed in Fig. \ref{FigLoops}. The flow of $\delta$ has been discussed in the previous section. It is easy to see (App. \ref{AppRG}) that the coupling $x$, which is marginal at the Gaussian fixed point, only receives corrections according to
\begin{align}
 \label{field13} \dot{x} = -\eta x.
\end{align}
At an interacting fixed point with $\eta >0$ the coupling is then attracted towards $x_{\star}=0$. This justifies setting $x=0$ in Eq. (\ref{field7}). In contrast to Eq. (\ref{field12}), $\eta$ is not exceptionally small so that a nonzero $x$ diminishes quickly.

The existence of a nontrivial fixed point close to $d=4$ dimensions with anomalous fermion scaling and charge renormalization for a three-dimensional quadratic band touching system has first been pointed out by Abrikosov \cite{abrikosov}. This approach to an NFL is ingenious in its simplicity, as it basically relies on the ``chirality'' of the band dispersion, i.e. the presence of positive and negative eigenenergies (which implies $e^2_\star>0$ and $\eta \sim e^2_\star >0$), and the frequency independence of the photon propagator (which implies $z=2-\eta <2$.) This has to be contrasted with an otherwise similar system of ultracold atoms at resonance with $\mu=0$, where the lower band is missing, and the dimer propagator is Galilean invariant, implying anomalous boson scaling, but $\eta=0$ and $z=2$ \cite{PhysRevA.73.033615,nikolic,PhysRevA.89.053630}. Abrikosov further points out that a stable fixed point of $\delta$ can only be at $\delta_\star=0$, and thus sets $\delta=0$. As we have seen above, this statement is true, but it is still reasonable to consider the modifications of the Abrikosov NFL fixed point when taking into account a nonzero value of $\delta$.

The fermion anomalous dimension is given by
\begin{align}
 \label{field14} \eta = \frac{2}{15}\Bigl[(1-\delta)f_{1\rm e}(\delta)+(1+\delta)f_{1\rm t}(\delta)\Bigr] e^2.
\end{align}
Close to $\delta_\star=0$ it reads
\begin{align}
 \label{field15} \eta \simeq \frac{4}{15} e^2 - \frac{4}{105} e^2\delta.
\end{align}
The numerical coefficient $4/15$ should be compared with the one in Eq. (\ref{field12}). 
Equations (\ref{field12}) and (\ref{field15}) are consistent with the results of Ref. \cite{PhysRevB.93.205138}, which have been obtained in a perturbative expansion in $\delta$. The comparison is facilitated by setting $y=0$ in the reference, and adjusting some couplings and prefactors, which leads to identical loop contributions to obtain the fermion self-energy. The dynamic critical exponent is given by
\begin{align}
 \label{field16} z=2-\eta.
\end{align}

\begin{figure}[t]
\centering
\includegraphics[width=7.5cm]{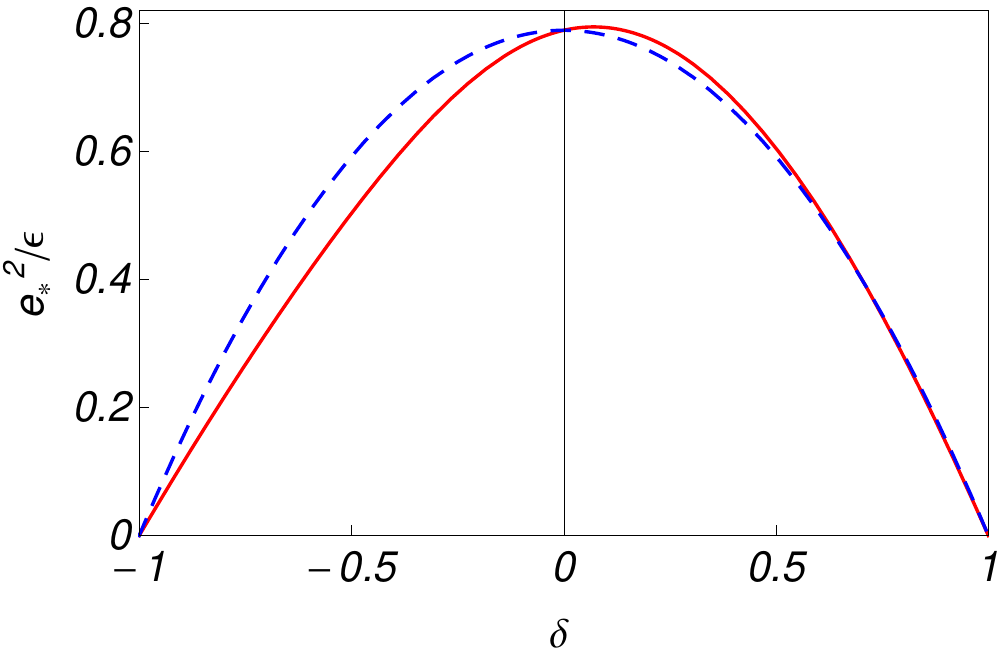}
\caption{Fixed point of the charge. The (effective) attractive infrared fixed point of $e^2$ close to four dimensions, divided here by $\vare=4-d$, survives for all fixed $|\delta|<1$. The solid red line shows the result from Eq. (\ref{field18}) and the dashed blue line is $\frac{15}{19}(1-\delta^2)$, which amounts to setting $f_{\star}(\delta)\approx 1$ and yields an excellent approximation. For strong anisotropy the theory becomes weakly coupled.}
\label{Fige2star}
\end{figure}

The flow equation for the charge is then given by
\begin{align}
 \label{field17} \dot{e}^2 =\frac{\mbox{d}e^2}{\mbox{d}\log b} = (4-d-\eta)e^2 - \frac{f_{e^2}(\delta)}{1-\delta^2}e^4.
\end{align}
The function $f_{e^2}(\delta)$ is bounded from below by $f_{e^2}(0.13)=0.987$. We find the Abrikosov fixed point of the charge for small $\vare=4-d$ to be
\begin{align}
 \label{field18} e^2_\star = \frac{15}{19}(1-\delta^2)f_\star(\delta) \vare.
\end{align}
Again, $f_\star(\delta)$ is positive and of order unity for all $\delta$, with $f_\star(0)=1$. We thus conclude that the Abrikosov fixed point persists for all values of $|\delta|<1$. However, as $|\delta|\to1$, the fixed point becomes weakly coupled. This behavior is visualized in Fig. \ref{Fige2star}.

The presence of the prefactor $(1-\delta^2)$ in the fixed point charge $e^2_\star$ renders the problem perturbative for strong anisotropy, even when working with $\vare=1$. A similar observation has been made in Ref. \cite{savary} at the anisotropic fixed point with $\delta_\star=-1$. In contrast to the $\vare$-expansion applied here, the fixed point in the reference is controlled by a $1/N$-expansion in three dimensions. The findings of the two investigations differ in that we do not find the anisotropic fixed point to be stable. This will be discussed further in Sec. \ref{SecConcl}.

\section{Short-range interactions}\label{SecShort}

 We restrict the discussion of short-range interactions to those that can be expressed in terms of local four-fermion terms. Every such term can be written as a contribution
\begin{align}
 \label{short1} L \sim g(\psi^\dagger M \psi)(\psi^\dagger N \psi)
\end{align}
to the Lagrangian, with coupling $g$ and some matrices $M,N\in\mathcal{X}$, where $\mathcal{X}$ is the set of complex Hermitean $4\times 4$ matrices.  In particular, even the terms of the form $ (\psi^\dagger M \psi^*)(\psi^{\rm t} N \psi) $ can be brought into the form of Eq. (\ref{short1}) by a Fierz transformation \cite{PhysRevB.93.205138}.

In order to cover all possible terms of the form (\ref{short1}), it is sufficient to restrict to contributions $g_{AB}(\psi^\dagger \Sigma^A\psi)(\psi^\dagger \Sigma^B\psi)$, where $\{\Sigma^A\}_{A=1,\dots,16}$ is an $\mathbb{R}$-basis of $\mathcal{X}$. Furthermore, rotation or cubic symmetry impose severe constraints on the possible values of $g_{AB}$. One possible choice of basis consists in $\Sigma^A \to \Gamma^A$ with
\begin{align}
 \label{short3} \{\Gamma^A\} = \{ \mathbb{1}_4,\ \gamma_a,\ \gamma_{ab} \},
\end{align}
where $\gamma_{ab}=\rmi \gamma_a\gamma_b$, $a,b =1,\dots,5$,  and we require $a<b$. (There are ten such matrices.) This choice is particularly convenient due to the computational simplifications arising from the Clifford algebra. However, the matrices $\gamma_{ab}$ do not have definite transformation properties under rotations of the cubic symmetry group. For our purposes it is convenient to introduce a basis which is manifestly cubic invariant. We denote it by
\begin{align}
 \label{short4} \{\Sigma^A\} = \{ \mathbb{1}_4,\ \mathJ_i,\ \gamma_a,\ W_\mu \}
\end{align}
with $i=1,2,3$ and $\mu=1,\dots,7$. Note that, indeed, both $\{\Gamma^A\}$ and $\{\Sigma^A\}$ consist of 16 elements each.

In this section we first construct irreducible spin tensors to derive the basis $\{\Sigma^A\}$, then study the fixed point structure of the RG flow including these couplings, and eventually investigate the instabilities associated with these fixed points (expressed through a divergent susceptibility at the transition).

\subsection{Irreducible spin tensors}

To obtain the basis (\ref{short4}) we construct all possible $4\times 4$ operators that transform as tensors under $\text{SO(3)}$. For this let $\mathcal{R}\in\text{SO}(3)$ be a rotation matrix, $\mathcal{R}^{\rm t}\mathcal{R}=\mathbb{1}_3$. We define a tensor $T$ of rank $\ell$ as any object labelled by indices $(i_1,\dots,i_\ell)$ that transforms as
\begin{align}
 \label{irr1} T_{i_1\dots i_\ell} \mapsto \mathcal{R}_{i_1j_1}\cdots\mathcal{R}_{i_\ell j_\ell} T_{j_1\dots j_\ell}
\end{align}
under a change of coordinates $x_i\mapsto \mathcal{R}_{ij}x_j$. In particular, we will be interested in the case that $T$ is a $4\times 4$ matrix. Recall that the rank of a tensor can be reduced by one or two units, respectively, by contracting it with $\vare_{ijk}$ or $\delta_{ij}$ according to
\begin{align}
 \label{irr2} T^\prime_{ki_3\dots i_\ell} &= \vare_{ijk}  T_{iji_3\dots i_\ell},\\
 \label{irr3} T^\prime_{i_3\dots i_\ell} &= \delta_{ij} T_{iji_3\dots i_\ell}.
\end{align}
If such a contraction yields zero we say that the tensor is irreducible. From Eqs. (\ref{irr2}) and (\ref{irr3}) it is then clear that irreducible tensors are precisely the symmetric and traceless tensors \cite{BookHess}. Here, we say that a tensor is traceless if all its partial traces yield zero, i.e., it vanishes whenever two indices are contracted.

In order to construct the basis $\{\Sigma^A\}$ with definite transformation properties under rotations we apply the following recipe: The  $j=3/2$ spin matrices $J_i$ transform as a vector under $\text{SO}(3)$, and the product $J_{i_1}\cdots J_{i_\ell}$ transforms as a tensor of rank $\ell$. The corresponding irreducible tensors of rank $\ell$ can be computed from these products through symmetrization and the subtraction of suitable  traces. One may wonder whether this procedure yields irreducible tensors of arbitrary rank. However, by means of the Cayley--Hamilton theorem it is easy to show that there are no irreducible spin tensors with rank $\ell> 2j$, see App. \ref{AppCH}. Hence, in our case, we obtain irreducible tensors of rank $\ell=0,1,2,3$, which can be used to build the basis $\{\Sigma^A\}$ introduced above.

Due to the non-commutativity of the spin matrices, $[J_i,J_j]=\rmi \vare_{ijk} J_k$, products of the type $J_{i_1}\cdots J_{i_\ell}$ are not symmetric. We define the symmetrized rank $\ell=2,3$ tensors
\begin{align}
 \label{irr4} \bar{S}_{ij} &= J_iJ_j +J_jJ_i,\\
 \label{irr5} \bar{B}_{ijk} &= J_iJ_jJ_k +\text{permutations of }ijk.
\end{align}
Next, irreducible tensors $S$, $B$ are constructed from the quantities with overbar by subtracting the partial traces such that $\delta_{ij}S_{ij}=\delta_{ij}B_{ijk}=0$. With a suitable ansatz and by making use of $J_iJ_kJ_i=\frac{11}{4}J_k$ we arrive at
\begin{align}
 \label{irr7} S_{ij} &= \bar{S}_{ij} -\frac{5}{2}\delta_{ij}\mathbb{1}_4,\\
 \label{irr8}  B_{ijk} &= \bar{B}_{ijk} - \frac{41}{10}\Bigl(\delta_{ij}J_k+\delta_{ik}J_j+\delta_{jk}J_i\Bigr).
\end{align}
These are the irreducible spin tensors of rank $\ell=2,3$ for spin $j=3/2$.

In App. \ref{AppTens} we show in detail that the irreducible tensors $S_{ij}$ and $B_{ijk}$ can be expressed as
\begin{align}
 \label{irr11} S_{ij} &= S_a \Lambda^a_{ij},\\
 \label{irr12} B_{ijk} &= B_\mu E^\mu_{ijk}
\end{align}
with $a=1,\dots,5$, $\mu=1,\dots,7$, and orthogonal basis tensors $\Lambda^a_{ij}$ and $E^\mu_{ijk}$. Since the spin tensors are matrix-valued, the components $S_a$ and $B_\mu$ are matrices as well. The desired basis $\{\Sigma^A\}$ in Eq. (\ref{short4}) is now constructed from the 1+3+5+7=16 elements $\mathbb{1}_4$, $J_i$, $S_a$, and $B_\mu$ after a proper normalization to satisfy
\begin{align}
 \label{short2} \mbox{tr}(\Sigma^A\Sigma^B)=4\delta^{AB}.
\end{align}
We define
\begin{align}
 \label{irr13} \mathJ_i &= \frac{2}{\sqrt{5}}J_i,\\
 \label{irr14} \gamma_a &=\frac{1}{\sqrt{3}}S_a,\\
 \label{irr15} W_\mu &=\frac{2}{3\sqrt{3}}B_\mu.
\end{align}
In particular, the five components $\gamma_a$ of the second rank tensor are the $\gamma$-matrices introduced in the context of the Luttinger Hamiltonian in Eq. (\ref{field4}). We explicitly have
\begin{align}
 \label{irr16} \gamma_1 &= \frac{1}{\sqrt{3}}(J_x^2-J_y^2) = \sigma_1 \otimes \mathbb{1},\\
 \label{irr17} \gamma_2 &= J_z^2-\frac{5}{4}\mathbb{1}_4 = \sigma_3\otimes \sigma_3,\\
 \label{irr18} \gamma_3 &= \frac{1}{\sqrt{3}}\{J_x,J_z\} = \sigma_3 \otimes \sigma_1,\\
 \label{irr19} \gamma_4 &= \frac{1}{\sqrt{3}}\{J_y,J_z\}=\sigma_3 \otimes \sigma_2,\\
 \label{irr20} \gamma_5 &=\frac{1}{\sqrt{3}}\{J_x,J_y\} =\sigma_2 \otimes \mathbb{1}.
\end{align}
Note that since  $J_y$ is purely imaginary, the matrices $\gamma_{1,2,3}$ are real, whereas $\gamma_{4,5}$ are imaginary. The seven components $W_\mu$ of the third-rank tensor are given by
\begin{align}
 \label{irr21} W_1 &= \frac{2\sqrt{5}}{3}\Bigl(J_{x}^3-\frac{41}{20}J_{x}\Bigr),\\
 \label{irr22} W_2 &= \frac{2\sqrt{5}}{3}\Bigl(J_{y}^3-\frac{41}{20}J_{y}\Bigr),\\
 \label{irr23} W_3 &= \frac{2\sqrt{5}}{3}\Bigl(J_{z}^3-\frac{41}{20}J_{z}\Bigr),\\
 \label{irr24} W_4 &= \frac{1}{\sqrt{3}}\{J_x,(J_y^2-J_z^2)\},\\
 \label{irr25} W_5 &= \frac{1}{\sqrt{3}}\{J_y,(J_z^2-J_x^2)\},\\
 \label{irr26} W_6 &= \frac{1}{\sqrt{3}} \{J_z,(J_x^2-J_y^2)\},\\
 \label{irr27} W_7 &= \frac{2}{\sqrt{3}}(J_xJ_yJ_z+J_zJ_yJ_x).
\end{align}
The operators $\mathbb{1}$ and $\gamma_a$ are even under time-reversal transformations \cite{PhysRevB.93.205138}, whereas $\mathJ_i$ and $W_\mu$ are odd. All operators feature inversion invariance. A nonzero expectation value of $\rho =\langle\psi^\dagger \mathbb{1}\psi\rangle$, $m_i=\langle\psi^\dagger\mathJ_i\psi\rangle$, and $\phi_a=\langle \psi^\dagger \gamma_a\psi\rangle$ corresponds to a nonzero density, magnetization, and nematic order, respectively. The order parameter $\chi_\mu = \langle\psi^\dagger W_\mu\psi\rangle$ constitutes a tensorial magnetization that does not point in a particular direction. We therefore suggest to refer to it as ``nemagnetic order". These observations are summarized in Table \ref{TabTensors}.

\begin{table}[t]
\begin{tabular}{|c||c|c|c|c|c|c|}
\hline $\Sigma^A$ & order & rank & \ \# \ &  $\mathcal{I}$ & $\mathcal{T}$ & cubic case \\
\hline \hline $\mathbb{1}$ & \begin{tabular}{c} density \\ $\rho = \langle \psi^\dagger \mathbb{1}\psi \rangle$ \end{tabular} & 0 &  1 & \ + \ & \ + \ & $\mathbb{1}$\\
\hline $\mathJ_i$ & \begin{tabular}{c} magnetic \\ $m_i = \langle \psi^\dagger \mathJ_i\psi \rangle$ \end{tabular} & 1 & 3 & + & $-$ & $\vec{\mathJ}=\begin{pmatrix} \mathJ_1 \\ \mathJ_2\\ \mathJ_3\end{pmatrix} $\\
\hline $\gamma_a$ & \begin{tabular}{c} nematic \\ $\phi_a = \langle \psi^\dagger \gamma_a\psi \rangle$ \end{tabular} & 2 & 5 & + & + & \begin{tabular}{c} $\vec{E}=\begin{pmatrix} \gamma_1\\ \gamma_2\end{pmatrix} $ \\ \hline $\vec{T}=\begin{pmatrix} \gamma_3\\ \gamma_4\\ \gamma_5\end{pmatrix}$ \end{tabular} \\
\hline \ $W_\mu$ \ & \ \begin{tabular}{c} nemagnetic \\ $\chi_\mu = \langle \psi^\dagger W_\mu\psi \rangle$ \end{tabular} \ & 3 & 7 & + & $-$ & \begin{tabular}{c} $\vec{W}=\begin{pmatrix}W_1\\ W_2\\W_3\end{pmatrix}$ \\ \hline $\vec{W}'=\begin{pmatrix} W_4\\ W_5 \\ W_6\end{pmatrix}$ \\ \hline $W_7$ (AIAO) \end{tabular} \\
\hline
\end{tabular}
\caption{Overview of local fermion bilinears  that feature full or cubic rotation invariance. We display the tensor rank with respect to $\text{SO}(3)$ and the number of independents components (\#) of these tensors. The scalar, vector, and rank-2 tensor order parameters $\langle \psi^\dagger\Sigma^A \psi \rangle$ constitute density, and the usual magnetic and nematic orders. We refer to a nonzero expectation value of $\chi_\mu=\langle \psi^\dagger W_\mu \psi\rangle$ as nemagnetic ordering. In particular, $\chi_7$ is the continuum version of the AIAO order on the pyrochlore lattice. We further indicate the behavior of these orders with respect to inversion ($\mathcal{I}$) and time-reversal ($\mathcal{T}$), where $+/-$ indicates even/odd transformation properties, and how the tensors split up into subgroups upon restricting rotations to the cubic group.}
\label{TabTensors}
\end{table}

\subsection{Local four-fermion couplings}

The most general local four-fermion interaction term in the rotation ($\delta=0$), inversion, and time-reversal invariant case is given by
\begin{align}
 \nonumber L_{\rm int} =\ &g_1(\psi^\dagger \psi)^2+g_J(\psi^\dagger \mathJ_i\psi)^2\\
 \label{rot1} &+g_2(\psi^\dagger \gamma_a\psi)^2+g_W(\psi^\dagger W_\mu\psi)^2.
\end{align}
The individual terms transform as scalar, vector, second- and third-rank tensors under $\text{SO}(3)$, respectively. For the latter two this becomes particularly transparent when writing $(\psi^\dagger \gamma_a\psi)^2 \propto (\psi^\dagger S_{ij}\psi)^2$ and $(\psi^\dagger W_\mu\psi)^2\propto (\psi^\dagger B_{ijk}\psi)^2$. There are two Fierz identities
\begin{align}
 \nonumber 0 &=5(\psi^\dagger\psi)^2 +(\psi^\dagger\mathcal{J}_i\psi)^2+ (\psi^\dagger\gamma_a\psi)^2+(\psi^\dagger W_\mu\psi)^2,\\
 \label{rot3}  0 &= \frac{1}{3}(\psi^\dagger\mathcal{J}_i\psi)^2- \frac{1}{7}(\psi^\dagger W_\mu\psi)^2,
\end{align}
which reveal that the expression (\ref{rot1}) contains a certain degree of redundancy that can be removed by eliminating two of the terms. Due to Eq. (\ref{rot3}), one of them needs to be either $(\psi^\dagger \mathJ_i\psi)^2$ or $(\psi^\dagger W_\mu\psi)^2$. Here we choose to eliminate both of them and thus arrive at the Fierz complete interaction term
\begin{align}
 \label{rot4} L_{\rm int} = g_1(\psi^\dagger \psi)^2+g_2(\psi^\dagger \gamma_a\psi)^2.
\end{align}
Together with the Lagrangian from Eq. (\ref{field2}) this constitutes the field theoretic setup considered in Ref. \cite{PhysRevLett.113.106401}.

To study the cubic symmetric case we introduce
\begin{align}
 \label{flow1} \vec{E}=\begin{pmatrix} \gamma_1 \\ \gamma_2 \end{pmatrix},\ \vec{T}= \begin{pmatrix} \gamma_3 \\ \gamma_4 \\ \gamma_5 \end{pmatrix},\ \vec{W}=\begin{pmatrix} W_1 \\ W_2 \\ W_3 \end{pmatrix},\ \vec{W}' = \begin{pmatrix} W_4 \\ W_5 \\ W_6\end{pmatrix}.
\end{align}
The most general cubic, inversion, and time-reversal symmetric local four-fermion term is given by
\begin{align}
 \label{flow2} L_{\rm int} = \sum_{i=1}^8 g_i L_i,
\end{align}
with
\begin{align}
 \label{flow3} L_1 &= (\psi^\dagger \psi)^2,\\
 \label{flow4} L_2 &= (\psi^\dagger \vec{E}\psi)^2,\\
 \label{flow5} L_3 &= (\psi^\dagger \vec{T}\psi)^2,\\
 \label{flow6} L_4 &= (\psi^\dagger \vec{\mathJ}\psi)^2,\\
 \label{flow7} L_5 &= (\psi^\dagger \vec{W}\psi)^2,\\
 \label{flow8} L_6 &= (\psi^\dagger \vec{W}'\psi)^2,\\
 \label{flow9} L_7 &= (\psi^\dagger W_7\psi)^2,
\end{align}
and
\begin{align}
 \label{flow10} L_{8} &= (\psi^\dagger \vec{\mathJ}\psi)\cdot(\psi^\dagger\vec{W}\psi).
\end{align}
Each of the terms $L_i$ transforms as a singlet under the action of the cubic group. There are also five Fierz identities among the $L_i$ (App. \ref{AppFierz}), allowing us to eliminate five couplings. We are thus left with \emph{three independent couplings}, which we choose to  be $g_{1,2,3}$. Hence
\begin{align}
 \label{flow16} L_{\rm int} =g_1(\psi^\dagger\psi)^2 + g_2(\psi^\dagger \vec{E}\psi)^2 + g_3 (\psi^\dagger \vec{T}\psi)^2
\end{align}
constitutes a Fierz complete interaction term. For $g_2=g_3$ it reduces to the expression in Eq. (\ref{rot4}). The short-range interaction terms $L_i$ can also be evoked for the study of other systems with four-component fermions, for instance having linear dispersion, such as Weyl semimetals \cite{PhysRevB.90.035126} or quantum critical antiperovskites \cite{PhysRevB.92.081304,PhysRevB.93.241113}.

Apart from the eight fermionic vertices that appear in Eq. (\ref{flow2}) one may wonder whether the combinations $L_9=(\psi^\dagger \vec{T}\psi)\cdot(\psi^\dagger\vec{\mathJ}\psi)$, $L_{10}=(\psi^\dagger \vec{T}\psi)\cdot(\psi^\dagger\vec{W}\psi)$, $L_{11}=(\psi^\dagger \vec{T}\psi)\cdot(\psi^\dagger\vec{W}'\psi)$, $L_{12}=(\psi^\dagger \vec{\mathJ}\psi)\cdot(\psi^\dagger\vec{W}'\psi)$, and $L_{13}=(\psi^\dagger \vec{W}\psi)\cdot(\psi^\dagger\vec{W}'\psi)$ can also be generated during the RG flow. This, however, is forbidden by cubic and time-reversal symmetry ($\mathcal{T}$), as can be seen as follows: Since $\vec{T}$ is even under $\mathcal{T}$, but $\vec{\mathJ}$, $\vec{W}$, and $\vec{W}'$ are odd, the terms $L_{9-11}$ explicitly break $\mathcal{T}$. Due to the original Lagrangian in Eq. (\ref{field2}) being time-reversal symmetric, no such term can be generated during the RG flow. The terms $L_{12,13}$ are forbidden by cubic symmetry. For this consider a rotation around the z-axis by $\pi/2$. The coordinate vector $\vec{x}=(x,y,z)^{\rm t}$ transforms as $\vec{x}\to(y,-x,z)^{\rm t}$. In the same way, $\vec{\mathJ}$ and $\vec{W}$ transform as $(\mathJ_1,\mathJ_2,\mathJ_3)^{\rm t}\to(\mathJ_2,-\mathJ_1,\mathJ_3)^{\rm t}$ and $(W_1,W_2,W_3)^{\rm t}\to (W_2,-W_1,W_3)^{\rm t}$, respectively. Accordingly, the term $L_8=(\psi^\dagger\vec{\mathJ}\psi)\cdot(\psi^\dagger\vec{W}\psi)$ is both time-reversal and cubic symmetric, and indeed emerges in the present RG analysis. On the other hand, $\vec{W}'$ transforms under the same rotation as $(W_4,W_5,W_6)\to(-W_5,W_4,-W_6)$, so that $L_{12,13}\to -L_{12,13}$. Again, since the original Lagrangian is cubic symmetric, the term $L_{12,13}$ cannot be generated during the RG evolution.

In the isotropic case, the four fermionic vertices appearing in Eq. (\ref{rot1}) are chosen such that they have distinct transformation properties under $\text{SO}(3)$. Accordingly, no mixing between these terms is possible due to symmetry. This changes in the cubic $\text{O}_{\rm h}$-invariant case, where we observe the term $L_8$ in Eq. (\ref{flow10}) to couple $\psi^\dagger \mathJ_i\psi$ and $\psi^\dagger W_i \psi$. In fact, for $\delta\neq 0$ the tensors $\mathJ_i$ and $W_i$ have exactly the same symmetries and are thus physically equivalent. Furthermore, every orthogonal pair of linear combinations of $\mathJ_i$ and $W_i$ may also be chosen for a basis. Here we define
\begin{align}
 \label{flow16c} U_i &= \frac{1}{\sqrt{5}}(2\mathJ_i-W_i),\\
 \label{flow16b} V_i &= \frac{1}{\sqrt{5}} (\mathJ_i+2W_i).
\end{align}
In Ref. \cite{PhysRevB.93.241113} these tensors are labelled $\vec{U}=\vec{\gamma}_{\rm s}$ and $\vec{V}=\vec{\gamma}_{\rm d}$, respectively. As further pointed out in the latter reference, the $V_i$ satisfy the three-dimensional Clifford algebra
\begin{align}
 \label{flow16d} \{ V_i,V_j\} = 2\delta_{ij}\mathbb{1}_4.
\end{align}
They are also generators of an $\text{SU}(2)$ algebra. Due to this extra symmetry, the linear combination $\vec{V} \propto \vec{\mathJ}+2\vec{W}$ is distinguished for $\delta \neq 0$. The second linear combination, $\vec{U}\propto 2\vec{\mathJ}-\vec{W}$, then follows as the orthogonal partner, but does not possess any further symmetries. Note that in terms of the original $J_i$ we have
\begin{align}
 U_i &= \frac{1}{6}(13 J_i-4J_i^3),\\
 \label{flow16e} V_i &= \frac{1}{3}(-7J_i+4J_i^3).
\end{align}
The pseudospin variable $\vec{I}$ introduced in Eq. (10) of Ref. \cite{savary} coincides with $(\psi^\dagger \vec{V}\psi)$ up to a prefactor. Further note that in the parametrization of Ref. \cite{murray}, where the magnetic channel is written as $\psi^\dagger(\cos\tilde{\alpha} J_i+\sin\tilde{\alpha} J_i^3)\psi$, $V_i$ corresponds to $\tilde{\alpha}=\arctan(-4/7)+\pi=2.62$, which unfortunately falls outside the range $\tilde{\alpha}\in[0,\pi/2]$ considered in the reference.

The Luttinger semimetal considered in this work may be understood as the low-energy effective field theory describing itinerant electrons on a pyrochlore lattice such as in Pyrochlore Iridates. The nemagnetic orders $\chi=\langle\psi^\dagger W_7\psi\rangle$ and $v_i=\langle\psi^\dagger V_i\psi\rangle$ can then be associated to particular orders of the local magnetic moments of electrons on the four corners of the tetrahedra forming the pyrochlore lattice. As shown in Ref. \cite{Goswami:2016fjn}, a nonzero expectation value of $\chi$ or one component of $\vec{v}$ corresponds to an all-in-all-out (AIAO) or spin ice (SI) ordering, respectively. (The reference uses the notation $\chi\to \varphi$ and $v_i\to M_i$.) The correspondence is facilitated by observing that
\begin{align}
 (\psi^\dagger W_7\psi)^2&=(\psi^\dagger \gamma_{12}\psi)^2,\\
 (\psi^\dagger \vec{V}\psi)^2&=(\psi^\dagger \gamma_{34}\psi)^2+(\psi^\dagger\gamma_{35}\psi)^2+(\psi^\dagger\gamma_{45}\psi)^2
\end{align}
from Eqs. (\ref{tapp23}) and (\ref{rg42}). For this reason we will refer to the tensor orders corresponding to $W_7$ and $V_i$ as AIAO and SI ordering, although, of course, these notions only make sense on the pyrochlore lattice.

\subsection{Renormalization group flow}

\begin{figure}[t]
\centering
\includegraphics[width=8.5cm]{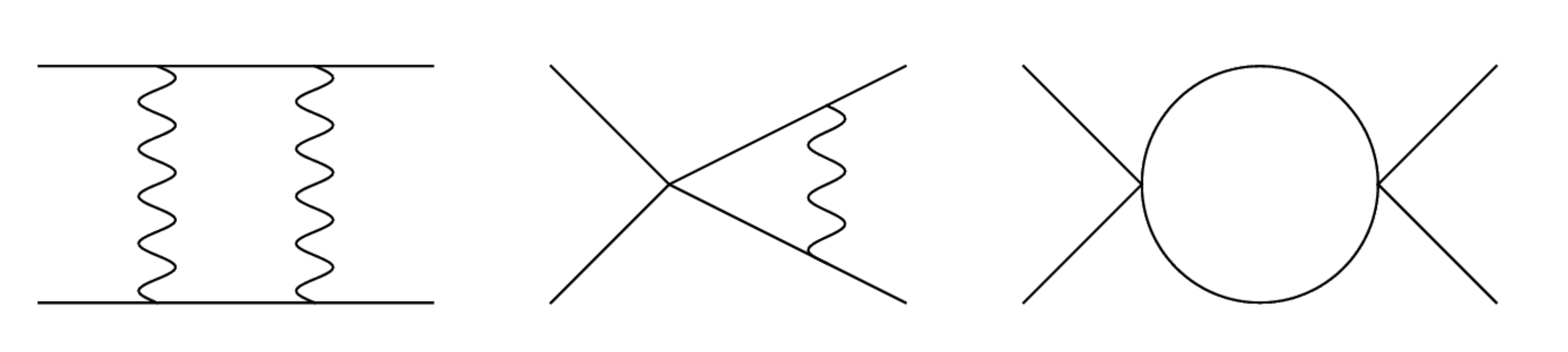}
\caption{Schematic loop contributions to the short-range interactions. The assignment of lines is as in Fig. \ref{FigLoops}, i.e., a straight (wiggly) line represents a fermion (photon) propagator. The box diagram to the left is proportional to $e^4$ and generates local short-range interactions even if they are initially absent. Once present, they contribute to the flow of the short-range couplings $\{g_i\}$ via the diagrams shown in the middle and to the right, which are proportional to $e^2g_i$ and $g_ig_j$, respectively.}
\label{FigLoop4Fermi}
\end{figure}

During the RG flow, short-range interactions are generated from long-range interactions by the diagram to the left in Fig. \ref{FigLoop4Fermi}. More explicitly, we have a nonvanishing term proportional to $e^4$ in Eqs. (\ref{flow19}) and (\ref{flow20}) below for the flow of $g_2$ and $g_3$. This generates $g_2$ and $g_3$, and they eventually also generate $g_1$. Once the $\{g_i\}$ are present, they couple via the remaining two diagrams in the figure, and thereby lead to a sufficiently rich fixed point structure of the flow.

The flow equations for the couplings $g_{1,2,3}$ are given by
\begin{align}
 \label{flow17} \dot{g}_i =(z-d)g_i +f_1(\delta)\cdot \Delta g_i
\end{align}
with $z=2-\eta$ as in Eqs. (\ref{field14}) and (\ref{field16}) and
\begin{align}
  \nonumber \Delta g_1 =\ & -\frac{4}{5}F_-\Bigl(g_1+\frac{e^2}{2}\Bigr)g_2-\frac{6}{5}F_+\Bigl(g_1+\frac{e^2}{2}\Bigr)g_3\\
   \label{flow18} &-g_2^2-6g_2g_3-3g_3^2,\\
   \nonumber \Delta g_2 =\ & -\frac{1}{5}F_-\Bigl(g_1+\frac{e^2}{2}\Bigr)^2+\frac{1}{5}(5+3F_+)\Bigl(g_1+\frac{e^2}{2}\Bigr)g_2\\
 \label{flow19}  &-3g_2^2-3(1+F_+)g_2g_3-\frac{3}{5}(5+F_-)g_3^2,\\
   \nonumber \Delta g_3 =\ & -\frac{1}{5}F_+\Bigl(g_1+\frac{e^2}{2}\Bigr)^2 +\frac{2}{5}(5-F_+)\Bigl(g_1+\frac{e^2}{2}\Bigr)g_3\\
  \nonumber  & -\frac{1}{5}(5+2F_+)g_2^2-2(4-F_+)g_2g_3\\
 \label{flow20}  &-\frac{2}{5}(15-F_--F_+)g_3^2.
\end{align}
The anisotropy parameter $\delta$ enters through the functions $f_1$ and $F_\pm$. We define the latter by
\begin{align}
  \label{flow21} F_-\equiv F_-(\delta) &= \frac{(1-\delta)f_{2\rm e}(\delta)}{f_1(\delta)},\\
  \label{flow22} F_+\equiv F_+(\delta) &= \frac{(1+\delta)f_{2\rm t}(\delta)}{f_1(\delta)}.
\end{align}
We have $F_+=F_-=1$ for $\delta=0$ and
\begin{align}
 \label{flow23} 2F_-(\delta) +3F_+(\delta) =5
\end{align}
for all $\delta$. In particular, this implies
\begin{align}
 \label{flow24} &F_-(-1)=\frac{5}{2},\ F_+(+1)=\frac{5}{3}
\end{align}
in the limits of strong anisotropy, since obviously $F_-(+1)=F_+(-1)=0$ from the very definition.

The flow of the couplings $g_i$ is supplemented by the flow equation for $e^2$ given in Eq. (\ref{field17}), namely
\begin{align}
 \label{flow24b} \dot{e}^2 = (z+2-d)e^2 - \frac{f_{e^2}(\delta)}{1-\delta^2}e^4.
\end{align}
Since the flow equation for $e^2$ is not altered by the short-range interactions, any possible fixed point in the space of couplings $(g_1,g_2,g_3,e^2)$ necessarily has either $e^2=0$ or $e^2=e^2_\star$ with $e^2_\star$ from Eq. (\ref{field18}). Note that the $\beta$-functions (\ref{flow18})-(\ref{flow20}) depend on $g_1$ and $e^2$ only through the combination $g_1+\frac{e^2}{2}$. In App. \ref{AppRG} we show that this behavior results from the fact that the frequency integral of the squared fermion propagator vanishes.

In the isotropic limit ($\delta=0$ and $g_2=g_3$) we recover the flow equations of Ref. \cite{PhysRevLett.113.106401} given by
\begin{align}
 \nonumber \dot{g}_1 =\ &(z-d)g_1 -2\Bigl(g_1+\frac{e^2}{2}\Bigr)g_2-10g_2^2,\\
 \nonumber \dot{g}_2 =\ &(z-d)g_2 -\frac{1}{5}\Bigl(g_1+\frac{e^2}{2}\Bigr)^2+\frac{8}{5}\Bigl(g_1+\frac{e^2}{2}\Bigr)g_2-\frac{63}{5}g_2^2,\\
 \label{flow25} \dot{e}^2 =\ & (z+2-d)e^2 - e^4.
\end{align}
Here we use a different convention for defining the renormalized couplings $g_i$ and $e^2$ than in the reference, see the comment below Eq. (\ref{scal4}) for a mapping. The flow equations (\ref{flow25}) only support the Abrikosov fixed point for $d>d_{\rm c}=3.26$. In $d_{\rm c}$ dimensions it annihilates with a quantum critical point, and consequently is absent for lower dimensions.

As the anisotropy $\delta$ is varied within the interval $\delta\in[-1,1]$, various new fixed points in the space $\{G_i\}=(g_1,g_2,g_3,e^2)$ appear as solutions of the RG flow equations. These fixed points typically have several relevant directions and their impact on the phase structure will be discussed below. In order to uniquely identify the Abrikosov fixed point in this zoo of fixed points we define it as the one having only irrelevant directions. (In the four-dimensional coupling space $\{G_i\}$ this corresponds to four negative eigenvalues of the stability matrix $\mathcal{M}_{ij}=\partial \beta_i/\partial G_j|_\star$ at the fixed point.) The fixed point defined in this manner indeed satisfies $e^2=e^2_\star>0$ and thus leads to NFL behavior.

\begin{figure}[t]
\centering
\includegraphics[width=8.5cm]{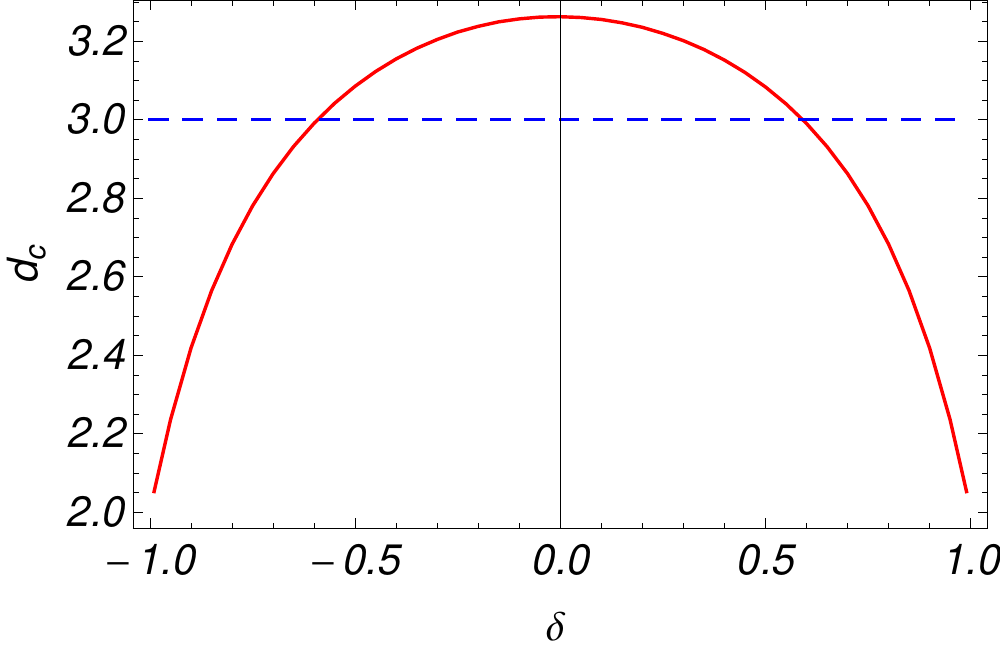}
\caption{Critical dimension $d_{\rm c}$ for survival of the Abrikosov fixed point. The curve crosses three dimensions for $|\delta|=\delta_{\rm c}=0.59$ (dashed line). To get a qualitative understanding of why anisotropy supports NFL behavior, observe that the charge fixed point is located at $e^2_\star\approx \frac{15}{19}(1-\delta^2)\vare$. A large anisotropy thus acts like a small $\vare$.}
\label{FigDC}
\end{figure}

We find that the Abrikosov fixed point survives in three spatial dimensions if the anisotropy is increased beyond a critical value according to $|\delta|\geq\delta_{\rm c}=0.59$. Interestingly, this value holds for both signs of $\delta$. We plot the result for the critical dimension for survival of the Abrikosov fixed point $d_{\rm c}(\delta)$ in Fig. \ref{FigDC}. The function $d_{\rm c}(\delta)$ is found numerically to be almost perfectly symmetric with respect to $\delta\to -\delta$. The slight asymmetry might be real or due to the numerical determination of the functions $f_i(\delta)$. The fact that the fixed point annihilation takes place at a lower critical dimension $d_c$ can be understood by recalling that the Abrikosov fixed point in the anisotropic case is located at $e^2_\star \approx \frac{15}{19}(1-\delta^2)\vare$. Accordingly, a larger $\delta$ is analogous to a smaller $\vare$, and thus anisotropy assists the NFL fixed point.

\subsection{Instabilities and anisotropy-induced fixed points}

So far we have linked the Abrikosov fixed point to NFL behavior of the system. In order to understand the phase structure that is implied by the other fixed points in the space of couplings we compute the order parameter susceptibilities at the remaining fixed points.

For this note that a fixed point of the RG flow corresponds to a (possibly fine-tuned) second order phase transition of the system. At such a phase transition, fluctuations of the order parameter become critical, which is indicated by a divergent order parameter susceptibility. Given a certain fixed point located at $(g_1,g_2,g_3,e^2)_\star$, it is nontrivial to deduce its ordering tendency, especially in our case where the couplings $(g_1,g_2,g_3)$ are obtained from eight couplings after applying five Fierz transformations. However, upon computing the susceptibility for each individual order parameter, we can deduce the  instability associated with every fixed point.

To compute the susceptibility of an order parameter of interest, $\Phi$, which shall be parametrized by a matrix $M$ through $\Phi=\langle\psi^\dagger M\psi\rangle$, we add  an additional source term
\begin{align}
 \label{sus1} L_\Phi \sim \Delta (\psi^\dagger M \psi)
\end{align}
to the Lagrangian, and study whether it is enhanced or suppressed  during the RG flow. In particular, a sufficiently strong divergence of $\Delta$ signals an instability. To see this note that the scaling dimension of $\Delta$, in the case that $z=2-\eta$, is given by $[\Delta]=z$, see App. \ref{AppScal}. Let $k$ be a momentum scale such that the infrared fixed point is approached for $k\to 0$. (For instance, we may choose $k\sim r^\nu$ with critical exponent $\nu$ and mass of the order parameter $r\to 0$ at the transition.) The free energy density close to the fixed point then has the scaling form
\begin{align}
 \label{sus2} \mathcal{F} = k^{d+z} H \Bigl(\frac{\Delta}{k^{z+\eta_\Phi}}\Bigr),
\end{align}
where $H(\cdot)$ is some scaling function, and $\eta_\Phi$ is obtained from the RG flow of $\Delta$ via $\dot{\Delta}=(z+\eta_\Phi)\Delta$. The susceptibility of $\Phi$ close to the fixed point is then given by
\begin{align}
 \label{sus3} \chi = \frac{\partial^2\mathcal{F}}{\partial \Delta^2} = k^{d-z-2\eta_\phi} H''(\infty)\ \text{as}\ k\to0.
\end{align}
We observe that a divergent susceptibility requires a sufficiently large $\eta_\Phi$ given by
\begin{align}
 \label{sus4} \eta_\Phi > \frac{d-z}{2}.
\end{align}
In the absence of a small parameter, more than one or none of the order parameter susceptibilities may satisfy this criterion. In this case we will identify the one with the largest exponent as the most likely to be the leading instability \cite{LukasPaper}.

In order to study insulating order parameter susceptibilities we use the source terms
\begin{align}
  \nonumber L_{\mathbb{1}} &= \Delta (\psi^\dagger \mathbb{1}\psi),\ L_{E_a} = \Delta (\psi^\dagger E_a \psi),\\
  \nonumber L_{T_a} &= \Delta (\psi^\dagger T_a \psi),\ L_{\mathJ_i} = \Delta (\psi^\dagger \mathJ_i \psi),\\
  \nonumber L_{W_i} &= \Delta (\psi^\dagger W_i\psi),\ L_{U_i} = \Delta (\psi^\dagger U_i \psi),\\
 \nonumber L_{V_i} &= \Delta (\psi^\dagger V_i \psi),\ L_{W'_\mu} = \Delta (\psi^\dagger W'_\mu \psi),\\
  \label{sus5} L_{W_7} &= \Delta (\psi^\dagger W_7 \psi).
\end{align}
We also study the susceptibilities with respect to superconducting order parameters by means of
\begin{align}
\label{sus5b} L_{\gamma_{45}}^{(\rm sc)} = \Delta (\psi^\dagger \gamma_{45}\psi^*),\ L_{\gamma_a\gamma_{45}}^{(\rm sc)} = \Delta (\psi^\dagger \gamma_a\gamma_{45}\psi^*).
\end{align}
They represent s-wave and d-wave superconducting orders \cite{PhysRevB.93.205138}. In general, coupling one of the terms (\ref{sus5})-(\ref{sus5b}) to the Lagrangian only generates exactly the same term to linear order in $\Delta$. This is dictated by the behavior under cubic, inversion, and time-reversal symmetry transformations of the individual terms, see also the discussion below Eq. (\ref{flow16}). However, an exception is given by the magnetic vertices labelled with $i=1,2,3$. They are, in fact, fully equivalent in the case of $\delta\neq 0$. Accordingly, introducing either $L_{\mathJ_i}$ or $L_{W_i}$ to the Lagrangian generates both $L_{\mathJ_i}$ and $L_{W_i}$. In order to find the most unstable direction we therefore consider the general term $L_{M_i}= \Delta (\psi^\dagger M_i\psi)$ with
\begin{align}
 \label{sus6} M_i =\alpha \mathJ_i + \beta W_i,
\end{align}
where $\alpha,\beta\in\mathbb{R}$ are such that $\alpha^2+\beta^2=1$. We tune $\alpha$ and $\beta$ such that coupling $L_{M_i}$ only generates $L_{M_i}$. This yields two possible solutions for $(\alpha,\beta)$, the more strongly divergent one of them being the leading instability. The corresponding analysis is performed in App. \ref{AppRGSus}, where we show that the two solutions for $M_i$ are precisely given by $U_i$ and $V_i$ defined in Eqs. (\ref{flow16c}) and (\ref{flow16b}). The fact that $(\psi^\dagger V_i \psi)$ does not generate $(\psi^\dagger U_i\psi)$, and vice versa, can be understood from the enhanced symmetry of the matrices $V_i$ as formulated in Eq. (\ref{flow16d}).

For our purposes we only need to focus on fixed points with a small number of relevant directions, defined as the number of positive eigenvalues of the stability matrix at the fixed point. We refer to a fixed point as quantum critical or bicritical if it has one or two relevant directions. To see why quantum critical points (QCPs) can be important consider an RG trajectory connecting a QCP (Q) to the fully infrared attractive Abrikosov fixed point (A) along the relevant direction of (Q). In coupling space, this line can be parametrized by some effective coupling constant $\tilde{g}$, with (A) and (Q) located at $\tilde{g}=0$ and $\tilde{g}=\tilde{g}_{\rm c}>0$, respectively. For small $\tilde{g}<\tilde{g}_{\rm c}$, the RG flow will be attracted to (A), and the ground state is an NFL. However, if $\tilde{g}$ exceeds the critical value $\tilde{g}_{\rm c}$, the RG flow is repelled from (Q) -- in the opposite direction. This runaway flow towards strong coupling signals an instability of the system towards an ordered ground state. The supercritical value of $\tilde{g}$ may be realized in actual materials through strong on-site interactions. The relevance of some bicritical points in our setup stems from the fact that a bicritical point in the space $(g_1,g_2,g_3,e^2)$ corresponds to a QCP in the plane spanned by $(g_1,g_2,g_3)$ for $e^2=0$, because charge is always a relevant direction. Hence by setting $e^2=0$ we can in this way study QCPs of charge neutral systems.

\begin{figure}[t!]
\centering
\begin{minipage}{0.46\textwidth}
\includegraphics[width=8.5cm]{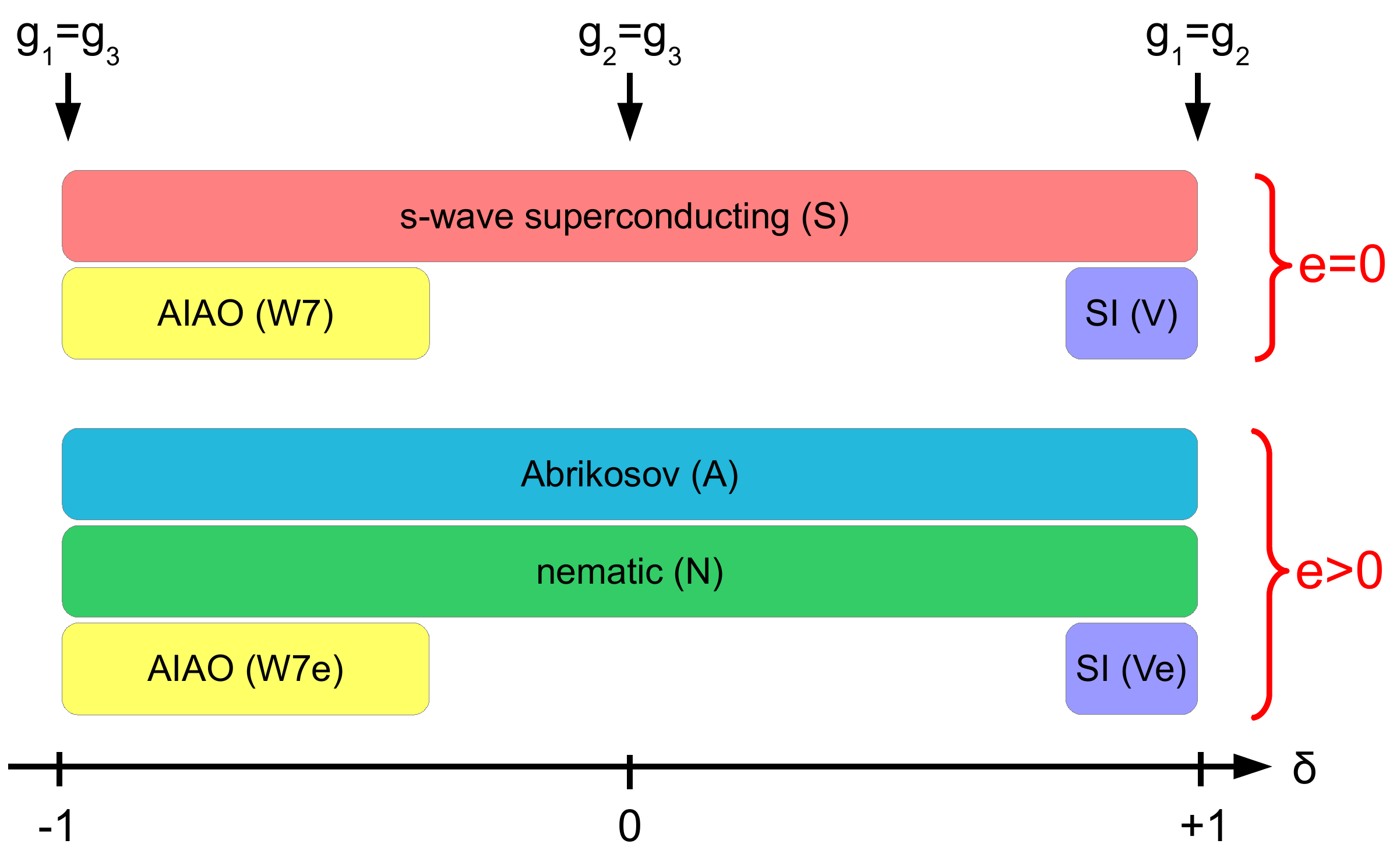}
\caption{Schematic fixed point structure of the Luttinger semimetal for $0<\vare\ll 1$ as a function of the (assumed constant) anisotropy parameter $\delta\in[-1,1]$. We only show the fixed points with a small number of relevant directions, namely QCPs of the neutral system (upper panel) and QCPs of the charged system (lower panel). The Abrikosov fixed point is fully attractive. For sufficiently strong anisotropy, new pairs of fixed points appear that are related to second order phase transitions into a $W_7$ or $V_i$ ordered state, respectively. On the pyrochlore lattice, the latter two orders correspond to all-in-all-out (AIAO) or spin ice (SI) configurations of local magnetic moments of the electrons on the tetrahedra. We also indicate the emergent fixed ratios of the couplings $g_i$ at the fixed points $\delta_\star=0,\pm1$ of the anisotropy.}
\label{FigFPa}\vspace{3mm}
\includegraphics[width=8.5cm]{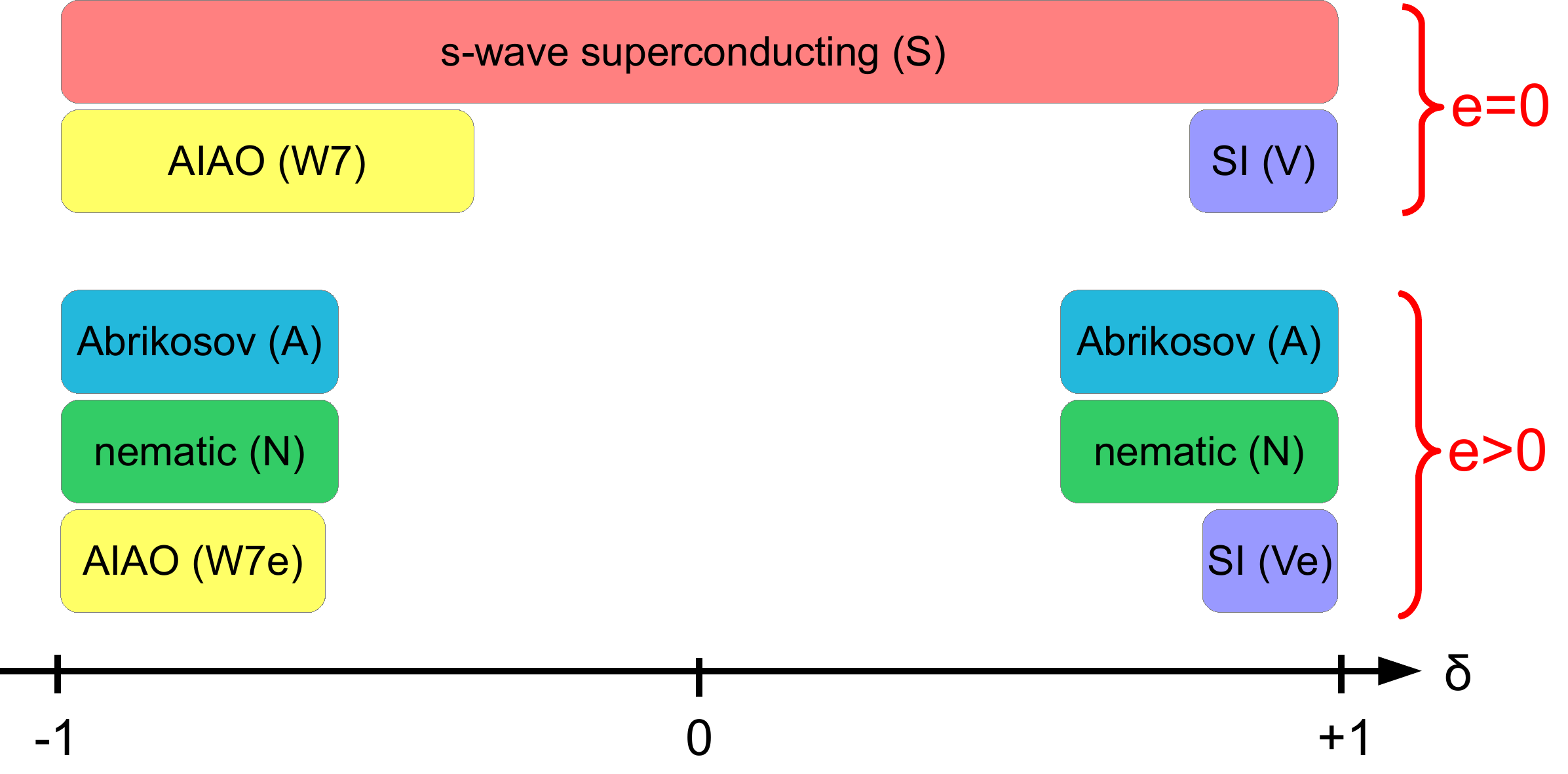}
\caption{Extrapolation of the above picture to three dimensions ($\vare=1$). We observe the nematic and Abrikosov fixed points to annihilate for $|\delta|<\delta_{\rm c}=0.59$. Also the survival of the anisotropy-induced fixed points is slightly influenced by a large value of $\vare$, but they can always be induced through a sufficiently strong anisotropy.}
\label{FigFPb}
\end{minipage}
\end{figure}

The fixed points of the systems with one or two relevant directions can be divided into two sets. The first set comprises the Gaussian fixed point (G), an s-wave superconducting fixed point of the neutral system (S), and the nematic (N) and Abrikosov fixed points (A) which participate in the collision scenario. These fixed points are always present for small $\vare>0$, irrespective of the value of $\delta\in[-1,1]$. Whether they survive the extrapolation to $\vare=1$ depends to some extent on the value of $\delta$. The second set consists of qualitatively new fixed points that show up for sufficiently strong anisotropy. They represent critical points towards $W_7$ order for $\delta \to -1$ and $V_i$ order for $\delta\to 1$. Their survival for $\vare= 1$ also depends on the value of $\delta$, but can always be enforced by a sufficiently strong anisotropy. This behavior is visualized in Figs. \ref{FigFPa} and \ref{FigFPb}.

A common feature of all fixed points -- \emph{independent of their associated diverging channel} -- is the following: At the fixed points of the anisotropy, $\delta_\star=0,\pm1$, the three couplings $(g_1,g_2,g_3)$ are not independent. In fact, from the flow equations (\ref{flow18})-(\ref{flow20}) one easily sees that even if all three couplings are different at some stage of the RG, they are attracted towards $g_2-g_3\to0$ (for $\delta=0$), $g_1-g_3\to0$ (for $\delta=-1$), and $g_1-g_2\to0$ (for $\delta=1$) in the infrared. For example, the superconducting fixed point (S) for $\delta=0$ and $d=3$ is located at
\begin{align}
 \label{sus7} (S)_{\delta=0}:\  (g_1,g_2,g_3,e^2)_\star = (-0.10,-0.09,-0.09,0)
\end{align}
and the linearized flow of $g_-=(g_2-g_3)$ at the fixed point reads $\dot{g}_-\simeq -0.87 g_-$. Accordingly, $g_-$ diminishes in the infrared. In contrast, in the anisotropic limits the fixed point (S) is located at
\begin{align}
 \nonumber (S)_{\delta=-1}&:\ (g_1,g_2,g_3,e^2)_\star = (-0.09,-0.07,-0.09,0),\\
 \label{sus8} (S)_{\delta=+1}&:\ (g_1,g_2,g_3,e^2)_\star = (-0.12,-0.12,-0.10,0),
\end{align}
respectively. If $\delta$ is only close to any of its three fixed points, the corresponding $g_i$'s are approximately equal.

After this remark, we begin by discussing the first set of fixed points. Besides the Gaussian fixed point (G), the s-wave superconducting fixed point (S) with $e^2=0$ is present for all $\delta$ and all $0<\vare\leq 1$. Since (S) is bicritical, it is mostly of importance for charge neutral systems. The corresponding critical behavior, including oscillatory corrections to scaling and exceptionally slow flow towards isotropy, has been discussed in Ref. \cite{PhysRevB.93.205138} by the present authors. The quantum critical nematic and fully attractive Abrikosov fixed points with $e^2=e^2_\star>0$, (N) and (A), are present for small $\vare>0$, but they annihilate each other above $d\geq 3$ for $|\delta|\leq0.59$ as discussed in the previous section. It is always (N) that collides with (A). Further, as $\delta\neq 0$, (N) changes its divergence character: Whereas the leading instability is towards $T_a$ for $\delta <0$, it is towards $E_a$ for $\delta>0$. Of course, for $\delta=0$ these two are related by the full rotational symmetry.

The second set of fixed points only appears for sufficiently strong anisotropy. To understand their nature, two important things should be noted. For one, since the parameter space spanned by $(g_1,g_2,g_3,e^2)$ is vast, the new fixed points do not interfere with any of the ones of the first set. In particular, they do not influence the fixed point collision scenario. Further, since the charge fixed point value $e^2_\star$ is small for large anisotropy, it is natural to expect that a zero of the beta functions located at $(g_1,g_2,g_3,e^2)_\star$ implies another zero at approximately $(g_1,g_2,g_3,0)_\star$ with $e^2=0$. This is indeed true for the additional fixed points found here. For most values of $\delta$ (but not all) there exist always two anisotropy-induced fixed points -- a QCP with $e^2_\star>0$ describing a quantum phase transition of the charged system, and a qualitatively similar bicritical point with $e^2=0$, which is a QCP for the neutral system. Since both fixed points are even mostly quantitatively similar we can simplify the discussion of their nature by considering the neutral charge fixed point with $e^2=0$.

For $\delta\leq -0.30$, a pair of fixed points with instability towards $W_7$ appears for small $\vare>0$. We name the fixed points with $e^2>0$ and $e^2=0$ (W7e) and (W7), respectively. When extrapolated to $\vare=1$, (W7e) survives for $\delta\leq -0.62$, whereas (W7) is more robust and survives for $\delta\leq -0.30$. In order to analyze the properties of (W7) for $d=3$, we set $e^2=0$ and $\delta=-1$ in the flow equations. In particular, this implies $f_1(-1)=1.094$, $F_+=0$, and $F_-=\frac{5}{2}$. The fixed point is then located at
\begin{align}
 \label{sus9} (W7)_{\delta=-1}:\ (g_1,g_2,g_3,e^2)_\star = (0.16,-0.21,0.16,0).
\end{align}
We see that, indeed, this satisfies $g_1=g_3$. The only divergent susceptibility corresponds to $W_7$ and is given by
\begin{align}
 \label{sus10} \eta_{W_7} = 2f_1(2g_1-g_2)_\star = 1.15.
\end{align}
Note that the value of $f_1$ is not important for $e=0$ since it can be absorbed into the couplings by means of $g_i \to f_1 g_i$. The fixed point values $(g_1,g_2,g_3,e^2)_\star$ of (W7e) approach those of (W7) for $\delta \to -1$, and coincide with the ones quoted in Eq. (\ref{sus9}) in the fully anisotropic limit.

For $\delta\geq 0.91$, a pair of fixed points towards ordering in $V_i$ appears for small $\vare>0$. Again, it is comprised of a quantum critical and a bicritical point, and we label them (Ve) and (V) in analogy to the previous paragraph. When extrapolated to $\vare=1$, (Ve) and (V) survive for $\delta\geq 0.94$ and $\delta\geq 0.91$, respectively. To better understand (V) in $d=3$, we set $e^2=0$ and $\delta=1$ in the flow equations. This implies $f_1(1)=0.813$, $F_+=\frac{5}{3}$, and $F_-=0$. We obtain the fixed point at
\begin{align}
 \label{sus11} (V)_{\delta=+1}:\ (g_1,g_2,g_3,e^2)_\star = (0.37,0.37,-0.28,0).
\end{align}
Also here, $g_1=g_2$ is satisfied in accordance with our previous statements. The leading instability is towards $V_i$ with exponent
\begin{align}
 \label{sus12} \eta_{V_i}= \frac{2}{3}f_1(3g_1-g_3)_\star = 0.75.
\end{align}
There is a subleading divergence in the $T_a$ channel with $\eta_{T_a}=-\frac{2}{3}f_1(g_1+5g_3)_\star=0.56$. For $\delta \to 1$, the couplings ($g_1,g_2,g_3,e^2)_\star$ of (Ve) approach those of (V), and coincide with Eq. (\ref{sus11}) for full anisotropy.

An interesting aspect of the limit $|\delta|\to1$ is that the fixed points (S) and (N) approach each other. This is facilitated by $e^2_\star\to0$ for $|\delta|\to1$. With the same substitutions as in the previous paragraphs it is easy to see that
\begin{align}
 \nonumber [(S)=(N)]_{\delta=-1}&:\ \eta_{\gamma_{45}}^{(\rm sc)}=\eta_{T_a} = -2f_1(2g_1+g_2)_\star =0.54,\\
 \label{sus13}  [(S)=(N)]_{\delta=+1}&:\ \eta_{\gamma_{45}}^{(\rm sc)}=\eta_{E_a} = -3f_1(g_1+g_3)_\star =0.55,
\end{align}
for $\delta\to\mp1$, respectively. The coincidence of the two divergent susceptibilities may hint at an enlarged symmetry group at the fixed point. We leave that aspect for future investigations.

\section{Discussion}\label{SecConcl}

In this work we have investigated the RG fixed point structure of a three-dimensional Luttinger semimetal, where short-range interactions are generated from long-range Coulomb forces. We found the anisotropy parameter $\delta$ to be an exceptionally slow direction in the RG flow, which motivated us to neglect the flow of $\delta$ while considering the impact of quantum fluctuations onto the remaining couplings of the theory. For sufficiently strong anisotropy, Abrikosov's NFL fixed point survives in three dimensions, and new ``nemagnetic" fixed points appear in parameter space. These fixed points are associated to quantum phase transitions that are driven by a sufficiently strong microscopic short-range interaction. The new fixed points trigger ordering in channels that have a very clear interpretation for electrons in Pyrochlore Iridates, namely AIAO and SI order.

In this section we first critically review the validity of our results and the underlying approximations. We then relate our findings to earlier theoretical investigations of similar systems in the literature, mostly in the context of Pyrochlore Iridates. Finally, we deduce the critical field theories that are related to the anisotropic quantum critical points (W7e) and (Ve).

The RG analysis presented here is based on an $\vare$-expansion close to four dimensions, which is the critical dimension of the Coulomb coupling $e$ for a system with quadratic dispersion. In particular, for small $0<\vare\ll 1$ this gives a controlled perturbative handle on Abrikosov's NFL fixed point. On the other hand, the critical dimension for the short-range interactions would be two (see for instance Refs. \cite{PhysRevB.82.115431,PhysRevB.85.235408,PhysRevB.89.201110,2014arXiv1409.8675L,PhysRevB.91.134509} for studies of two-dimensional QBT systems), so obviously some compromise needs to be made to study the generation of short-range interactions in three dimensions and especially the fixed point collision scenario. Furthermore, our analysis of instabilities is based on formulas that are perturbative in the couplings $\{g_i\}$ and thus cannot faithfully capture the flow to strong coupling or additional competition effects between the couplings that set in once one coupling gets large.

A rather strong point, however, can be made about the fixed point collision scenario which determines the fate of Abrikosov's NFL ground state. Within the $\vare$-expansion the collision happens for $\vare\simeq 0.74$ in the isotropic system and consequently the ground state in three dimensions is an ordered state \cite{PhysRevLett.113.106401}. The very same conclusion is found in a different approach to the system using Dyson--Schwinger equations in three dimensions for large fermion number. In fact, the corresponding analysis in Ref. \cite{PhysRevB.93.165109} finds the ground state to be a topological excitonic insulator with nematic order in the physical limit. Further evidence for the collision scenario is provided in Ref. \cite{LukasPaper} from different RG approaches. To understand what happens for $\delta\neq 0$ recall that Abrikosov's fixed point is always stable for sufficiently small $\vare$. For sufficiently strong anisotropy, the charge fixed point $e^2_\star$ is attracted towards weak coupling, as can be seen from the prefactor $\sim \frac{1}{1-\delta^2}$ on the right hand side of Eq. (\ref{field17}). This prefactor results from additional line nodes in the dispersion of the fully anisotropic system that make the polarization diagram (Fig. \ref{FigLoops} b) diverge for $|\delta|\to 1$. Put differently, for small $0<1-\delta^2\ll 1$ the anisotropy acts as an infrared regulator that can be used to perturbatively control the equations in three dimensions. The resulting fixed point charge $e^2_\star\propto (1-\delta^2)\vare$ is small even when $\vare=1$. Hence a large anisotropy acts like a small $\vare$ (or a large number of fermions) and thus aids the survival of Abrikosov's fixed point. This way it is a very natural finding that the NFL ground state is realized for sufficiently strong anisotropy.

Concerning the question of reliability of our findings on nemagnetic quantum phase transitions in the anisotropic system we point out that they are in concord with the works of Savary, Moon, Balents (SMB, Ref. \cite{savary}) and Goswami, Roy, Das Sarma (GRDS, Ref. \cite{Goswami:2016fjn}) on anisotropic three-dimensional Luttinger semimetals. We emphasize, however, that in contrast to the mentioned references our analysis treats all ordering patterns in an unbiased fashion, and it is an outcome of the competition of long- and short-range interactions that anisotropy induces instabilities towards AIAO or SI order. It is appealing that these orders are also dictated to some extent by experimental findings in the Pyrochlore Iridates.

The analysis of SMB employs RG to study the long-range interacting system in three dimensions, perturbatively controlling the equations with a large number of fermions $N$. Further, it is based on a Yukawa theory involving both fermions and bosons, where fluctuations of the AIAO-like order parameter $\chi=\langle \psi^\dagger W_7\psi\rangle$ are incorporated and the existence of the corresponding quantum critical point is derived from the self-consistency of the equations. The approach is thus distinct from our $\vare$-expansion of the purely fermionic theory, but many common observations on the intriguing role of anisotropy in Luttinger semimetals can be made.  A stable fixed point is found by SMB for $\delta=-1$ and $x=0$ (corresponding to $c_1/c_2=0$ and $c_0/c_1=0$ in the reference). Further, an additional logarithmic divergence, e.g. in the anomalous dimension $\eta \sim 1/(N |\log c_1/c_2|^2)$, results in an effectively weakly coupled theory in terms of the charge. Remarkably, the possibility of a stable fixed point with $\delta_\star=-1$ implies that the fully anisotropic limit may arise in Luttinger semimetals as an emergent phenomenon, even though realistic microscopic Luttinger parameters are likely to yield $|\delta|<1$. From our results, the AIAO nature of the phase transition and the small charge $\eta \sim e^2_\star \sim (1-\delta^2)\vare$ are also visible. Although we find the fixed point at $\delta=-1$ to be unstable, the discrepancy may be attributed to the fact that SMB also take into account the boson-fermion-loop contribution to the fermion self energy, i.e., a second diagram besides the one in Fig. \ref{FigLoops} a), which may change the sign of the $\beta$-function for $\delta$ close to the fixed point and thereby stabilize it. In our fermionic approach the associated contribution would be a two-loop (sunset) diagram. Still, the flow of the anisotropy derived by SMB is very slow close to the fixed point. Since the philosophy of the present work is to regard $\delta$ as a constant anyway, the question of its slight relevance or irrelevance becomes less important.

In the study of GRDS in the context of Pr$_2$Ir$_2$O$_7$ the tendencies towards AIAO and SI order have been addressed in terms of the order parameter susceptibilities at the Gaussian fixed point. (The comparison to our results is facilitated by writing $\delta=\frac{m_2-m_1}{m_1+m_2}$ with $m_{1,2}$ as introduced in the reference.) Assuming a Yukawa coupling of both orders to the fermions it is seen that a smaller critical coupling is required for AIAO (SI) when $m_2<m_1$ ($m_2>m_1$). This agrees with our finding of AIAO and SI fixed points appearing for $\delta\to-1$ and $\delta\to +1$, respectively. Whereas GRDS focus on the charge neutral system, we find that QCPs towards AIAO and SI order are present both for zero or nonzero charge. It is remarkable that the SI order, which is a very natural and plausible configuration of spins on a pyrochlore lattice, emerges in our continuum field theoretic approach. Recall how the SI instability is obtained in the present work: We investigate all possible susceptibilities, in particular those towards a nonzero expectation value of $\langle \psi^\dagger (\alpha \mathJ_i+\beta W_i)\psi\rangle$, where $\alpha^2+\beta^2=1$. It turns out that the strongest divergence is reached for this particular channel when $\beta=2\alpha$, which precisely corresponds to SI order. In particular, this value of $(\alpha,\beta)$ is independent of $\delta$, which is not obvious from the general expressions for the susceptibilities in Eqs. (\ref{rg46}). In fact, the emergence of this particular combination of $(\alpha,\beta)$ is very likely rooted in an enhanced symmetry of the underlying field theory, as we elaborate in the next paragraph. Note eventually that the analysis in Ref. \cite{murray} does not rule out a stable fixed point for $\delta=+1$, as it does not explore this regime of anisotropy.

The fixed points (N), (S), (W7) and (W7e), (V) and (Ve), describe quantum phase transitions that can be driven by sufficiently strong short-range interactions in real materials. It is thus interesting to study their individual character by also taking into account order parameter fluctuations. A detailed study of (N) in terms of an $\vare$-expansion is performed in Ref. \cite{PhysRevB.92.045117}. A remarkable feature of this theory is the existence of an interaction term that is cubic in the order parameter, leading to characteristic corrections to critical exponents beyond mean-field theory. The superconducting quantum critical point (S) is investigated in Ref. \cite{PhysRevB.93.205138}, where non-Fermi liquid behavior, oscillatory corrections to scaling, and an exceptionally small flow towards isotropy are observed. The field theory describing the fixed point (W7e) for $\delta\to-1$ is given by
\begin{align}
 \nonumber L =\ &\psi^\dagger(\partial_\tau + d_1\gamma_1+d_2\gamma_2+\rmi a)\psi + \frac{1}{2e^2}(\nabla a)^2\\
 \label{dis1} &+ g_\chi\ \chi (\psi^\dagger \gamma_{12}\psi) + \frac{1}{2}(\nabla\chi)^2+\frac{1}{2}r\chi^2,
\end{align}
where $\chi$ is a real scalar. The related quantum critical physics is discussed in Ref. \cite{savary}. As pointed out in the same reference, the pseudospin $\vec{I}=\int\mbox{d}^3x\ \psi^\dagger [\gamma_{34},\gamma_{35},\gamma_{45}]^{\rm t}\psi$ is a conserved quantity of this theory, representing an internal SU(2) symmetry. In the opposite limit of strong anisotropy $\delta\to+1$ the field theory for (Ve) reads
\begin{align}
 \nonumber L =\ &\psi^\dagger(\partial_\tau + d_3\gamma_3+d_4\gamma_4+d_5\gamma_5+\rmi a)\psi + \frac{1}{2e^2}(\nabla a)^2\\
 \label{dis2} &+ g_v\ \vec{v}\cdot(\psi^\dagger [\gamma_{34}, \gamma_{35}, \gamma_{45} ]^{\rm t} \psi) + \frac{1}{2}(\nabla v_i)^2+\frac{1}{2}rv_i^2
\end{align}
with real $v_i$. A conserved quantity is given by $\int\mbox{d}^3x\ \psi^\dagger\gamma_{12}\psi$, which is related to an internal U(1) symmetry. The symmetry enhancement in the Lagrangian (\ref{dis2}) may explain why the particular combination $(\alpha,\beta)$ for SI order emerges in the limit $\delta\to+1$: The linear combination $\vec{V}\propto \vec{\mathJ}+2\vec{W}$ is dictated by the fact that this is precisely $\vec{V} \propto [\gamma_{34}, \gamma_{35}, \gamma_{45} ]^{\rm t}$ (up to signs), see Eqs. (\ref{rg42}). Similarly, $W_7 =\gamma_{12}$ may be selected by the same mechanism for $\delta\to-1$. We are thus left with an apparent intimate relation between the algebra of the Luttinger Hamiltonian and typical spin configurations on the pyrochlore lattice that deserves closer investigation in future work.

\begin{center}
 \textbf{Acknowledgements}
\end{center}

\noindent We thank L. Janssen and M. M. Scherer for inspiring discussions. IB acknowledges funding by the DFG under Grant No. BO 4640/1-1. IFH is supported by NSERC of Canada.

\begin{appendix}

\section{Renormalization group equations}\label{AppRG}

\subsection{Scaling dimensions}\label{AppScal}

We determine the scaling dimensions of the couplings involved in our analysis. The presentation is kept concise and we suggest to consult App. B.1 of Ref. \cite{PhysRevB.93.205138} for a more comprehensive account.

The effective Lagrangian $\bar{L}$ entering the effective action $\Gamma=\int \mbox{d}\bar{\tau} \mbox{d}^dr \bar{L}$ at a given RG time $b_0$ can be written as
\begin{align}
 \nonumber \bar{L} =\ &\bar{\psi}^\dagger \Bigl(\bar{S}\partial_{\bar{\tau}} -\bar{x} \nabla^2 +A_\psi \sum_a d_a\gamma_a +\bar{\delta} \sum_a s_ad_a\gamma_a +\rmi \bar{a}\Bigr)\bar{\psi}\\
 \label{scal1} &+\frac{1}{2\bar{e}^2} (\nabla\bar{a})^2+\bar{g}(\bar{\psi}^\dagger M \bar{\psi})^2.
\end{align}
Here $M$ is a dimensionless $4\times 4$ matrix such that the last term symbolizes a generic short-range coupling. Our goal is to bring the Lagrangian into the form of the main text by means of rescalings that do not affect the physics. For this map $\bar{\psi}\to \hat{\psi}=A_\psi^{1/2}\bar{\psi}$ and $\bar{a} \to a = A_\psi^{-1}\bar{a}$. We then have
\begin{align}
 \nonumber \bar{L} =\ &\hat{\psi}^\dagger \Bigl(S\partial_{\bar{\tau}}- x \nabla^2 + \sum_a d_a\gamma_a +\delta \sum_a s_ad_a\gamma_a +\rmi a\Bigr)\hat{\psi}\\
 \label{scal2} &+\frac{A_\psi^2}{2\bar{e}^2} (\nabla a)^2+\frac{\bar{g}}{A_\psi^2}(\hat{\psi}^\dagger M \hat{\psi})^2
\end{align}
with $S=\bar{S}/A_\psi$, $x=\bar{x}/A_\psi$, and $\delta=\bar{\delta}/A_\psi$. Next rescale imaginary time according to $\bar{\tau}=S\tau$. The effective action remains form-invariant, $\Gamma=\int \mbox{d}\tau \mbox{d}^dr L$, with $\hat{\psi} \to \psi = S^{1/2}\hat{\psi}$ and
\begin{align}
 \nonumber L=\ &\psi^\dagger \Bigl(\partial_{\tau}-x \nabla^2 + \sum_a d_a\gamma_a +\delta \sum_a s_ad_a\gamma_a +\rmi a\Bigr)\psi\\
 \label{scal3} &+\frac{A_\psi^2S}{2\bar{e}^2} (\nabla a)^2+\frac{\bar{g}}{A_\psi^2S}(\psi^\dagger M \psi)^2.
\end{align}
This motivates to introduce the rescaled couplings $e^2$ and $g$ by means of
\begin{align}
 \label{scal4} e^2 = \frac{\bar{e}^2}{A_\psi^2 S} \frac{\text{S}_3\Lambda^{d-4}}{(2\pi)^3},\ g= \frac{\bar{g}}{A_\psi^2 S}\frac{\text{S}_3\Lambda^{d-2}}{(2\pi)^3},
\end{align}
where $\text{S}_3=4\pi$ is the surface area of the unit sphere.  By re-defining $e^2\to \frac{1}{2}e^2$ and $g_i\to \frac{1}{4}g_i$, our results map to the considerations in Ref. \cite{PhysRevLett.113.106401}.

The canonical scaling dimensions of the (generalized) running couplings $\bar{\lambda}$ entering Eq. (\ref{scal1}) are given by
\begin{align}
 [\Gamma] &=0,\\
 [\bar{x}] &= [\bar{\delta}] =0,\\
 [\bar{\tau}] &= -2,\ [\mbox{d}^dr]=-d,\\
 [\bar{\psi}] &=d/2,\ [\bar{a}]=2,\\
 [\bar{e}^2] &=4-d,\\
 [\bar{g}]&= 2-d.
\end{align}
At a nontrivial fermionic fixed point we find that loop contributions lead to the flow equations $\dot{A}_\psi = \eta A_\psi$ and $\dot{S}=(z-2)S$. This may formally be written as $[A_\psi]=\eta$ and $[S]=z-2$. Equivalently, we can rescale all couplings according to Eq. (\ref{scal3}), so that $A_\psi=S=1$ holds for all $b$, and the scaling dimensions of the rescaled couplings are obtained from
\begin{align}
 \label{scal5} [\lambda] = \Bigl[ \bar{\lambda}A_\psi^{d_1}S^{d_2}\Bigr] = [\bar{\lambda}]+d_1[A_\psi] +d_2[S].
\end{align}
In this way we arrive at
\begin{align}
 [x] &= [\delta] = -\eta,\\
 [\tau]  &= -z,\\
 [\psi]&= \frac{d+\eta+z-2}{2},\\
 [a] &= 2-\eta,\\
 [e^2] &= 6-d-z-2\eta,\\
 [g] &= 4-d-z-2\eta.
\end{align}
Under the assumption that $z=2-\eta$, which is valid for the present work, these formulas simplify further.

Let us also consider the scaling of $\Delta$ considered in the susceptibility analysis. For this we couple a term
\begin{align}
 \label{scal6} \bar{L}_\Phi = \bar{\Delta}(\bar{\psi}^\dagger M \bar{\psi}) = \frac{\bar{\Delta}}{A_\psi}(\hat{\psi}^\dagger M \hat{\psi})
\end{align}
to the Lagrangian. The canonical dimension of $\bar{\Delta}$ is given by $[\bar{\Delta}]=2$. The effective action (or free energy) related to this term is given by $\Gamma_\Phi = \int\mbox{d}^dr \mbox{d}\bar{\tau}\bar{L}_\Phi = \int\mbox{d}^dr \mbox{d}\tau L_\Phi$ with
\begin{align}
 \label{scal7} L_\Phi = \Delta (\psi^\dagger M \psi),\ \Delta = \frac{\bar{\Delta}}{A_\psi}.
\end{align}
Hence the scaling dimension of $\Delta$ is given by $[\Delta]=2-\eta$, which equals $z$ in our case. We define the free energy density $\mathcal{F}$ by means of $\Gamma+\Gamma_\Phi=\int\mbox{d}^dr \mbox{d}\tau \mathcal{F}$ so that $[\mathcal{F}]=d+z$. The scaling form of $\mathcal{F}$ is thus given by
\begin{align}
 \label{scal8} \mathcal{F} = k^{d+z} H\Bigl(\frac{\Delta}{k^{z+\eta_\Phi}}\Bigr),
\end{align}
as discussed further in Eq. (\ref{sus2}).

\subsection{Propagators}

We write $Q=(q_0,\textbf{q})$ to collect frequencies and momenta. Further define
\begin{align}
 \int_Q (\dots) = \int_{-\infty}^\infty\frac{\mbox{d} q_0}{2\pi} \int_{\Lambda/b\leq q\leq\Lambda} \frac{\mbox{d}^dq}{(2\pi)^d}(\dots),
\end{align}
where the momentum integration on the right hand side is restricted to a momentum shell.

The fermion propagator is given by
\begin{align}
 \label{prop1} G_\psi^Q = \frac{1}{\mbox{det}_F^Q}\Bigl(-\rmi q_0\mathbb{1}_4 +\sum_{a} (1+\delta s_a)d_a\gamma_a\Bigr)
\end{align}
with $d_a\equiv d_a(\textbf{q})$   and
\begin{align}
 \label{prop2} \mbox{det}_F^Q &=  q_0^2 +\sum_{b=1}^5 (1+\delta s_b)^2d_b^2\\
 &= q_0^2 +(1-\delta)^2 q^4 +12\delta\sum_{i<j} q_i^2q_j^2.
\end{align}
This denominator follows from the expression for $H^2$ given in Eq. (\ref{field8}).

The photon propagator is given by
\begin{align}
 \label{prop3} G_a^Q = G_a(\textbf{q}) = \frac{\bar{e}^2}{q^2}.
\end{align}
It is frequency-independent to leading order.

\subsection{Fermion self-energy}

The fermion self-energy is given by
\begin{align}
 \label{ferm1} \Sigma_\psi(P) = \int_Q G_a^{Q+(1-v)P}G_\psi^{-Q+vP}.
\end{align}
Herein, $v$ is an arbitrary real parameter. Due to translation invariance the result should not depend on $v$. However, the momentum shell breaks translation invariance and thus we obtain a slight $v$-dependence of the result when momentum integrals are evaluated for $d<4$. Still, varying $v$ in the physical range $v\in[0,1]$ does not change any qualitative aspects of the RG flow. In particular, it cannot lead to a sign change in $\eta$ or $\dot{\delta}$. For the results presented in the main text, we choose $v=0$ and evaluate the angular integral for $d=3$. Since the analysis for $v\neq 0$ is rather long-winded we limit the derivation to the case of $v=0$. We will, however, comment at the end of the section on the $v$-dependence of the results.

The loop in Eq. (\ref{ferm1}) does not generate a contribution to the frequency dependence $\sim \rmi p_0$ of fermions. Indeed, we have
\begin{align}
 \label{ferm2} \frac{1}{\rmi} \frac{\partial \Sigma_\psi(P)}{\partial p_0}\Bigr|_{P=0} = \bar{e}^2 v \mathbb{1}_4 \int_Q \frac{q_0^2-\sum_b(1+\delta s_b)^2d_b^2}{q^2(q_0^2+\sum_b(1+\delta s_b)^2d_b^2)^2}.
\end{align}
This expression vanishes upon frequency integration. Hence we have
\begin{align}
 z=2-\eta,
\end{align}
with the anomalous dimension $\eta$ to be determined below.

We henceforth set $p_0=0$ in the loop and have
\begin{align}
 \label{ferm3} \Sigma_\psi(\textbf{p}) &=  \int_Q G_a^{Q+\textbf{p}}\frac{\sum_a (1+\delta s_a)d_a(\textbf{q})\gamma_a}{\mbox{det}_F^{Q}},
\end{align}
because the term linear in $\rmi q_0$ vanishes. For the photon propagator we write
\begin{align}
 \label{ferm4} G_a^{Q+\textbf{p}} = \frac{\bar{e}^2}{q^2+2(\textbf{q}\cdot\textbf{p})+p^2}.
\end{align}
Due to $v=0$, this is the only term contributing a $p$-dependence to Eq. (\ref{ferm3}). The term quadratic in the components $p_i$ can be determined according to
\begin{align}
 \label{ferm5} &\frac{1}{2} \frac{\partial^2\Sigma(t\textbf{p})}{\partial t^2}\Bigr|_{t=0} \\
 \nonumber &= -\frac{\bar{e}^2}{2}\int_{\textbf{q}} \frac{p^2q^2-4(\textbf{q}\cdot\textbf{p})^2}{q^6X^{1/2}}\sum_a(1+\delta s_a)d_a(\textbf{q})\gamma_a\\
 \nonumber &= \frac{2(d-1)\bar{e}^2}{d}\sum_{a,c}(1+\delta s_a)d_c(\textbf{p})\gamma_a  \int_{\textbf{q}} \frac{d_a(\textbf{q})d_c(\textbf{q})}{q^6X^{1/2}},
\end{align}
with $X=\sum_b(1+\delta s_b)^2d_b^2$. We used
\begin{align}
 \label{ferm6} (\textbf{q}\cdot\textbf{p})^2 &= \frac{1}{d}\Bigl[(d-1)d_c(\textbf{q})d_c(\textbf{p})-q^2p^2\Bigr].
\end{align}
In a rotation invariant setting, or for $\delta=0$, we could have now used a formula like
\begin{align}
 \label{ferm7} \int_{\textbf{q}} d_a(\textbf{q}) d_c(\textbf{q}) \chi(q^2) = \frac{2\delta_{ac}}{(d-1)(d+2)}\int_{\textbf{q}} q^4\chi(q^2)
\end{align}
for any function $\chi(q^2)$ having finite support. However, $X$ is not a function of $q^2$. Consequently, the strongest statement that can be made at this point is that the integral on the right hand side of Eq. (\ref{ferm5}) is proportional to $\delta_{ac}$. We arrive at
\begin{align}
 \nonumber \Sigma_\psi(\textbf{p}) - \Sigma_\psi(0) =\ & \frac{2(d-1)\bar{e}^2}{d}\sum_{a}(1+\delta s_a)d_a(\textbf{p})\gamma_a  \\
 \label{ferm8} &\times \int_{\textbf{q}} \frac{d_a^2}{q^6X^{1/2}}+\mathcal{O}(p^3)
\end{align}
We neglect the constant contribution $\Sigma_\psi(0)$ in the following. In particular, using the definitions of $f_{1 \rm e,t}$ from Eqs. (\ref{cub2}) and (\ref{cub5}), and setting $d=3$, we arrive at
\begin{align}
 \nonumber \Sigma_\psi(\textbf{p}) =\ & \frac{4\bar{e}^2}{15}\Bigl[(1-\delta)f_{1\rm e}(\delta)\sum_{a=1,2}d_a(\textbf{p})\gamma_a\\
 \label{ferm9} &+(1+\delta)f_{1\rm t}(\delta)\sum_{a=3,4,5}d_a(\textbf{p})\gamma_a\Bigr]\int_{\textbf{q}}\frac{1}{q^4}+\mathcal{O}(p^3).
\end{align}

We now determine $\eta$ and $\dot{\delta}$ from Eq. (\ref{ferm9}). From Eq. (\ref{scal4}) we see that $\frac{d}{d\log b}\bar{e}^2\int_{\textbf{q}} \frac{1}{q^4}=e^2$. Further, we can read off the one-loop contributions $\Delta\eta$ and $\Delta\delta$ that appear in the final expressions for $\eta$ and $\dot{\delta}$ by writing
\begin{align}
 \label{ferm10} \dot{\Sigma}_\psi(\textbf{p}) = \dot{A}_{1,2} \sum_{a=1,2} d_a(\textbf{p})\gamma_a + \dot{A}_{3,4,5} \sum_{a=3,4,5}d_a(\textbf{p})\gamma_a.
\end{align}
We find
\begin{align}
 \label{ferm11} \dot{A}_{1,2} &= \frac{4}{15} e^2 (1-\delta)f_{1\rm e}(\delta),\\
 \label{ferm12} \dot{A}_{3,4,5} &= \frac{4}{15} e^2 (1+\delta)f_{1\rm t}(\delta).
\end{align}
Accordingly, we have
\begin{align}
 \nonumber \eta = \Delta \eta &= \frac{1}{2}(\dot{A}_{1,2}+\dot{A}_{3,4,5})\\
 \label{fermi13} &= \frac{2}{15}\Bigl[(1-\delta)f_{1\rm e}(\delta)+(1+\delta)f_{1\rm t}(\delta)\Bigr]e^2,\\
 \nonumber \Delta \delta &= \frac{1}{2}(-\dot{A}_{1,2}+\dot{A}_{3,4,5})\\
 \label{ferm14} &= \frac{2}{15}\Bigl[-(1-\delta)f_{1\rm e}(\delta)+(1+\delta)f_{1\rm t}(\delta)\Bigr]e^2.
\end{align}
Note that the flow equation for $\delta$ is given by
\begin{align}
 \nonumber \dot{\delta} &= -\eta \delta +\Delta \delta\\
 \label{ferm15} &= -\frac{2}{15}(1-\delta^2)\Bigl[f_{1\rm e}(\delta)-f_{1\rm t}(\delta)\Bigr]e^2.
\end{align}

When expanding Eq. (\ref{ferm1}) for finite $v$, the result depends on $v$ when evaluating the angular integral in three dimensions, whereas the result is independent of $v$ when evaluating it in four dimensions. The slight $v$-dependence is not problematic for our analysis which does not aspire to be quantitatively precise, but it is important to understand whether qualitative modifications of the RG flow can occur. In particular, we are interested in whether (i) the flow of $\delta$ is still exceptionally slow for $v\neq0$, and (ii) whether the fixed point structure of $\dot{\delta}$ remains invariant. For this purpose it is sufficient to consider $\eta$ and $\dot{\delta}$ to linear order in $\delta$. By means of a calculation along the lines of Eqs. (B54)-(B77) in Ref. \cite{PhysRevB.93.205138} one can show that in three dimensions
\begin{align}
 \eta &= F(v)\Bigl(\frac{4}{15}e^2-\frac{4}{105}e^2\delta\Bigr)+\mathcal{O}(\delta^2),\\
 \dot{\delta} &= -F(v)\frac{8}{105}e^2\delta+\mathcal{O}(\delta^2)
\end{align}
with $F(v)=1-(1/2)v-(1/8)v^2$. The function $F(v)$ is positive and of order unity in the range $v\in[0,1]$. Accordingly, the flow is not qualitatively changed. In particular, since the function $\dot{\delta}(\delta)$ has negative slope at $\delta_\star=0$ for all $v$, and given the topology of the function seen in Fig. \ref{FigAniso}, it is also clear that the fixed point structure is not modified: The stable fixed point is at $\delta_\star=0$, whereas $\delta_\star=\pm1$ are repulsive. Note further that when evaluating the momentum integrals in four dimensions (including using nine $d_a$-functions) one finds
\begin{align}
 \eta|_{\rm 4D} &= \frac{1}{6}e^2 -\frac{2}{45}e^2\delta+\mathcal{O}(\delta^2),\\
 \dot{\delta}|_{\rm 4D} &= -\frac{1}{30}e^2\delta+\mathcal{O}(\delta^2)
\end{align}
for all $v$. These results confirm that the flow of $\delta$ is indeed exceptionally slow.

\subsection{Photon self-energy}

The photon self-energy is given by
\begin{align}
 \label{phot1} \Sigma_a(P) = -\int_Q \mbox{tr}(G_\psi^{Q+(1-v)P}G_\psi^{Q-vP}).
\end{align}
We have, again, introduced a real parameter $v$ that allows to distribute the external momentum onto the two fermion lines.

The photon self-energy cannot acquire a frequency dependence that is relevant close to $d=4$ dimensions since the photon is described by a real field. Indeed, Eq. (\ref{phot1}) has the symmetry $\Sigma_a(p_0,\textbf{p}) = \Sigma(-p_0,\textbf{p})$, which forbids a term $\Sigma_a\sim \rmi p_0$, and the leading term is $\Sigma_a \sim c^2p_0^2$. We can thus neglect the external frequency in the computation, i.e., $p_0=0$ in the following.

Setting the external frequency to zero we obtain
\begin{align}
 \nonumber &\Sigma_a(\textbf{p}) =- 4 \int_Q \frac{1}{\mbox{det}_F^{Q+(1-v)\textbf{p}}\mbox{det}_F^{Q-v\textbf{p}}}\\
 \label{phot2} &\times\Bigl[-q_0^2+\sum_a(1+\delta s_a)^2d_a(\textbf{q}+(1-v)\textbf{p})d_a(\textbf{q}-v\textbf{p})\Bigr]
\end{align}
For a derivative expansion of this expression we write
\begin{align}
 \label{phot3} \mbox{det}_F^{Q+\textbf{p}} = \mbox{det}_F^Q + D_1 +D_2 +D_3+\mathcal{O}(p^3)
\end{align}
with
\begin{align}
 \label{phot4} D_1 &= 4A_d \sum_a(1+\delta s_a)^2d_a(\textbf{q})(q_i\Lambda^a_{ij}p_j),\\
 \label{phot5} D_2 &= 2\sum_a(1+\delta s_a)^2 d_a(\textbf{q})d_a(\textbf{p}),\\
 \label{phot6} D_3 &= 4A_d^2 \sum_a(1+\delta s_a)^2(q_i\Lambda^a_{ij}p_j)(q_k\Lambda^a_{kl}p_l).
\end{align}
Note that the coefficients $D_i$ have definite scaling properties under $\textbf{p}\mapsto t \textbf{p}$. We used
\begin{align}
 \label{phot7} d_a(\textbf{p}) = A_d q_i \Lambda^a_{ij}q_j
\end{align}
with $A_d=\sqrt{\frac{d}{2(d-1)}}$. We have
\begin{align}
 \label{phot8} \mbox{det}_F^{Q+(1-v)\textbf{p}} &= \mbox{det}_F^Q +(1-v)D_1 +(1-v)^2(D_2+D_3),\\
 \label{phot9} \mbox{det}_F^{Q-v\textbf{p}} &= \mbox{det}_F^Q -vD_1 +v^2(D_2+D_3).
\end{align}
In the same way we expand
\begin{align}
 \nonumber &\sum_a(1+\delta s_a)^2d_a(\textbf{q}+(1-v)\textbf{p})d_a(\textbf{q}-v\textbf{p}) \\
 \nonumber &= X +(1-2v)\frac{1}{2}D_1 +[(1-v)^2+v^2]\frac{1}{2}D_2 -v(1-v) D_3\\
 \label{phot10} &+\mathcal{O}(p^3)
\end{align}
with $X=\sum_a(1+\delta s_a)^2 d_a^2$. Note that $\mbox{det}_F^Q=q_0^2+X$. We then find for the quadratic part
\begin{align}
 \label{phot14} \frac{1}{2}\frac{\partial^2\Sigma_a(t\textbf{p})}{\partial t^2}\Bigr|_{t=0} = -\frac{1}{8} \int_{\textbf{q}} \Bigl(\frac{D_1^2}{X^{5/2}}-\frac{4D_3}{X^{3/2}}\Bigr)
\end{align}
\emph{independently of $v$}. In the following we evaluate this integral for $d=3$.

Due to the absence of rotation invariance, only a limited number of simplifications is possible when computing integrals of the type (\ref{phot14}). If $\chi_{\rm cub}(\textbf{q})$ is a function cubic in the components $q_i$, then $\int_{\textbf{q}} q_iq_j\chi_{\rm cub}(\textbf{q}) = \delta_{ij} \int_{\textbf{q}} q_i^2 \chi_{\rm cub}(\textbf{q})$. Further, we have $\int_{\textbf{q}}q_{i_1}\cdots q_{i_n}\chi_{\rm cub}(\textbf{q})=0$ for $n$ odd. In addition, for every fixed $i$, we have
\begin{align}
 \label{phot15} \int_{\textbf{q}} q_i^n\ \chi_{\rm cub}(\textbf{q}) = \frac{1}{3} \int_\textbf{q} (q_x^n+q_y^n+q_z^n) \chi_{\rm cub}(\textbf{q}).
\end{align}
We apply these relations to compute
\begin{align}
 \nonumber \int_{\textbf{q}} \frac{D_3}{X^{3/2}} &= 4 A_3^2 \sum_a(1+\delta s_a)^2 \Lambda^a_{ij}\Lambda^a_{kl} p_j p_l \int_{\textbf{q}} \frac{q_iq_k}{X^{3/2}}\\
 \nonumber &=  \sum_a(1+\delta s_a)^2  (\Lambda^a\Lambda^a)_{jl} p_jp_l \int_{\textbf{q}} \frac{q^2}{X^{3/2}}\\
 \nonumber &= \frac{2}{3} p^2\sum_a(1+\delta s_a)^2  \frac{f_2(\delta)}{(1-\delta^2)} \int_{\textbf{q}} \frac{1}{q^4}\\
 \label{phot16} &=\frac{2}{3}p^2\Bigl[2(1-\delta)^2+3(1+\delta)^2\Bigr]\frac{f_2(\delta)}{(1-\delta^2)} \int_{\textbf{q}} \frac{1}{q^4}.
\end{align}
Note that $4A_3^2=3$.

In deriving Eq. (\ref{phot16}) we used
\begin{align}
 \label{phot17} p_j(\Lambda^a\Lambda^b)_{jl}p_l &= \frac{2}{d} p^2\delta^{ab} + \frac{1}{2}J_{abc}\Lambda^c_{jl}p_jp_l
\end{align}
with $J_{abc}=\mbox{tr}(\Lambda^a\Lambda^b\Lambda^c)$ and $\sum_a J_{aac}=\sum_a s_aJ_{aac}=0$. For the latter relations see Ref. \cite{PhysRevB.93.205138}, App C.2. To prove Eq. (\ref{phot17}) first note that only the symmetric part of the matrix $\Lambda^a\Lambda^b$ enters the product on the left hand side of the equation. More explicitly, we have
\begin{align}
 \label{phot18} \Lambda^a_{ji}\Lambda^b_{il}p_jp_l &= \frac{1}{2}\underbrace{(\Lambda^a_{ji}\Lambda^b_{il}+\Lambda^a_{li}\Lambda^b_{ij})}_{M_{jl}}p_jp_l.
\end{align}
The matrix $M_{jl}$ is symmetric in $jl$. Hence it can be decomposed according to
\begin{align}
 \label{phot19} M_{jl} = \frac{1}{d}\mbox{tr}(M) \delta_{jl}+\frac{1}{2}\mbox{tr}(M\Lambda^c)\Lambda^c_{jl}
\end{align}
with
\begin{align}
 \label{phot20} \mbox{tr}(M) &= 2 \Lambda^a_{ji}\Lambda^b_{ij} = 4\delta^{ab},\\
 \label{phot21} \mbox{tr}(M\Lambda^c) &= 2 \Lambda^a_{ji}\Lambda^b_{il}\Lambda^c_{lj} = 2J_{abc},
\end{align}
which coincides with Eq. (\ref{phot17}). This formula is valid for arbitrary dimension $d$.

To evaluate the remaining integral in Eq. (\ref{phot14}) we verify via direct computation that
\begin{widetext}
\begin{align}
 \nonumber D_1^2 &= 16A_3^2\sum_{a,b}(1+\delta s_a)^2(1+\delta s_b)^2d_ad_b (q_i\Lambda^a_{ij}p_j)(q_k\Lambda^b_{kl}p_l)\\
 \nonumber &=16A_3^2\frac{4}{3}\Bigl[(1-2\delta+\delta^2) \sum_i p_i q_i \cdot q_i^2 +(1+4\delta+\delta^2) \sum_{i\neq j} p_iq_i\cdot q_j^2\Bigr]^2\\
 \nonumber &= 16\Bigl[(1-2\delta+\delta^2) \sum_i p_i q_i \cdot q_i^2 +(1+4\delta+\delta^2) \sum_{i} p_iq_i\cdot (q^2-q_i^2)\Bigr]^2\\
 \nonumber &=16\Bigl[ \sum_i p_iq_i\Bigl(-6\delta q_i^2+(1+4\delta+\delta^2)q^2\Bigr)\Bigr]^2\\
 \label{phot22} &=16\Bigl[ 36\delta^2 \sum_{i,j}p_ip_jq_i^3q_j^3 -12\delta (1+4\delta+\delta^2)q^2\sum_{i,j} p_ip_jq_i^3q_j +(1+4\delta+\delta^2)^2q^4 \sum_{i,j} p_ip_jq_iq_j\Bigr].
\end{align}
\end{widetext}
Upon multiplication with the cubic function $\chi_{\rm cub}(\textbf{q})=X^{-5/2}$ and integration over $\textbf{q}$, all of these terms are proportional to $\delta_{ij}$. Hence we arrive at
\begin{align}
 \nonumber \int_{\textbf{q}}\frac{D_1^2}{X^{5/2}} &= \frac{16}{3} p^2 \Bigl[ 36\delta^2 \int_{\textbf{q}}\frac{q_x^6+q_y^6+q_z^6}{X^{5/2}} -12\delta (1+4\delta+\delta^2)\\
 \label{phot23} &\times \int_{\textbf{q}}\frac{q^2(q_x^4+q_y^4+q_z^4)}{X^{5/2}}+(1+4\delta+\delta^2)^2 \int_{\textbf{q}} \frac{q^6}{X^{5/2}}\Bigr].
\end{align}
We apply
\begin{align}
 \label{phot24} q^2(d_1^2+d_2^2) &= q_x^6+q_y^6+q_z^6-3q_x^2q_y^2q_z^2,\\
 \label{phot24b} q^4 &= q_x^4+q_y^4+q_z^4 +2\sum_{i<j}q_i^2q_j^2,\\
  \label{phot24c}   d_3^2+d_4^2+d_5^2 &= 3\sum_{i<j}q_i^2q_j^2,\\
 \label{phot25} d_3d_4d_5 &= 3\sqrt{3} q_x^2q_y^2q_z^2,
\end{align}
to write
\begin{align}
 \nonumber &\int_{\textbf{q}}\frac{D_1^2}{X^{5/2}}  \\
 \nonumber &=\frac{16}{3} p^2 \Bigl[ 36\delta^2 \int_{\textbf{q}}\frac{q^2(d_1^2+d_2^2)}{X^{5/2}}-12\delta (1+4\delta+\delta^2) \int_{\textbf{q}}\frac{q^6}{X^{5/2}}\\
 \nonumber &+12\delta (1+4\delta+\delta^2)\frac{2}{3} \int_{\textbf{q}}\frac{q^2(d_3^2+d_4^2+d_5^2)}{X^{5/2}}\\
 \label{phot26}&+\frac{36}{\sqrt{3}}\delta^2\int_{\textbf{q}}\frac{d_3d_4d_5}{X^{5/2}}+(1+4\delta+\delta^2)^2 \int_{\textbf{q}} \frac{q^6}{X^{5/2}}\Bigr].
\end{align}
At last we use $\sum_a d_a^2=q^4$ to bring this into the more symmetric form
\begin{align}
 \nonumber \int_{\textbf{q}}\frac{D_1^2}{X^{5/2}}  =\ &\frac{16}{3} p^2 \Bigl[ (1-\delta)^4 \int_{\textbf{q}}\frac{q^2(d_1^2+d_2^2)}{X^{5/2}}\\
 \nonumber &+[(1+\delta)^4-4\delta^2] \int_{\textbf{q}}\frac{q^2(d_3^2+d_4^2+d_5^2)}{X^{5/2}}\\
 \nonumber &+\frac{36}{\sqrt{3}}\delta^2\int_{\textbf{q}}\frac{d_3d_4d_5}{X^{5/2}}\Bigr]\\
 \nonumber =\ &\frac{16}{3} p^2 \Bigl[ \frac{2}{5} \frac{(1-\delta)^2}{(1-\delta^2)} f_{3\rm e}(\delta)+\frac{3}{5}\frac{(1+\delta)^2}{(1-\delta^2)} f_{3\rm t}(\delta)\\
 \nonumber &-\frac{12}{5}\frac{\delta^2}{(1-\delta)(1+\delta)^3}f_{3\rm t}(\delta)\\
 \label{phot27}&+\frac{36}{35}\frac{\delta^2}{(1+\delta)^3}f_{345}(\delta)\Bigr]\int_{\textbf{q}}\frac{1}{q^4}.
\end{align}

We conclude that the $p^2$-contribution to the photon self-energy is given by
\begin{align}
 \label{phot28} \frac{1}{2}\frac{\partial^2\Sigma_a(t\textbf{p})}{\partial t^2}\Bigr|_{t=0} &= p^2 \frac{f_{e^2}(\delta)}{(1-\delta^2)}\int_{\textbf{q}}\frac{1}{q^4}
\end{align}
with
\begin{align}
 \nonumber f_{e^2}(\delta) &=\frac{1}{3} \Bigl(2(1-\delta)^2+3(1+\delta)^2\Bigr) f_2(\delta)\\
 \nonumber &- \frac{2}{3} \Biggl(\frac{2}{5} (1-\delta)^2 f_{3\rm e}(\delta)+\frac{3}{5}(1+\delta)^2 f_{3\rm t}(\delta)\\
 \label{phot29} &-\frac{12}{5}\frac{\delta^2}{(1+\delta)^2}f_{3\rm t}(\delta)+\frac{36}{35}\frac{\delta^2(1-\delta)}{(1+\delta)^2}f_{345}(\delta)\Biggr).
\end{align}
The function $f_{e^2}(\delta)$ is finite for all $\delta$ (although at first glance one would expect a singularity for $\delta \to -1$) and satisfies $f_{e^2}(0)=1$. Accordingly, the one-loop correction to the charge is given by
\begin{align}
 \label{phot30} \Delta e^2 = -\frac{f_{e^2}(\delta)}{(1-\delta^2)} e^4,
\end{align}
which leads to the flow equation
\begin{align}
 \label{phot31} \dot{e}^2 = (4-d-\eta)e^2 -\frac{f_{e^2}(\delta)}{(1-\delta^2)} e^4.
\end{align}
Inserting $\eta$ from Eq. (\ref{fermi13}) yields the fixed points $e^2=0$ and
\begin{align}
\label{phot32} e^2_\star = \frac{15}{19} (1-\delta^2) f_\star(\delta) \vare
\end{align}
with $\vare=4-d$ and
\begin{align}
 \label{phot33} f_\star(\delta) = \frac{19}{2(1-\delta^2)[(1-\delta)f_{1\rm e}+(1+\delta)f_{1\rm t}]+15f_{e^2}}.
\end{align}
Note that $f_\star(0)=1$. We plot $f_{e^2}(\delta)$ and $f_\star(\delta)$ in Fig. \ref{Figfi}.

\subsection{Short-range interactions}
We derive the renormalization of short-range interactions with the help of a set of master formulas that apply to arbitrary four-fermion theories. By specializing to the propagators and vertices which are of interest here, we obtain the running of couplings $g_{1,2,3}$ in Eq. (\ref{flow16}).

To introduce our notation we start with a few general remarks on the Wilsonian RG scheme. The partition function of the theory is found from
\begin{align}
 \label{rg1} Z = \int \mbox{D}\phi \ e^{-\int_x L_{\rm kin}(\phi) -\int_x L_{\rm int}(\phi)},
\end{align}
where $\phi=(\psi,a)$ is a schematic field variable, $x=(\tau,\textbf{x})$, $L_{\rm kin}$ is the kinetic part of the Lagrangian quadratic in the fields, and $L_{\rm int}$ is the interaction part involving more powers of fields. Introducing a momentum shell and dividing the field into fast and slow movers according to $\phi=\phi_<+\phi_>$, the quadratic part factorizes, and we have
\begin{align}
 \label{rg2} Z = Z_>\int\mbox{D}\phi_<\ e^{-\int_x L_{\rm kin}(\phi_<)} \langle e^{-\int_x L_{\rm int}(\phi_<+\phi_>)}\rangle_>
\end{align}
with average over fast modes
\begin{align}
 \label{rg3} \langle \mathcal{O} \rangle_> = \frac{1}{Z_>} \int\mbox{D}\phi_>\ \mathcal{O} e^{-\int_x L_{\rm kin}(\phi_>)}
\end{align}
and $Z_>=\langle 1\rangle_>$. The term $\langle e^{-\int_x L_{\rm int}(\phi_<+\phi_>)}\rangle_>$ generates contributions with different powers of $\phi_<$ and $\phi_>$. We eventually write
\begin{align}
 \label{rg4} Z = Z_>\int\mbox{D}\phi_< e^{-\int_x L_{\rm eff}(\phi_<)},
\end{align}
where $L_{\rm eff}(\phi)$ is the effective Lagrangian after the fast modes have been integrated out.

In our case, the interaction part of the Lagrangian is given by
\begin{align}
 \label{rg5} L_{\rm int}(\psi,a) = \psi^\dagger \rmi a \psi + \sum_i g_i (\psi^\dagger M_i\psi)^2,
\end{align}
where the matrices $M_i$ comprise the short-range interactions. Upon insertion into Eq. (\ref{rg2}), the one-loop contributions to the four-fermion interactions (i.e. those terms involving precisely four slow modes $\psi_<$) are generated by means of
\begin{align}
 \nonumber &\langle e^{-\int_x L_{\rm int}} \rangle_> = \frac{1}{2}\sum_{i,j}g_ig_j \int_{xy} \langle (\psi^\dagger M_i\psi)_x^2(\psi^\dagger M_j\psi)_y^2 \rangle_>\\
 \nonumber &-\frac{1}{6}3\sum_i g_i \int_{xyz} \langle (\psi^\dagger \rmi a \psi)_x(\psi^\dagger \rmi a \psi)_y(\psi^\dagger M_i\psi)^2_z\rangle_>\\
 \label{rg6} &+\frac{1}{24} \int_{xyzw} \langle (\psi^\dagger \rmi a \psi)_x (\psi^\dagger \rmi a \psi)_y (\psi^\dagger \rmi a \psi)_z (\psi^\dagger \rmi a \psi)_w\rangle_>.
\end{align}
Herein, the first, second, and third line constitute the contributions proportional to $g_ig_j$, $e^2g_i$, and $e^4$, respectively. The contractions can be evaluated by noting that the fast modes in Eq. (\ref{rg3}) are described by a quadratic action, hence the propagators read
\begin{align}
 \label{rg7}\langle \psi_{>}(x)\psi_{>}^\dagger(y)\rangle _> &= G_\psi(x,y),\\
 \label{rg8} \langle a_>(x)a_>(y) \rangle_> &= G_a(x,y),
\end{align}
or, after Fourier transformation,
\begin{align}
 \label{rg9}\langle \psi_{>}(Q)\psi_{>}^\dagger(Q')\rangle _> &= G_\psi^Q\cdot\delta(Q-Q'),\\
 \label{rg10} \langle a_>(Q)a_>(Q') \rangle_> &= G_a^Q\cdot\delta(Q+Q')
\end{align}
with $G_\psi^Q$ and $G_a^Q$ from Eqs. (\ref{prop1}) and (\ref{prop3}), respectively.

Let $M,N$ be any of the matrices $\{M_i\}$. The relevant contractions are then given by
\begin{widetext}
\begin{align}
 \nonumber \int_{xy} \langle (\psi^\dagger M \psi)_x^2(\psi^\dagger N \psi)^2_y \rangle_> =\ & -4 \int_Q(\psi^\dagger M\psi)(\psi^\dagger N\psi) \mbox{tr}(G_\psi^QNG_\psi^QM)+4 \int_Q(\psi^\dagger M\psi)(\psi^\dagger N G_\psi^Q M G_\psi^Q N \psi)\\
 \nonumber &+4\int_Q (\psi^\dagger MG_\psi^QNG_\psi^QM\psi)(\psi^\dagger N\psi)+4\int_Q  (\psi^\dagger MG_\psi^QN\psi)(\psi^\dagger N G_\psi^QM\psi)\\
 \label{rg11} &+2\int_Q  (\psi^\dagger MG_\psi^QN \psi)(\psi^\dagger MG_\psi^{-Q}N\psi)+2\int_Q  (\psi^\dagger NG_\psi^QM \psi)(\psi^\dagger NG_\psi^{-Q}M \psi),
\end{align}
\begin{align}
 \nonumber  \int_{xyz} \langle(\rmi \psi^\dagger a \psi)_x(\rmi \psi^\dagger a \psi)_y (\psi^\dagger M \psi)^2_z \rangle_> =\ & -2\int_Q G_a^Q\Biggl[2 (\psi^\dagger G_\psi^{Q}MG_\psi^{Q}\psi)(\psi^\dagger M\psi)+ (\psi^\dagger G_\psi^{-Q}M\psi)(\psi^\dagger G_\psi^Q M\psi)\\
 \label{rg12}&+ 2 (\psi^\dagger G_\psi^QM\psi)(\psi^\dagger M G_\psi^Q\psi)+(\psi^\dagger MG_\psi^{-Q}\psi)(\psi^\dagger MG_\psi^Q\psi)\Biggr].
\end{align}
We relabelled $\psi_<(x)\to \psi$ and omitted the integral $\int_x$ on the right hand side. Using Eq. (\ref{fierz2}) we can decompose the result into contributions $\sim(\psi^\dagger \Sigma^A\psi)(\psi^\dagger \Sigma^B\psi)$. We have
\begin{align}
 \nonumber \int_{xy} \langle (\psi^\dagger M \psi)_x^2(\psi^\dagger N \psi)_y^2 \rangle_> =\ & \frac{-1}{16}\Biggl[-4 \int_Q\mbox{tr}(G^Q_\psi NG^Q_\psi M)\mbox{tr}(M\Sigma^AN\Sigma^B)+4 \int_Q\mbox{tr}(M\Sigma^AN G^Q_\psi M G^Q_\psi N\Sigma^B)\\
 \nonumber &+4\int_Q \mbox{tr}(MG^Q_\psi NG^Q_\psi M\Sigma^AN\Sigma^B)+4\int_Q \mbox{tr}(MG^Q_\psi N\Sigma^AN G^Q_\psi M\Sigma^B)\\
 \label{rg13} &+2\int_Q  \mbox{tr}(MG^Q_\psi N\Sigma^AMG^{-Q}_\psi N\Sigma^B)+2\int_Q  \mbox{tr}(NG^Q_\psi M\Sigma^ANG^{-Q}_\psi M\Sigma^B)\Biggr](\psi^\dagger\Sigma^A\psi)(\psi^\dagger\Sigma^B\psi),
\end{align}
\begin{align}
 \nonumber  \int_{xyz} \langle(\rmi \psi^\dagger a \psi)_x(\rmi \psi^\dagger a \psi)_y (\psi^\dagger M \psi)^2_z \rangle_> =\ & \frac{1}{8}\int_Q G_a^Q\Biggl[2\ \mbox{tr}(G_\psi^{Q}MG_\psi^{Q}\Sigma^AM\Sigma^B)+\mbox{tr}(G_\psi^{-Q}M\Sigma^A G_\psi^Q M\Sigma^B) \\
\label{rg14}  &+ 2\ \mbox{tr}(G_\psi^QM\Sigma^AM G_\psi^Q\Sigma^B) +\mbox{tr}(MG_\psi^{-Q}\Sigma^AMG_\psi^Q\Sigma^B)\Biggr](\psi^\dagger\Sigma^A\psi)(\psi^\dagger\Sigma^B\psi).
\end{align}
\end{widetext}
 The contribution of order $e^4$ is found to be
\begin{align}
 \nonumber &\int_{xyzw}\langle (\psi^\dagger \rmi a \psi)_x (\psi^\dagger \rmi a \psi)_y (\psi^\dagger \rmi a \psi)_z (\psi^\dagger \rmi a \psi)_w \rangle_>\\
 \label{rg15} &=  12 \int_Q (G_a^Q)^2 \Bigl[ (\psi^\dagger G_\psi^Q\psi)^2 + (\psi^\dagger G_\psi^Q\psi)(\psi^\dagger G_\psi^{-Q}\psi)\Bigr].
\end{align}
In deriving these results, we made no further assumptions on the structure of the fermion propagator $G_\psi^Q$ or the vertices $M,N$. Hence the expressions are valid for general four-fermion theories with differing kinetic terms or number of fermion components. Note, however, that the displayed results are only meaningful after the appropriate Fierz identities of the theory have been applied to eliminate the contained redundancies.

We denote the running of the couplings $g_{i=1,\dots,8}$ before applying Fierz by
\begin{align}
 \label{rg16} \dot{g}_i^{(0)} = (z-d)g_i +f_1(\delta)\cdot\Delta g_i^{(0)},
\end{align}
and the actual flow equations for $g_{1,2,3}$ after Fierz by
\begin{align}
  \label{rg17} \dot{g}_i = (z-d)g_i +f_1(\delta)\cdot\Delta g_i.
\end{align}
From the Fierz identities (\ref{fierz17})-(\ref{fierz21}) derived below we obtain
\begin{align}
 \nonumber \Delta g_1 &= \Delta g_1^{(0)}-\frac{3}{2}(\Delta g_4^{(0)}+\Delta g_5^{(0)}+\Delta g_6^{(0)})-\frac{1}{2}\Delta g_7^{(0)},\\
 \nonumber \Delta g_2 &= \Delta g_2^{(0)}-\frac{3}{10}\Delta g_4^{(0)}-\frac{6}{5}\Delta g_5^{(0)}+\frac{1}{2}\Delta g_7^{(0)}-\frac{6}{5}\Delta g_8^{(0)},\\
 \nonumber \Delta g_3 &= \Delta g_3^{(0)}-\frac{3}{10}\Delta g_4^{(0)}+\frac{3}{10}\Delta g_5^{(0)}-\frac{1}{2}\Delta g_6^{(0)}-\frac{1}{2}\Delta g_7^{(0)}\\
 \label{rg18} &+\frac{4}{5}\Delta g_8^{(0)}.
\end{align}
Performing the frequency and cubic momentum integrals in the same fashion as for the self-energies, we are left with
\begin{widetext}
\begin{align}
 \nonumber \Delta g_1 =\ & \frac{1}{5}(-5+2F_-+3F_+)g_1^2-\frac{2}{5}(-5+4F_-+3F_+)g_1g_2-\frac{3}{5}(-5+2F_-+5F_+)g_1g_3-g_2^2-6g_2g_3-3g_3^2\\
 \label{rg19} &+\frac{1}{10}(5-2F_--3F_+)e^2g_1-\frac{2}{5}F_-e^2g_2-\frac{3}{5}F_+e^2g_3,\\
 \nonumber \Delta g_2 =\ & -\frac{1}{5}F_-g_1^2+\frac{1}{5}(5+3F_+)g_1g_2 -\frac{1}{5}(5+4F_-+6F_+)g_2^2-3(1+F_+)g_2g_3-\frac{3}{5}(5+F_-)g_3^2\\
 \label{rg20} &-\frac{1}{5}F_-e^2g_1+\frac{1}{10}(5+3F_+)e^2g_2 -\frac{1}{20} F_-e^4,\\
 \nonumber \Delta g_3 =\ & -\frac{1}{5}F_+g_1^2 +\frac{1}{5}(5+2F_-+F_+)g_1g_3 -\frac{1}{5}(5+2F_+)g_2^2-\frac{2}{5}(10+4F_-+F_+)g_2g_3-\frac{2}{5}(5+3F_-+5F_+)g_3^2\\
 \label{rg21} &-\frac{1}{5}F_+e^2g_1 + \frac{1}{10}(5+2F_-+F_+)e^2g_3-\frac{1}{20} F_+e^4,
\end{align}
\end{widetext}
with $F_{\pm}(\delta)$ as defined in Eqs. (\ref{flow21}) and (\ref{flow22}). The $\beta$-functions simplify further upon using $2F_-(\delta)+3F_+(\delta)=5$ for all values of $\delta$. This yields the flow equations (\ref{flow18})-(\ref{flow20}) discussed in the main text.

The contributions $\Delta g_i$ in Eqs. (\ref{rg19})-(\ref{rg21}) have the particular feature that $g_1$ and $e^2$ only appear in the combination
\begin{align}
 \label{rg22} g_1 +\frac{e^2}{2}.
\end{align}
This can be traced back to the momentum shell regularization scheme, the photon propagator being independent of frequency, $G_a^Q = \frac{\bar{e}^2}{q^2}$, and the square of the fermion propagator vanishing upon frequency integration, namely
\begin{align}
 \label{rg23} \int_{q_0} (G_\psi^Q)^2 = 0.
\end{align}
Indeed, this is satisfied for $G_\psi^Q$ from Eq. (\ref{prop1}), but does, for instance, also hold for a Dirac particle.

To prove this statement we first consider the terms that are linear in $g_1+\frac{e^2}{2}$. Set $N=\mathbb{1}$ in Eq. (\ref{rg11}) and let $g_i\neq g_1$. We then have
\begin{widetext}
\begin{align}
 \nonumber g_ig_1\int_{xy} \langle (\psi^\dagger M \psi)_x^2(\psi^\dagger \psi)^2_y \rangle_> &=g_ig_1\Biggl[ -4 \int_Q(\psi^\dagger M\psi)(\psi^\dagger \psi) \mbox{tr}(G_\psi^QG_\psi^QM)+4 \int_Q(\psi^\dagger M\psi)(\psi^\dagger  G_\psi^Q M G_\psi^Q  \psi)\\
 \nonumber &+4\int_Q (\psi^\dagger MG_\psi^QG_\psi^QM\psi)(\psi^\dagger \psi)+4\int_Q  (\psi^\dagger MG_\psi^Q\psi)(\psi^\dagger  G_\psi^QM\psi)\\
 \label{rg24} &+2\int_Q  (\psi^\dagger MG_\psi^Q \psi)(\psi^\dagger MG_\psi^{-Q}\psi)+2\int_Q  (\psi^\dagger G_\psi^QM \psi)(\psi^\dagger G_\psi^{-Q}M \psi)\Biggr].
\end{align}
Now use Eq. (\ref{rg23}) to simplify this according to
\begin{align}
 \nonumber g_ig_1\int_{xy} \langle (\psi^\dagger M \psi)_x^2(\psi^\dagger \psi)^2_y \rangle_>  &=2g_ig_1\Biggl[ 2 \int_Q(\psi^\dagger M\psi)(\psi^\dagger  G_\psi^Q M G_\psi^Q  \psi)+2\int_Q  (\psi^\dagger MG_\psi^Q\psi)(\psi^\dagger  G_\psi^QM\psi)\\
 \label{rg25} &+\int_Q  (\psi^\dagger MG_\psi^Q \psi)(\psi^\dagger MG_\psi^{-Q}\psi)+\int_Q  (\psi^\dagger G_\psi^QM \psi)(\psi^\dagger G_\psi^{-Q}M \psi)\Biggr].
\end{align}
\end{widetext}
This is structurally identical to the right hand side of Eq. (\ref{rg12}). Upon re-exponentiating, Eq. (\ref{rg25}) for $g_i\neq g_1$ gets multiplied by $-1$, and Eq. (\ref{rg12}) gets multiplied by $\frac{1}{2}$. Accordingly, we can only have contributions $\sim g_i (g_1+\frac{e^2}{2})$. To show the appearance of this combination also to quadratic order, set $g_i=g_j=g_1$ ($M=N=\mathbb{1}$) in Eq. (\ref{rg11}). We then find
\begin{align}
 \nonumber \ &g_1^2\int_{xy} \langle (\psi^\dagger \psi)_x^2(\psi^\dagger \psi)^2_y \rangle_> \\
 \nonumber &=2g_1^2\Biggl[ 2 \int_Q(\psi^\dagger \psi)(\psi^\dagger  G_\psi^Q G_\psi^Q  \psi)+2\int_Q  (\psi^\dagger G_\psi^Q\psi)(\psi^\dagger  G_\psi^Q\psi)\\
 \nonumber &+\int_Q  (\psi^\dagger G_\psi^Q \psi)(\psi^\dagger G_\psi^{-Q}\psi)+\int_Q  (\psi^\dagger G_\psi^Q \psi)(\psi^\dagger G_\psi^{-Q}\psi)\Biggr]\\
 \label{rg26} &= 4g_1^2\Biggl[ \int_Q  (\psi^\dagger G_\psi^Q\psi)(\psi^\dagger  G_\psi^Q\psi)+\int_Q  (\psi^\dagger G_\psi^Q \psi)(\psi^\dagger G_\psi^{-Q}\psi)\Biggr].
\end{align}
This is obviously structurally identical to the right hand side of Eq. (\ref{rg15}). When re-exponentiated, Eq. (\ref{rg11}) for $g_i=g_j$ gets multiplied by $-\frac{1}{2}$ and Eq. (\ref{rg15}) gets multiplied by $-\frac{1}{24}$. Hence we can only have terms
\begin{align}
\sim \Bigl(2 g_1^2 +\frac{1}{2}e^4\Bigr) = 2 \Bigl(g_1+\frac{e^2}{2}\Bigr)^2+\dots
\end{align}
We conclude that Coulomb interactions can be included by replacing $g_1 \to g_1+\frac{e^2}{2}$ in the $\beta$-functions for the short-range interactions, and, of course, by including the renormalization of the fermion self-energy.

\subsection{Susceptibilities}\label{AppRGSus}
In this section we compute the order parameter susceptibilities for both insulating and superconducting terms. For this we couple terms $L_\Phi=\Delta(\psi^\dagger M \psi)$ or $L_\Phi^{(\rm sc)}=\Delta(\psi^\dagger M \psi^*)$ to the Lagrangian and determine the resulting flow equations for $\Delta$. We give general master formulas in a similar fashion to the discussion of the flow of short-range couplings. We further analyze in detail the particular role of magnetic order in the cubic case.

We first consider the influence of an insulating term $L_\Phi=\Delta(\psi^\dagger M \psi)$ with test vertex $M$. Writing the interaction part of the Lagrangian as in Eq. (\ref{rg5}) we find the relevant contributions to the path integral to be given by
\begin{align}
 \nonumber &\langle e^{-\int_x (L_\Phi+L_{\rm int})} \rangle_> = \frac{1}{2}2\Delta\sum_{i}g_i\int_{xy} \langle (\psi^\dagger M\psi)_x(\psi^\dagger M_i\psi)_y^2 \rangle_>\\
 \label{rg27} &-\frac{1}{6}3\Delta \int_{xyz} \langle (\psi^\dagger \rmi a \psi)_x(\psi^\dagger \rmi a \psi)_y(\psi^\dagger M\psi)_z\rangle_>.
\end{align}
The individual contractions that contribute to $\dot{\Delta}$ can then be computed with the help of the master formulas
\begin{align}
 \nonumber \int_{xy} \langle (\psi^\dagger M\psi)_x(\psi^\dagger N \psi)_y^2\rangle=\ & -2(\psi^\dagger N\psi) \int_Q \mbox{tr}(MG_\psi^QNG_\psi^Q) \\
 \label{rg28} &+ 2\int_Q \psi^\dagger(NG_\psi^QMG_\psi^QN)\psi,
\end{align}
and
\begin{align}
 \nonumber \int_{xyz}  &\langle(\psi^\dagger \rmi a \psi)_x(\psi^\dagger \rmi a \psi)_y (\psi^\dagger M\psi)_z \rangle\\
 \label{rg29}  &= -2 \int_Q G_a^Q \psi^\dagger(G_\psi^{Q}MG_\psi^Q)\psi.
\end{align}
Again, using Eq. (\ref{fierz2}), we can express $M$ and $N$ in terms of $\{\Sigma^A\}$. Upon coupling a superconducting vertex $L_\Phi^{(\rm sc)}=\Delta(\psi^\dagger M \psi^*)$ we obtain corrections from the same terms as in Eq. (\ref{rg27}), given we replace $(\psi^\dagger M \psi)\to(\psi^\dagger M \psi^*)$. The master formulas for the corresponding contractions read
\begin{align}
 \label{rg30} & \int_{xy} \langle (\psi^\dagger M\psi^*)_x(\psi^\dagger N \psi)_y^2 \rangle  = 2\int_Q \psi^\dagger N G_\psi^{-Q} M [G_\psi^Q]^t N^t \psi^*
\end{align}
and
\begin{align}
 \nonumber \int_{xyz} &\langle (\psi^\dagger \rmi a\psi)_x(\psi^\dagger\rmi a \psi)_y(\psi^\dagger M \psi^*)_z \rangle\\
 \label{rg31} &= -2 \int_Q G_a^Q \psi^\dagger  G_\psi^{-Q} M [G_\psi^Q]^t \psi^*.
\end{align}
We used that $M^{\rm t}=-M$ for superconducting vertices $M$. With the same manipulations as in the previous section it is possible to show that both the insulating and superconducting susceptibilities depend on $g_1$ and $e^2$ only through the combination $g_1+\frac{e^2}{2}$.

By adding a term $L_{\Phi}$ with particular symmetry properties to the Lagrangian, we generate all possible terms that are allowed within the symmetry constraints. In particular, choosing the test vertex to be $M=\Sigma^A$, we typically only generate a contribution proportional to $L_\Phi$ through Eq. (\ref{rg27}). For instance, for $M=W_7$ we schematically write
\begin{align}
  \label{rg32} \Delta (\psi^\dagger W_7 \psi) \stackrel{\text{loops}}{\Longrightarrow} \eta_{W_7} \Delta (\psi^\dagger W_7 \psi),
\end{align}
with the prefactor giving the exponent of the susceptibility. The corresponding results are summarized in Eqs. (\ref{rg46})-(\ref{rg51}) below.

A peculiar situation arises, however, when coupling terms with $M=\mathJ_i$ or $M=W_i$. For $\delta=0$ they are distinguished through their transformation properties under $\text{SO}(3)$ via the tensor rank. On the other hand, for $\delta\neq 0$ their symmetry pattern is completely identical and both represent magnetic ordering. In fact, the contributions that are generated through Eq. (\ref{rg27}) read
\begin{align}
  \nonumber \Delta (\psi^\dagger \mathJ_i \psi) &\stackrel{\text{loops}}{\Longrightarrow} \Delta\Bigl[  \eta_{\mathJ_i}  (\psi^\dagger \mathJ_i \psi)+ a (\psi^\dagger W_i \psi)\Bigr],\\
 \label{rg33} \Delta (\psi^\dagger W_i \psi) &\stackrel{\text{loops}}{\Longrightarrow} \Delta\Bigl[  a (\psi^\dagger \mathJ_i \psi)+ \eta_{W_i}  (\psi^\dagger W_i \psi)\Bigr].
\end{align}
Hence, enhancing $\mathJ_i$ will always generate a contribution to $W_i$, and vice versa. The leading instability is then generically a linear combination of both. To find the leading instability we consider the test vertex $L_{M_i}=\Delta(\psi^\dagger M_i\psi)$ with
\begin{align}
 \label{rg34} M_i = \alpha \mathJ_i + \beta W_i.
\end{align}
We choose $\alpha^2+\beta^2=1$ to ensure $\mbox{tr}(M_iM_j)=4\delta_{ij}$, generalizing Eq. (\ref{short2}). Further restrictions on $\alpha$ and $\beta$ appear upon requiring that $L_{M_i} \stackrel{\text{loops}}{\Longrightarrow} \eta_{M_i} L_{M_i}$ holds true. Using the above equations we easily see that
\begin{align}
 \label{rg35} L_{M_i} \stackrel{\text{loops}}{\Longrightarrow} \Delta \Bigl(\eta_{\mathJ_i}+\frac{\beta a}{\alpha}\Bigr)L_{M_i} + h \Delta (\psi^\dagger W_i\psi)
\end{align}
with
\begin{align}
 \label{rg36} h = \beta(\eta_{W_i}-\eta_{\mathJ_i})+\frac{a}{\alpha}(\alpha^2-\beta^2).
\end{align}
A stable divergence of a particular channel $M_i$ then requires the self-consistency condition $h=0$ to be satisfied.

In the isotropic case ($\delta=0$ and $g_2=g_3$) we have $a=0$ and $\eta_{\mathJ_i}=\eta_{W_\mu}$ for all $i$ and $\mu$, see Eqs. (\ref{rg45}) below. Hence, there is no mixing between $L_{\mathJ_i}$ and $L_{W_i}$, and the condition $h=0$ is satisfied. For the anisotropic case let $a\neq 0$. In order to solve $h=0$ for $\alpha$ and $\beta$ we introduce $y=\frac{\eta_{W_i}-\eta_{\mathJ_i}}{a}$ so that we are left with the condition  $y\alpha\beta+\alpha^2-\beta^2=0$. We show below that in fact $y=3/2$ for all $\delta\neq 0$, which limits the solutions to $\beta=2\alpha$ or $\beta=-\alpha/2$. After a proper normalization we are left with two bilinears $L_{M_i^{(\pm)}}$, which we label according to $M_i \to U_i,\ V_i$. They read
\begin{align}
 \label{rg37} U_i &= \frac{1}{\sqrt{5}} \Bigl(2\mathJ_i-W_i\Bigr),\\
 \label{rg38} V_i &= \frac{1}{\sqrt{5}}\Bigl(\mathJ_i +2 W_i\Bigr),
\end{align}
and satisfy  $\mbox{tr}(V_i V_j )=\mbox{tr}(U_i U_j)=4\delta_{ij}$ and $\mbox{tr}(V_i U_j)=0$. The associated susceptibility exponents are found from Eq. (\ref{rg35}) to be $\eta_{M_i} =\eta_{\mathJ_i}+\frac{2\beta}{3\alpha}(\eta_{W_i}-\eta_{\mathJ_i})$, where we exploited again $y=3/2$. We then find
\begin{align}
 \label{rg39} \eta_{U_i} &= \frac{1}{3}(4\eta_{\mathJ_i}-\eta_{W_i}),\\
   \label{rg40}  \eta_{V_i} &= \frac{1}{3}(4\eta_{W_i}-\eta_{\mathJ_i}).
\end{align}
As pointed out in Ref. \cite{PhysRevB.93.241113} with $V_i=\gamma_{{\rm d},i}$, the components of $\vec{V}$ satisfy the three-dimensional Clifford algebra
\begin{align}
 \label{rg41} \{ V_i , V_j \} = 2\delta_{ij} \mathbb{1}_4.
\end{align}
Due to this extra symmetry, it is natural that coupling a term $L_{V_i}$ to the Lagrangian only generates terms again proportional to $L_{V_i}$. Note further that
\begin{align}
 \nonumber V_1 &= \gamma_{35},\\
 \label{rg42} V_2 &=-\gamma_{45},\\
 \nonumber V_3 &=-\gamma_{34}
\end{align}
and
\begin{align}
 \nonumber U_1 &= \frac{1}{2}\gamma_{14}+\frac{\sqrt{3}}{2}\gamma_{24},\\
 \label{rg43} U_2 &=\frac{1}{2}\gamma_{13}-\frac{\sqrt{3}}{2}\gamma_{23},\\
 \nonumber U_3&=-\gamma_{15}.
\end{align}

We now summarize the order parameter susceptibility exponents obtained from Eq. (\ref{rg27}) for the system considered here. In the isotropic case ($\delta=0$, $g_2=g_3$) we obtain
\begin{align}
 \nonumber \eta_{1} &= 0,\\
 \nonumber \eta_{\gamma_a} &= \frac{4}{5} \Bigl(g_1+\frac{1}{2}e^2\Bigr)-\frac{28}{5}g_2,\\
 \nonumber \eta_{\gamma_{ab}}  &= \frac{2}{5} \Bigl(g_1+\frac{1}{2}e^2\Bigr)+\frac{2}{5}g_2,\\
 \nonumber \eta^{(\rm sc)}_{\gamma_{45}}&= -\Bigl(g_1+\frac{1}{2}e^2\Bigr)-5g_2,\\
 \label{rg44} \eta^{(\rm sc)}_{\gamma_a\gamma_{45}}&= -\frac{1}{5}\Bigl(g_1+\frac{1}{2}e^2\Bigr)+\frac{3}{5}g_2.
\end{align}
In particular, we have
\begin{align}
 \label{rg45} \eta_{\mathJ_i} = \eta_{W_i} = \eta_{\gamma_{ab}}.
\end{align}
For the cubic symmetric system with general $\delta$ we find
\begin{align}
 \nonumber \eta_{\mathbb{1}} =& 0,\\
 \nonumber \eta_{E_a} =& \frac{1}{10}f_1(5+3F_+) \Bigl[g_1+\frac{e^2}{2}-4g_2-3g_3\Bigr],\\
 \nonumber \eta_{T_a} =&  \frac{1}{10}f_1(5+2F_-+F_+)\Bigl[g_1+\frac{e^2}{2}-2g_2-5g_3\Bigr],\\
 \nonumber \eta_{\mathJ_i} =&  \frac{2}{25}f_1\Bigl[ 5\Bigl(g_1+\frac{e^2}{2}\Bigr)+2F_+g_2+(5-2F_+)g_3\Bigr],\\
 \nonumber \eta_{W_i} =&  \frac{1}{50}f_1\Bigl[5(1+3F_+)\Bigl(g_1+\frac{e^2}{2}\Bigr)+32F_+g_2\\
 \nonumber &\ +(5-17F_+)g_3\Bigr],\\
 \nonumber \eta_{W'_\mu} =&  \frac{1}{10}f_1(5-F_+)\Bigl[g_1+\frac{e^2}{2}+g_3\Bigr],\\
 \label{rg46} \eta_{W_7} =&  \frac{2}{5}f_1F_-\Bigl[g_1+\frac{e^2}{2}-2g_2+3g_3\Bigr]
\end{align}
for the insulating channels, and
\begin{align}
 \nonumber \eta^{(\rm sc)}_{\gamma_{45}} &= -f_1\Bigl[g_1+\frac{e^2}{2}+2g_2+3g_3\Bigr],\\
 \nonumber \eta^{(\rm sc)}_{\gamma_a\gamma_{45}} &= \frac{1}{10}f_1(5-3F_+)\Bigl[-\Bigl(g_1+\frac{e^2}{2}\Bigr)+3g_3\Bigr]\ (a=1,2),\\
 \label{rg47}\eta^{(\rm sc)}_{\gamma_a\gamma_{45}} &= \frac{1}{5}f_1F_+\Bigl[-\Bigl(g_1+\frac{e^2}{2}\Bigr)+2g_2+g_3\Bigr]\ (a=3,4,5)
\end{align}
for the superconducting ones. The isotropic limits (\ref{rg44}) are recovered for $\delta \to 0$.

The mixing term $a$ introduced in Eq. (\ref{rg33}) is given by
\begin{align}
 \nonumber a =&\ -\frac{2}{25}f_1\Bigl[(F_--F_+)\Bigl(g_1+\frac{e^2}{2}\Bigr)-4F_+g_2\\
 \label{rg48} &+(F_-+3F_+)g_3\Bigr].
\end{align}
Obviously, $a=0$ for $\delta=0$. ($F_{\pm}=1$ and $g_2=g_3$ in the isotropic limit.) For $\delta\neq 0$, the term is nonzero and we verify by inserting the above expressions that
\begin{align}
 \label{rg49} y = \frac{\eta_{W_i}-\eta_{\mathJ_i}}{a}\Bigr|_{\delta\neq 0} = \frac{3}{2}.
\end{align}
We used that $2F_-+3F_+=5$ for all $\delta$. Remarkably, $y=3/2$ holds true for all possible values of the couplings $(g_1,g_2,g_3,e^2)$. From Eqs. (\ref{rg39}) and (\ref{rg40}) we deduce
\begin{align}
 \label{rg51} \eta_{U_i} &= \frac{1}{10} f_1(5-F_+)\Bigl(g_1+\frac{e^2}{2}+g_3\Bigr),\\
   \label{rg50}  \eta_{V_i} &= \frac{2}{5}f_1 F_+\Bigl(g_1+\frac{e^2}{2}+2g_2-g_3\Bigr).
\end{align}

\section{Tensor decomposition}\label{AppTens}

\subsection{Symmetric tensor bases}

We construct tensor bases for symmetric and symmetric traceless tensors. The dimension of the vector space of $d$-dimensional symmetric tensors of rank $\ell$, and thus the number of basis elements, is $ \binom{d+\ell-1}{\ell}$. Hence we have
\begin{align}
 \label{tens1} &\ell=1:\ \binom{d}{1} = d,\\
 \label{tens2} &\ell=2:\ \binom{d+1}{2} = \frac{d(d+1)}{2},\\
 \label{tens3} &\ell=3:\ \binom{d+2}{3} = \frac{d(d+1)(d+2)}{6}.
\end{align}
In particular, in three dimensions there are 3 vectors, 6 symmetric second rank tensors, and 10 symmetric third rank tensors.

\emph{Rank 2.} We start with the case of rank $\ell=2$. A tensor basis for symmetric second rank tensors is given by
\begin{align}
 \label{tens4} \bar{E}^{(l,m)}_{ij} = e^{(l)}_ie^{(m)}_j + e^{(l)}_j e^{(m)}_i,\ l\leq m.
\end{align}
The indices $i,j,l,m$ run from $1,\dots,d$. Herein $\textbf{e}^{(k)}$ is the unit vector pointing in $k$ direction. Of course, $e^{(l)}_i=\delta_{il}$. In three dimensions we have six possible index combinations for $(l,m)$, given by (1,1), (2,2), (3,3), (1,2), (1,3), (2,3). We can split the basis elements into one diagonal element and several symmetric traceless components. In $d=3$ dimensions we define
\begin{align}
 \label{tens5} \Lambda^0_{ij} &= \sqrt{\frac{2}{d}} \Bigl(e^{(1)}_ie^{(1)}_j+e^{(2)}_ie^{(2)}_j+e^{(3)}_i e^{(3)}_j\Bigr) = \sqrt{\frac{2}{d}}  \delta_{ij}\\
 \label{tens6} \Lambda^1_{ij} &= e^{(1)}_ie^{(1)}_j-e^{(2)}_ie^{(2)}_j,\\
 \label{tens7} \Lambda^2_{ij} &= \frac{1}{\sqrt{3}}\Bigl(-e^{(1)}_ie^{(1)}_j-e^{(2)}_ie^{(2)}_j+2e^{(3)}_ie^{(3)}_j\Bigr),\\
 \label{tens8} \Lambda^3_{ij} &=  e^{(1)}_ie^{(3)}_j+e^{(1)}_je^{(3)}_i,\\
 \label{tens9} \Lambda^4_{ij} &=  e^{(2)}_ie^{(3)}_j+e^{(2)}_je^{(3)}_i,\\
 \label{tens10} \Lambda^5_{ij} &=  e^{(1)}_ie^{(2)}_j+e^{(1)}_je^{(2)}_i.
\end{align}
The matrices $\Lambda^a$ coincide with the real Gell-Mann matrices and satisfy the \emph{orthogonality} condition
\begin{align}
 \label{tens11} \mbox{tr} (\Lambda^a\Lambda^b) &= 2\delta_{ab}
\end{align}
with $a,b=0,1,\dots,5$. We further have $\mbox{tr}(\Lambda^0\Lambda^0)=2$ and the orthogonality of $\Lambda^0$ and $\Lambda^a$ implies tracelessness according to
\begin{align}
 \mbox{tr}(\Lambda^a)\propto \mbox{tr} (\Lambda^0\Lambda^a) &= 0
\end{align}
Along the lines presented here, the Gell-Mann matrices are easily generalized to arbitrary dimension $d$, see, for instance, Ref. \cite{PhysRevB.92.045117}.

The matrices $\Lambda^a$ satisfy the \emph{completeness} condition
\begin{align}
 \label{tens12} \Lambda^0_{ij}\Lambda^0_{kl}+\Lambda^a_{ij}\Lambda^a_{kl} = \delta_{ik}\delta_{jl}+\delta_{il}\delta_{jk}.
\end{align}
To see why this is indeed the completeness relation, assume that we have a complete set of symmetric traceless matrices $\{E^a\}$ with $\mbox{tr}(E^aE^b)=2\delta^{ab}$. Then every symmetric $d \times d$ matrix $M$ can be written as
\begin{align}
 \label{tens13} M_{ij} = \frac{1}{d}\mbox{tr}(M) \delta_{ij} +\frac{1}{2}\mbox{tr}(ME^a)E^a_{ij}
\end{align}
For a general $d\times d$ matrix $N$, the right hand side of this equation can still be computed, but it only gives the symmetric part of $N$, namely
\begin{align}
 \label{tens14} \frac{1}{2}(N_{ij}+N_{ji}) = \frac{1}{d}\mbox{tr}(N) \delta_{ij} +\frac{1}{2}\mbox{tr}(NE^a)E^a_{ij},
\end{align}
or,
\begin{align}
 \label{tens15} \frac{1}{2} (\delta_{ik}\delta_{jl}+\delta_{il}\delta_{jk})N_{kl}=\Bigl(\frac{1}{d}\delta_{kl} \delta_{ij} +\frac{1}{2}E^a_{kl}E^a_{ij}\Bigr)N_{kl}.
\end{align}
In this equation, $N_{kl}$ is completely arbitrary, and thus
\begin{align}
 \label{tens16} \frac{1}{2} (\delta_{ik}\delta_{jl}+\delta_{il}\delta_{jk}) = \frac{1}{d} \delta_{lk}\delta_{ij}+\frac{1}{2}E^a_{lk}E^a_{ij}.
\end{align}
Writing $E^0_{kl}=\sqrt{\frac{2}{d}}\delta_{kl}$, we arrive at
\begin{align}
 \label{tens17} E^0_{ij}E^0_{kl}+E^a_{ij}E^a_{kl} = \delta_{ik}\delta_{jl}+\delta_{il}\delta_{jk},
\end{align}
which agrees with Eq. (\ref{tens12}) for $E^a=\Lambda^a$.

We conclude that every symmetric second rank tensor $\bar{S}_{ij}$ can be written as
\begin{align}
 \label{tens18} \bar{S}_{ij} = \frac{1}{d} S_0 \delta_{ij} + S_a \Lambda^a_{ij}
\end{align}
with, using Eq. (\ref{tens13}) for $M=\bar{S}$,
\begin{align}
 \label{tens19} S_0 &= \mbox{tr}(\bar{S}) = \bar{S}_{ii},\\
 \label{tens20} S_a &=  \frac{1}{2} \mbox{tr}(\bar{S}\Lambda^a) = \frac{1}{2}\bar{S}_{ij} \Lambda^a_{ij}.
\end{align}
In the second line we used that $\Lambda^a_{ij}$ is symmetric in $ij$.

\emph{Rank 3.} Next we turn to symmetric third rank tensors. A basis is given by
\begin{align}
 \label{tens21} \bar{E}^{(l,m,n)}_{ijk} &= e^{(l)}_ie^{(m)}_je^{(n)}_k + \text{permutations of }ijk
\end{align}
with $l\leq m\leq n$. In $d=3$ there are 10 such combinations given by (1,1,1), (2,2,2), (3,3,3), (1,1,2), (1,1,3), (1,2,2), (1,2,3), (1,3,3), (2,2,3), (2,3,3). As in the second rank case, we construct proper linear combinations of the $\bar{E}^{(l,m,n)}_{ijk}$ which constitute a suitably normalized basis for the trace(s) and traceless components of symmetric third rank tensors. Note that for such a tensor, called $\bar{B}_{ijk}$, there are $d$ traces $\delta_{ij}\bar{B}_{ijk}$ labelled by the index $k$. For $d=3$ we define
\begin{align}
 \nonumber F^1_{ijk} &= \sqrt{\frac{2}{5}}\Bigl(\delta_{ij}e^{(1)}_k+\delta_{ik}e^{(1)}_j+\delta_{jk}e^{(1)}_i\Bigr),\\
 \nonumber F^2_{ijk} &= \sqrt{\frac{2}{5}}\Bigl( \delta_{ij}e^{(2)}_k+\delta_{ik}e^{(2)}_j+\delta_{jk}e^{(2)}_i\Bigr),\\
 \label{tens22} F^3_{ijk} &= \sqrt{\frac{2}{5}}\Bigl( \delta_{ij}e^{(3)}_k+\delta_{ik}e^{(3)}_j+\delta_{jk}e^{(3)}_i\Bigr),
\end{align}
and
\begin{align}
 \nonumber E^1_{ijk} &= \sqrt{15}\Bigl(e^{(1)}_ie^{(1)}_je^{(1)}_k-\frac{1}{5}(\delta_{ij}e^{(1)}_k+\delta_{ik}e^{(1)}_j+\delta_{jk}e^{(1)}_k)\Bigr)\\
 \nonumber E^2_{ijk} &= \sqrt{15}\Bigl(e^{(2)}_ie^{(2)}_je^{(2)}_k-\frac{1}{5}(\delta_{ij}e^{(2)}_k+\delta_{ik}e^{(2)}_j+\delta_{jk}e^{(2)}_k)\Bigr),\\
 \nonumber E^3_{ijk} &= \sqrt{15}\Bigl(e^{(3)}_ie^{(3)}_je^{(3)}_k-\frac{1}{5}(\delta_{ij}e^{(3)}_k+\delta_{ik}e^{(3)}_j+\delta_{jk}e^{(3)}_k)\Bigr),\\
 \nonumber E^4_{ijk} &= \frac{1}{2}\Bigl(\bar{E}^{(1,2,2)}_{ijk}-\bar{E}^{(1,3,3)}_{ijk}\Bigr),\\
\nonumber  E^5_{ijk} &= \frac{1}{2}\Bigl(\bar{E}^{(2,3,3)}_{ijk}-\bar{E}^{(2,1,1)}_{ijk}\Bigr),\\
\nonumber  E^6_{ijk} &= \frac{1}{2}\Bigl(\bar{E}^{(3,1,1)}_{ijk}-\bar{E}^{(3,2,2)}_{ijk}\Bigr),\\
\label{tens23}  E^7_{ijk} &= \bar{E}^{(1,2,3)}_{ijk}.
\end{align}
We label the former and latter by indices $m=1,2,3$ and $\mu=1,\dots,7$, respectively. We have
\begin{align}
 F^m_{iik} =\sqrt{10} e^{(m)}_k.
\end{align}
The tensors (\ref{tens22}) and (\ref{tens23}) are \emph{orthogonal} according to
\begin{align}
 \label{tens24} F^m_{ijk} F^n_{ijk} &= 6\delta^{mn},\\
 \label{tens25} E^\mu_{ijk}E^\nu_{ijk} &= 6\delta^{\mu\nu},\\
 \label{tens26} F^m_{ijk}E^\mu_{ijk} &=0.
\end{align}
Note that $6=3!$.

Analogous to the second rank case we derive the completeness relation for symmetric third rank tensors. Let $N_{ijk}$ be a tensor with three indices, which need not be symmetric. The expression
\begin{align}
 \label{tens27} \frac{1}{6} (N_{ijk}F^{m'}_{ijk})F^{m'}_{lmn} + \frac{1}{6}(N_{ijk}E^\mu_{ijk})E^\mu_{lmn}
\end{align}
captures the symmetric part of $N_{ijk}$. Thus, generally, we have
\begin{align}
 \nonumber &\frac{1}{6} (N_{ijk}F^{m'}_{ijk})F^{m'}_{lmn} + \frac{1}{6}(N_{ijk}E^\mu_{ijk})E^\mu_{lmn}\\
 \label{tens28} &=\frac{1}{6}\Bigl(N_{lmn} + \text{permutations of }lmn\Bigr).
\end{align}
Since $N_{ijk}$ is arbitrary we conclude that \emph{completeness} in the space of symmetric third rank tensors is equivalent to
\begin{align}
 \label{tens29} F^{m'}_{ijk}F^{m'}_{lmn} + E^\mu_{ijk}E^\mu_{lmn} = \delta_{il}\delta_{jm}\delta_{kn} + \text{permutations of }ijk.
\end{align}
By a direct computation one verifies that this relation is satisfied for the tensors defined in Eqs. (\ref{tens22}) and (\ref{tens23}).

Thus we have shown that every symmetric third rank tensor $\bar{B}_{ijk}$ can be decomposed as
\begin{align}
 \label{tens30} \bar{B}_{ijk} = b_m F^m_{ijk} + B_\mu E^\mu_{ijk}
\end{align}
with
\begin{align}
 \label{tens31} b_m&= \frac{1}{6} \bar{B}_{ijk}F^m_{ijk},\\
 \label{tens32} B_\mu &= \frac{1}{6}\bar{B}_{ijk}E^\mu_{ijk}.
\end{align}

\subsection{Application to irreducible spin tensors}

We apply the results of the previous section to the spin $j=3/2$ matrices $J_i$. Although the results of this work, by construction, are representation independent, we choose at some points a particular representation for displaying explicit expressions for the algebraic objects. We define
\begin{align}
 \nonumber J_+ &= \begin{pmatrix} 0 & \sqrt{3} & 0 & 0\\ 0 & 0 & 2 & 0 \\ 0 & 0 & 0 & \sqrt{3} \\ 0 & 0 & 0 & 0 \end{pmatrix},\ J_- =\begin{pmatrix} 0 & 0 & 0 &0 \\ \sqrt{3} & 0 & 0 & 0 \\ 0 & 2 & 0 & 0 \\ 0 & 0 & \sqrt{3} & 0 \end{pmatrix},\\
  \label{tapp0a} J_z &=\begin{pmatrix} \frac{3}{2} & 0 & 0 & 0 \\ 0 & \frac{1}{2} & 0 & 0 \\ 0 & 0 & -\frac{1}{2} & 0 \\ 0 & 0 & 0 & -\frac{3}{2}\end{pmatrix}.
\end{align}
With $J_+=J_x+\rmi J_y$ and $J_-=J_x-\rmi J_y$ this implies
\begin{align}
 \label{tapp0b} J_x &= \frac{1}{2}(J_++J_-) =\begin{pmatrix} 0 & \frac{\sqrt{3}}{2} & 0 & 0 \\ \frac{\sqrt{3}}{2} & 0 & 1 & 0 \\ 0 & 1 & 0 & \frac{\sqrt{3}}{2} \\ 0 & 0 & \frac{\sqrt{3}}{2} & 0 \end{pmatrix},\\
 \label{tapp0c} J_y &=\frac{1}{2\rmi}(J_+-J_-) = \begin{pmatrix} 0 & -\rmi \frac{\sqrt{3}}{2} & 0 & 0 \\ \rmi \frac{\sqrt{3}}{2} & 0 & -\rmi &0 \\ 0 & \rmi & 0 & -\rmi \frac{\sqrt{3}}{2} \\ 0 & 0 & \rmi \frac{\sqrt{3}}{2} & 0\end{pmatrix}.
\end{align}
This representation corresponds to the common parametrization with quantization axis $\textbf{h}=\textbf{e}_z$.

Consider now the second rank operator valued tensor
\begin{align}
 \label{tapp1} \bar{S}_{ij} = J_iJ_j + J_jJ_i.
\end{align}
Using the definition in Eqs. (\ref{tens5}) we find $S_0=\frac{15}{2} \mathbb{1}_4$. This implies that the symmetric traceless tensor
\begin{align}
 \label{tapp8} S_{ij} = \bar{S}_{ij} - \frac{1}{d}S_0\delta_{ij} = S_a\Lambda^a_{ij}
\end{align}
coincides with
\begin{align}
 \label{tapp9} S_{ij} = J_iJ_j+J_jJ_i - \frac{5}{2}\mathbb{1}_4
\end{align}
defined in Eq. (\ref{irr7}). The coefficients $S_a$ of the matrix-valued tensor $S_{ij}$ are matrix-valued. For our purposes we choose the normalization
\begin{align}
 \label{tapp10} \gamma_a = \frac{1}{\sqrt{3}}S_a.
\end{align}
These matrices are the $\gamma$-matrices introduced in Eq. (\ref{field4}) which satisfy the Clifford algebra. (The latter property can easily be shown by direct computation.) We have $\mbox{tr}(\gamma_a \gamma_b)=4\delta_{ab}$ and $(\gamma_a)^\dagger=\gamma_a$. We have
\begin{align}
\label{tapp10b} S_{ij}=\sqrt{3}\gamma_a\Lambda^a_{ij}
\end{align}
with $S_a=S_{ij}\Lambda^a_{ij}$ such that the $\gamma_a$ are given by the expressions in Eqs. (\ref{irr16})-(\ref{irr20}). We can also express this in the components of $S_{ij}$ by using
\begin{align}
 \label{tapp16} \gamma_a = \frac{1}{2\sqrt{3}} S_{ij}\Lambda^a_{ij}
\end{align}
which gives
\begin{align}
 \label{tapp17} \gamma_1 &= \frac{1}{2\sqrt{3}}(S_{xx}-S_{yy}),\ \gamma_2 =  \frac{1}{2}S_{zz},\\
 \label{tapp18} \gamma_3 &= \frac{1}{\sqrt{3}} S_{xz},\ \gamma_4 = \frac{1}{\sqrt{2}}S_{yz},\ \gamma_5 = \frac{1}{\sqrt{3}}S_{xy}.
\end{align}
We used the symmetry $S_{ij}=S_{ji}$ and vanishing of the trace $S_{xx}+S_{yy}+S_{zz}=0$ of $S$.

Next consider the symmetric third rank spin tensor
\begin{align}
 \label{tapp19} \bar{B}_{ijk} &= J_iJ_jJ_k +\text{permutations of }ijk.
\end{align}
It can be decomposed according to
\begin{align}
 \label{tapp19b} \bar{B}_{ijk} = b_m F^m_{ijk} + B_\mu E^\mu_{ijk}.
\end{align}
Using $\delta_{ij}\bar{B}_{ijk} = \frac{41}{2}J_k$ we obtain
\begin{align}
 \label{tapp20} b_m = \frac{1}{6} \bar{B}_{ijk}F^m_{ijk} = \frac{41}{4}\sqrt{\frac{2}{5}} J_m.
\end{align}
Consequently, the irreducible component
\begin{align}
 \nonumber B_{ijk} &= \bar{B}_{ijk} -b_m F^m_{ijk} \\
 \nonumber &= \bar{B}_{ijk} - \frac{41}{4} \frac{2}{5} \Bigl(\delta_{ij}e^{(m)}_k+\delta_{ik}e^{(m)}_j+\delta_{jk}e^{(m)}_i\Bigr)J_m\\
 \label{tapp21} &= \bar{B}_{ijk} - \frac{41}{10} \Bigl(\delta_{ij}J_k+\delta_{ik}J_j+\delta_{jk}J_i\Bigr)
\end{align}
coincides with $B_{ijk}$ derived in Eq. (\ref{irr8}). We introduce the matrices $W_\mu$ such that they satisfy $\mbox{tr}(W_\mu W_\nu)=4\delta_{\mu\nu}$ according to
\begin{align}
 \label{tapp22} W_\mu = \frac{2}{3\sqrt{3}}B_\mu.
\end{align}
Applying Eqs. (\ref{tens23}), we  then arrive at Eqs. (\ref{irr21})-(\ref{irr27}).

It is instructive to relate $J_i$ and $W_\mu$ to the matrices $\gamma_{ab}$ with $a<b$. Applying $M=\frac{1}{4}\mbox{tr}(M\Gamma^A)\Gamma^A$ with $M=J_i,W_\mu$ and $\Gamma^A=\gamma_{ab}$ yields
\begin{align}
 J_x &=\frac{1}{2}\gamma_{14} +\frac{\sqrt{3}}{2}\gamma_{24}+\frac{1}{2}\gamma_{35},\\
 J_y &= \frac{1}{2}\gamma_{13}-\frac{\sqrt{3}}{2}\gamma_{23}-\frac{1}{2}\gamma_{45},\\
 J_z &= -\gamma_{15}-\frac{1}{2}\gamma_{34},
\end{align}
and
\begin{align}
 W_1 &= -\frac{1}{2\sqrt{5}}\gamma_{14}-\frac{1}{2}\sqrt{\frac{3}{5}}\gamma_{24}+\frac{2}{\sqrt{5}}\gamma_{35},\\
 W_2 &= -\frac{1}{2\sqrt{5}}\gamma_{13}+\frac{1}{2}\sqrt{\frac{3}{5}}\gamma_{23}-\frac{2}{\sqrt{5}}\gamma_{45},\\
 W_3 &= \frac{1}{\sqrt{5}}\gamma_{15}-\frac{2}{\sqrt{5}}\gamma_{34},\\
 W_4 &= \frac{\sqrt{3}}{2} \gamma_{14}-\frac{1}{2}\gamma_{24},\\
 W_5 &=-\frac{\sqrt{3}}{2}\gamma_{13}-\frac{1}{2}\gamma_{23},\\
 W_6 &= \gamma_{25},\\
 \label{tapp23} W_7 &= \gamma_{12}.
\end{align}

\subsection{Higher rank tensors and Cayley--Hamilton}\label{AppCH}
The procedure outlined in the previous section can also be applied to construct higher rank irreducible spin tensors. However, for a representation of the spin $j$ algebra, no irreducible spin tensor with rank $\ell>2j$ exists \cite{BookHess}. We recall the proof of this statement here explicitly for $j=3/2$ and then discuss the general case.

Consider the symmetric fourth rank $j=3/2$ spin tensor
\begin{align}
 \label{four1} \bar{K}_{ijkl} &= J_iJ_jJ_kJ_l +\text{permutations of }ijkl.
\end{align}
The tensor is easily made traceless by an appropriate ansatz and employing $J_iS_{kl}J_i=\frac{3}{4}S_{kl}$, which yields
\begin{align}
\nonumber  K_{ijkl} &=\bar{K}_{ijkl} - 5 \Bigl(\delta_{ij}S_{kl}+\delta_{ik}S_{jl}+\delta_{il}S_{jk}+\delta_{jk}S_{il}\\
 \label{four2} &+\delta_{jl}S_{ik}+\delta_{lk}S_{ij}\Bigr)-\frac{41}{2} \Bigl(\delta_{ij}\delta_{kl}+\delta_{ik}\delta_{jl}+\delta_{il}\delta_{jk}\Bigr)\mathbb{1}_4.
\end{align}
We then indeed have $\delta_{ij}K_{ijkl}=0$ for all $kl$. However, we even verify by explicit computation that
\begin{align}
 \label{four3} K_{ijkl} =0
\end{align}
for all indices $ijkl$. This finding can be understood as a consequence of the Cayley--Hamilton (CH) theorem.

The CH theorem states that every operator is a zero of its characteristic polynomial. For matrices $M$, as in our case, the characteristic polynomial is defined as
\begin{align}
\label{four4} p(T) = \mbox{det}(T \mathbb{1}-M),
\end{align}
where $T$ is a formal place holder. The zeros of $p(\lambda)$ for $\lambda\in \mathbb{C}$ are just the eigenvalues of $M$. According to the CH theorem, inserting $M$ into the polynomial yields the zero operator. To see this, let $M$ be Hermitean for simplicity and let it act in some vector space, solve the eigenvalue problem $M \textbf{e}^{(\lambda)}=\lambda \textbf{e}^{(\lambda)}$, and decompose any vector in this vector space as $\textbf{v}=\sum_\lambda c_\lambda\textbf{e}^{(\lambda)}$. But then
\begin{align}
 \label{four5} p(M) \textbf{v}= \sum_\lambda c_\lambda p(M)\textbf{e}^{(\lambda)}=\sum_\lambda c_\lambda p(\lambda) \textbf{e}^{(\lambda)} = 0.
\end{align}
As $\textbf{v}$ was arbitrary we conclude $p(M)=0$. This proves the CH theorem for Hermitean matrices.

To prove Eq. (\ref{four3}) via CH we follow Ref. \cite{BookHess}. For the spin $j=3/2$ matrices we have
\begin{align}
 \label{four6} \Bigl(\textbf{h}\cdot\textbf{J}-\frac{3}{2}\Bigr) \Bigl(\textbf{h}\cdot\textbf{J}-\frac{1}{2}\Bigr) \Bigl(\textbf{h}\cdot\textbf{J}+\frac{1}{2}\Bigr) \Bigl(\textbf{h}\cdot\textbf{J}+\frac{3}{2}\Bigr) =0,
\end{align}
where $\textbf{h}\in\mathbb{R}^3$ is an arbitrary quantization axis with $|\textbf{h}|=1$. This well-known fact is a manifestation of the CH theorem. Indeed, given the representation from Eq. (\ref{tapp0a}) with $\textbf{h}=\textbf{e}_z$, we see the characteristic polynomial of $\textbf{h}\cdot\textbf{J}=J_z$ to be
\begin{align}
 \label{four7} p(T) =\prod_{m=-j}^j(T-m).
\end{align}
Applying rotation invariance then yields Eq. (\ref{four6}) as the more general case. We multiply Eq. (\ref{four6}) by the irreducible fourth rank tensor $\widehat{h_ih_jh_kh_l}$ constructed from the product $h_ih_jh_kh_l$ along the lines introduced above, and integrate over all possible values of $\textbf{h}$ to find
\begin{align}
 \nonumber  0 &= \int\mbox{d}^3h\ \prod_{m=-j}^j\Bigl(\textbf{h}\cdot\textbf{J}-m\Bigr) \widehat{h_ih_jh_kh_l}\\
 \label{four8}&= \int\mbox{d}^3h\ (\textbf{h}\cdot\textbf{J})^4 \widehat{h_ih_jh_kh_l} \propto K_{ijkl}.
\end{align}
In the integration all terms except to those involving $(\textbf{h}\cdot\textbf{J})^4$ vanish as they are either odd in $\textbf{h}$ or would result in a partial traces of $\widehat{h_ih_jh_kh_l}$, which vanishes by construction. The remaining integration projects onto the symmetric and traceless part of $J_iJ_jJ_kJ_l$, which is precisely $K_{ijkl}$. The proof also works for different $j$ since in this case $p(T)$ has $2j+1$ factors so that upon multiplication with $h_{i_1}\cdots h_{i_{2j+1}}$ and subsequent integration, the irreducible tensor of rank $\ell=2j+1$ is projected out and seen to vanish. Further note that multiplying Eq. (\ref{four6}) with an arbitrary power of $(\textbf{h}\cdot\textbf{J})$ one can show in the same manner that all irreducible spin tensors with $\ell\geq 2j+1$ vanish.

\section{Fierz identities}\label{AppFierz}
In this section we derive Fierz identities for both the rotation symmetric and the anisotropic case. For a detailed discussion of Fierz identities in general see also Refs. \cite{PhysRevB.79.085116,PhysRevB.93.205138}. Let $\mathcal{X}$ again be the space of Hermitean $4\times4$ matrices, and let $\{\Sigma^A\}_{A=1,\dots,16}$ be an $\mathbb{R}$-basis of $\mathcal{X}$. Assume the basis to be orthogonal according to
\begin{align}
 \label{fierz1} \mbox{tr}(\Sigma^A\Sigma^B)=4\delta^{AB}.
\end{align}
Every matrix $M\in\mathcal{X}$ can then be expressed as
\begin{align}
 \label{fierz2} M = \frac{1}{4}\mbox{tr}(M\Sigma^A)\Sigma^A.
\end{align}
Further let  $\psi=(\psi_1,\psi_2,\psi_3,\psi_4)^{\rm t}$ be a four-component vector with anticommuting Grassmann variables $\psi_i$. Given $M,N\in\mathcal{X}$, general  four-fermion interaction terms can be decomposed as
\begin{align}
 \label{fierz3} (\psi^\dagger M\psi)(\psi^\dagger N\psi) = -\frac{1}{16}\mbox{tr}(M\Sigma^AN\Sigma^B)(\psi^\dagger \Sigma^A\psi)(\psi^\dagger \Sigma^B\psi),\\
 \label{fierz4} (\psi^\dagger M\psi^*)(\psi^{\rm t}N\psi) = \frac{1}{16}\mbox{tr}(M\Sigma^AN\Sigma^B)(\psi^\dagger(\Sigma^A)^{\rm t}\psi)(\psi^\dagger \Sigma^B\psi),
\end{align}
respectively. Now apply Eq. (\ref{fierz3}) for the basis vectors $M=\Sigma^A$ and $N=\Sigma^B$, or sums thereof. One would expect the trivial result to emanate. However, this is not the case, and the nontrivial relations obtained with this procedure are the Fierz identities.

We first consider the rotation invariant case. Let
\begin{align}
\label{fierz5}  X_1 &= (\psi^\dagger\psi)^2,\\
 \label{fierz6} X_2 &= (\psi^\dagger \gamma_a\psi)^2,\\
 \label{fierz7} X_3 &= (\psi^\dagger \mathJ_i \psi)^2,\\
 \label{fierz8} X_4 &= (\psi^\dagger W_\mu\psi)^2
\end{align}
be the four-fermion terms allowed by rotation symmetry. Applying Eq. (\ref{fierz3}) according to
\begin{align}
 \nonumber \sum_i (\psi^\dagger M_i\psi)(\psi^\dagger N_i\psi) =\ & -\frac{1}{16}\Bigl(\sum_i\mbox{tr}(M_i\Sigma^AN_i\Sigma^B)\Bigr)\\
 \label{fierz9} &\times (\psi^\dagger \Sigma^A\psi)(\psi^\dagger\Sigma^B\psi),
\end{align}
with $\{M_i,N_i\}$ containing the matrices that enter $X_{1,\dots,4}$, we arrive at
\begin{align}
 \label{fierz10} \textbf{X} = F \textbf{X}
\end{align}
with $\textbf{X}=(X_1,X_2,X_3,X_4)^{\rm t}$ and
\begin{align}
 \label{fierz11} F = \begin{pmatrix} -1/4 & -1/4 & -1/4 & -1/4 \\ -5/4 & 3/4 & -1/4 & -1/4 \\ -3/4 & -3/20 & -11/20 & 9/20 \\ -7/4 & -7/20 & 21/20 & 1/20 \end{pmatrix}.
\end{align}
We have $F^2=\mathbb{1}_4$. Denote $F'=F-\mathbb{1}_4$, which has matrix rank $\text{rank}(F')=2$. This constitutes the number of Fierz identities. After row reduction, $F'$ is given by
\begin{align}
 \label{fierz12} F' \sim \begin{pmatrix} 1 & 1/5 & 1/5 & 1/5 \\ 0 & 0 & 1 & -3/7 \\ 0 & 0 & 0 & 0 \\ 0 & 0 & 0 & 0 \end{pmatrix},
\end{align}
from which we read off the Fierz identities
\begin{align}
 \nonumber 0 &= (\psi^\dagger\psi)^2 + \frac{1}{5}(\psi^\dagger \gamma_a\psi)^2+\frac{1}{5}(\psi^\dagger\mathJ_i\psi)^2+\frac{1}{5}(\psi^\dagger W_\mu\psi)^2,\\
 \label{fierz13} 0 &= \frac{1}{3}(\psi^\dagger \mathJ_i\psi)^2 -\frac{1}{7}(\psi^\dagger W_\mu\psi)^2.
\end{align}
These relations allow to eliminate two couplings from the analysis of the rotation invariant case.

Next consider the cubic symmetric case with $\textbf{L}=(L_1,\dots,L_8)^{\rm t}$ and the $L_i$ from Eq. (\ref{flow3})-(\ref{flow10}). With the same procedure as in the rotation invariant setup we obtain
\begin{align}
 \label{fierz14} \textbf{L} = F\textbf{L}
\end{align}
with
\begin{widetext}
\begin{align}
\label{fierz15} F=\begin{pmatrix}
-1/4 & -1/4 & -1/4 & -1/4 & - 1/4 & -1/4 & -1/4 & 0\\
-1/2 & 0 & 1/2 & -1/10 & -2/5 & 0 & 1/2 & -2/5 \\
-3/4 & 3/4 & 1/4 & -3/20 & 3/20 &-1/4 & -3/4 & 2/5 \\
-3/4 & -3/20 & -3/20 & -11/20 & 9/20 & 9/20 & 9/20 & 0 \\
-3/4 & -3/5 & 3/20 & 9/20 & -1/10 & 3/10 & -9/20 & 3/5\\
-3/4 & 0 & -1/4 & 9/20 & 3/10 & -1/2 & 3/4 & -1/5\\
-1/4 & 1/4 & -1/4 & 3/20 & -3/20 & 1/4 & -1/4 & -2/5 \\
0 & -3/10 & 1/5 & 0 & 3/10 & -1/10 & -3/5 & -3/5
\end{pmatrix}.
\end{align}
\end{widetext}
We have $F^2=\mathbb{1}_8$. For $F'=F-\mathbb{1}_8$ we have $\text{rank}(F')=5$, implying five Fierz identities, and row reduction yields
\begin{align}
 \label{fierz16} F' \sim \begin{pmatrix}
1 & 0 & 1/3 & 0 & 0 & 0 & 2 & 5/3 \\
0 & 1 & -2/3 & 0 & 0 & 0 & 0 & 5/3 \\
0 & 0 & 0 & 1 & 0 & 0 & -3 & -3\\
 0 & 0 & 0 & 0 & 1 & 0 & -3 & -9/2\\
0 & 0 & 0 & 0 & 0 & 1 & -3 & -5/2\\
 0 & 0 & 0 & 0 & 0 & 0 & 0 & 0 \\
 0 & 0 & 0 & 0 & 0 & 0 & 0 & 0 \\
 0 & 0 & 0 & 0 & 0 & 0 & 0 & 0\end{pmatrix}.
\end{align}
We read off the Fierz identities
\begin{align}
 \label{fierz17} 0 &= L_1 +\frac{1}{3}L_3+2L_7 +\frac{5}{3}L_8,\\
 \label{fierz18} 0 &= L_2-\frac{2}{3}L_3+\frac{5}{3}L_8,\\
 \label{fierz19} 0 &= L_4-3L_7 -3L_8,\\
 \label{fierz20} 0 &= L_5 - 3L_7 -\frac{9}{2}L_8,\\
 \label{fierz21} 0 &= L_6 - 3L_7 -\frac{5}{2}L_8.
\end{align}
We may use them to eliminate $L_{4,\dots,8}$ by means of
\begin{align}
 L_4 &= -\frac{3}{2}L_1-\frac{3}{10}L_2-\frac{3}{10}L_3,\\
 L_5 &= -\frac{3}{2}L_1-\frac{6}{5}L_2+\frac{3}{10}L_3,\\
 L_6 &= -\frac{3}{2}L_1-\frac{1}{2}L_3,\\
 L_7 &= -\frac{1}{2}L_1+\frac{1}{2}L_2-\frac{1}{2}L_3,\\
 L_8 &= -\frac{3}{5}L_2+\frac{2}{5}L_3.
\end{align}

\section{Cubic integrals}\label{AppCub}

\begin{figure*}[t!]
\centering
\begin{minipage}{\textwidth}
\includegraphics[width=0.245\textwidth]{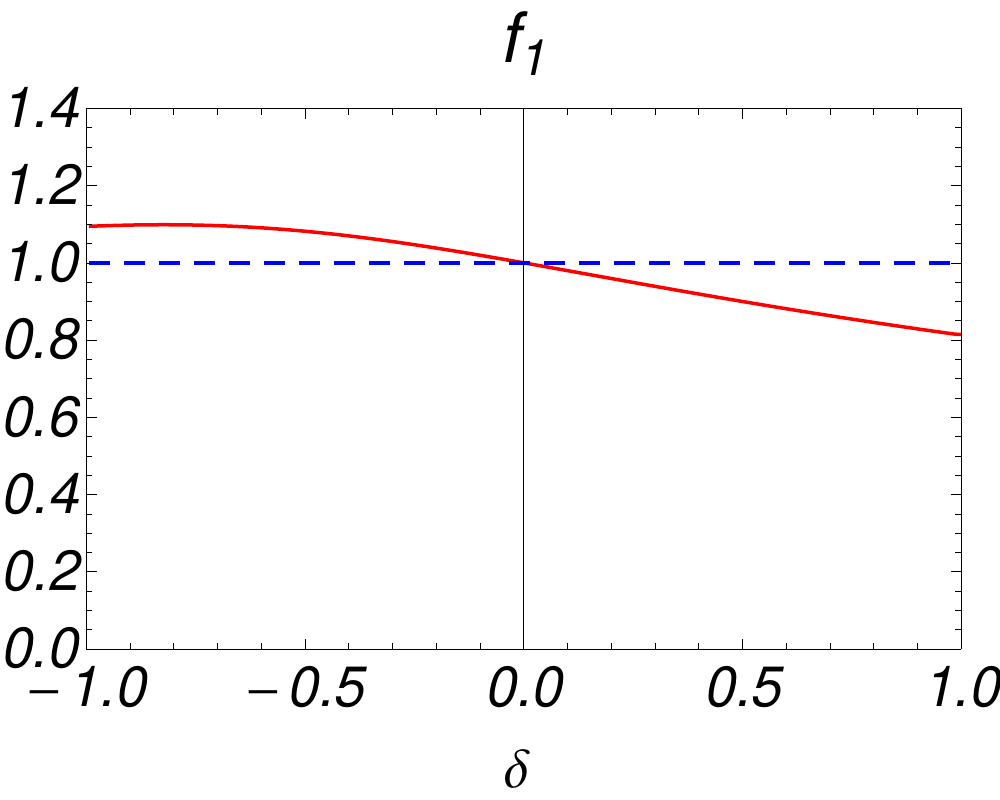}
\includegraphics[width=0.245\textwidth]{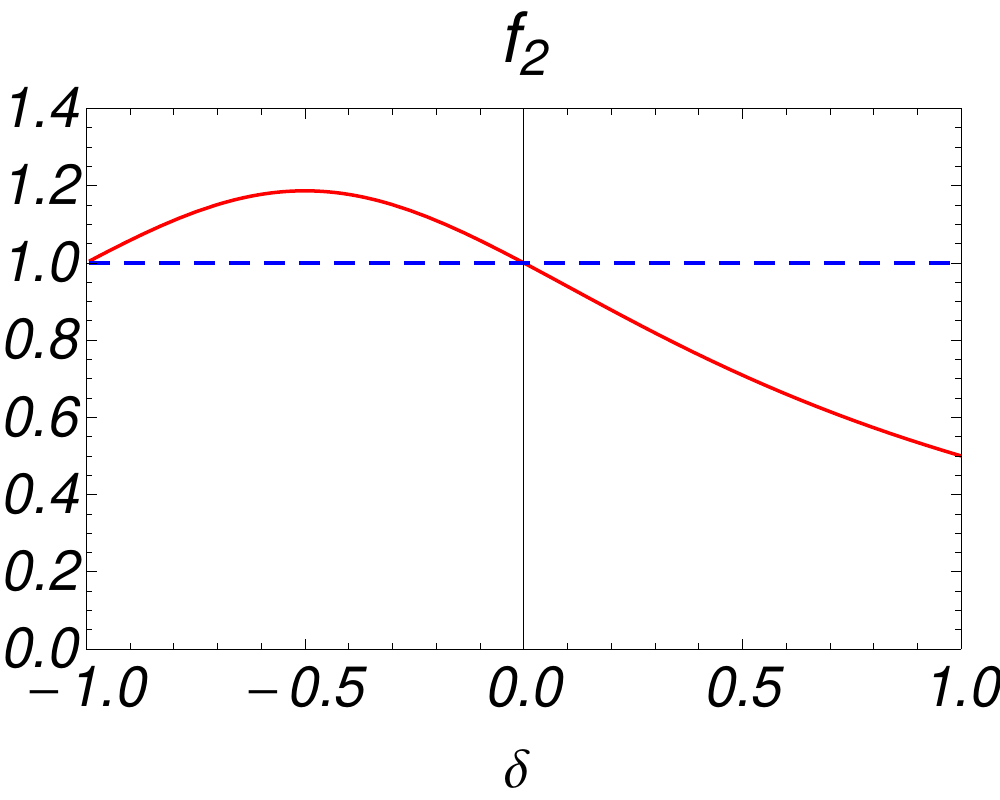}
\includegraphics[width=0.245\textwidth]{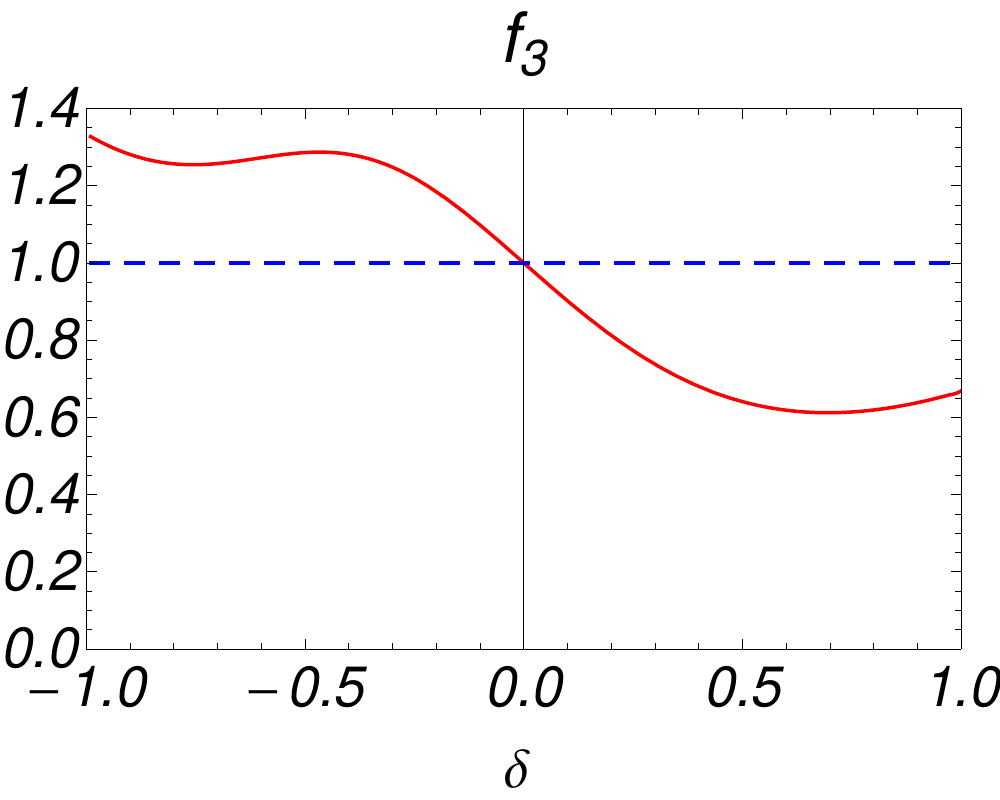}
\includegraphics[width=0.245\textwidth]{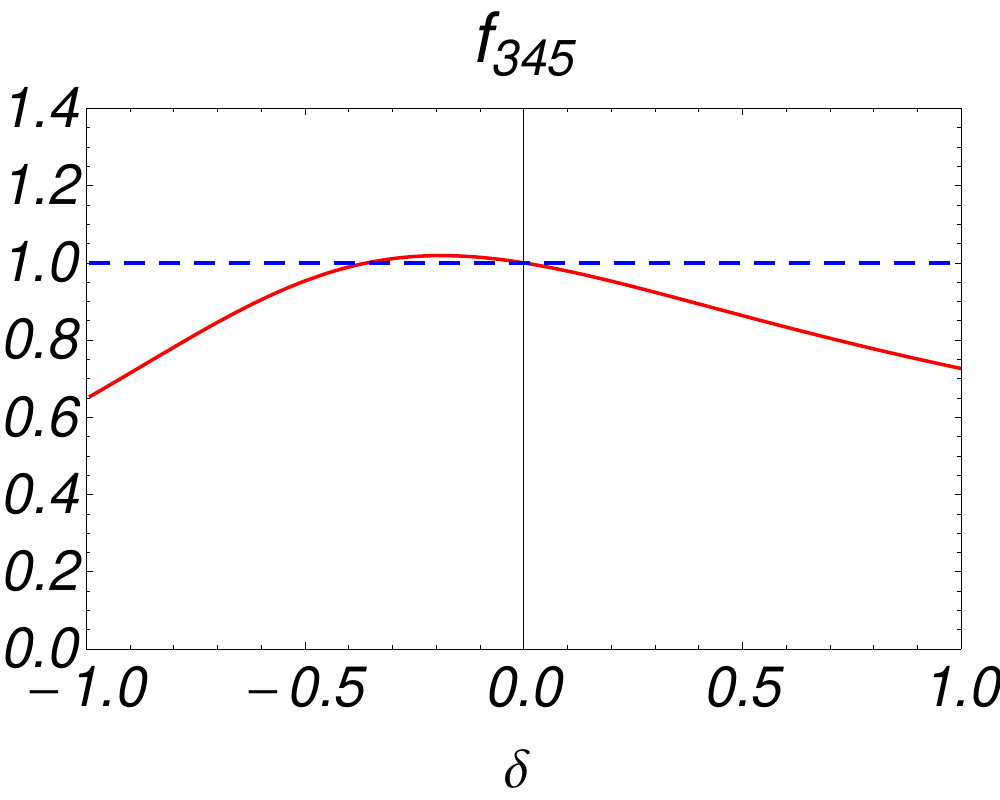}\\
\includegraphics[width=0.245\textwidth]{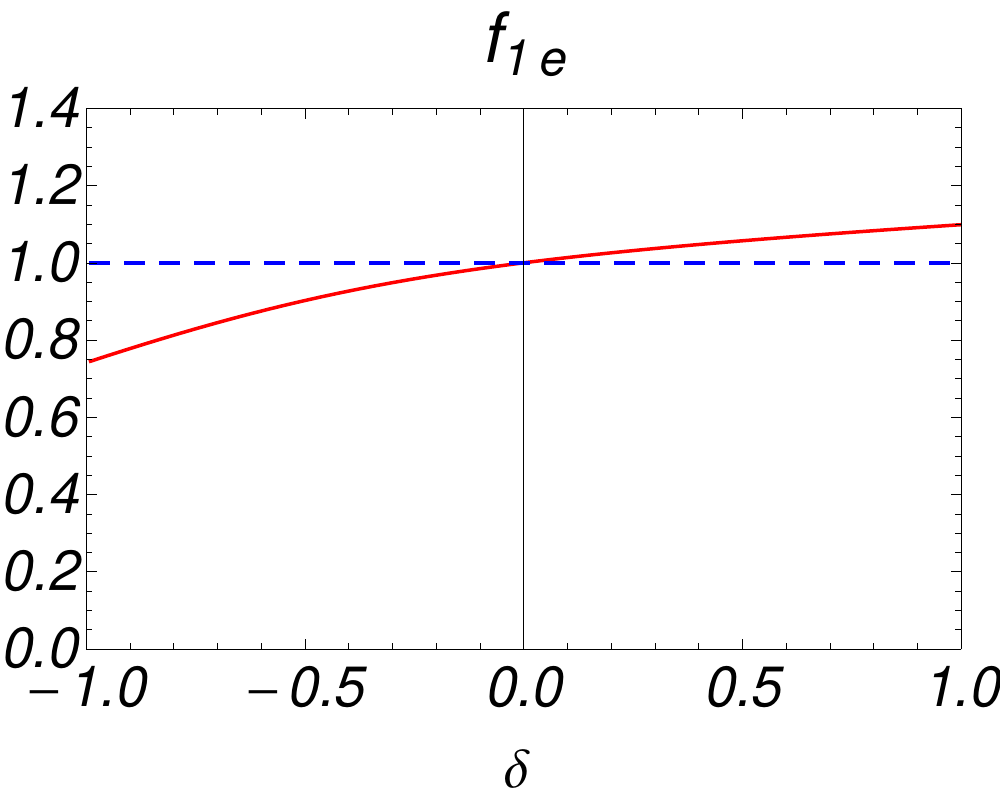}
\includegraphics[width=0.245\textwidth]{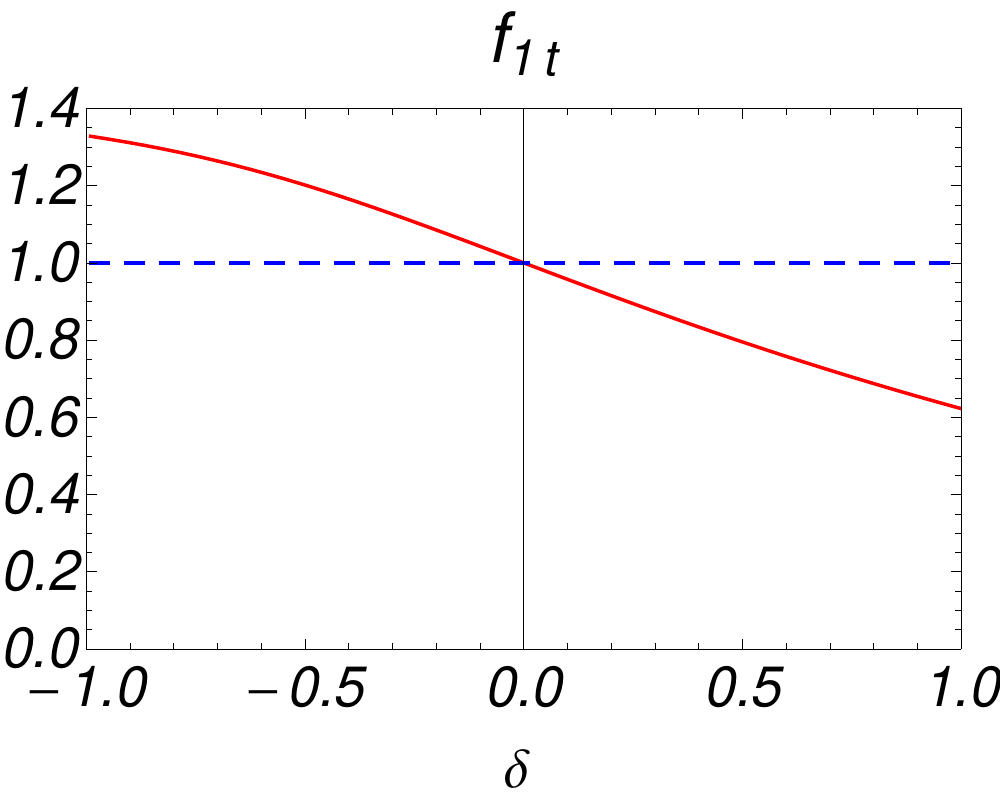}
\includegraphics[width=0.245\textwidth]{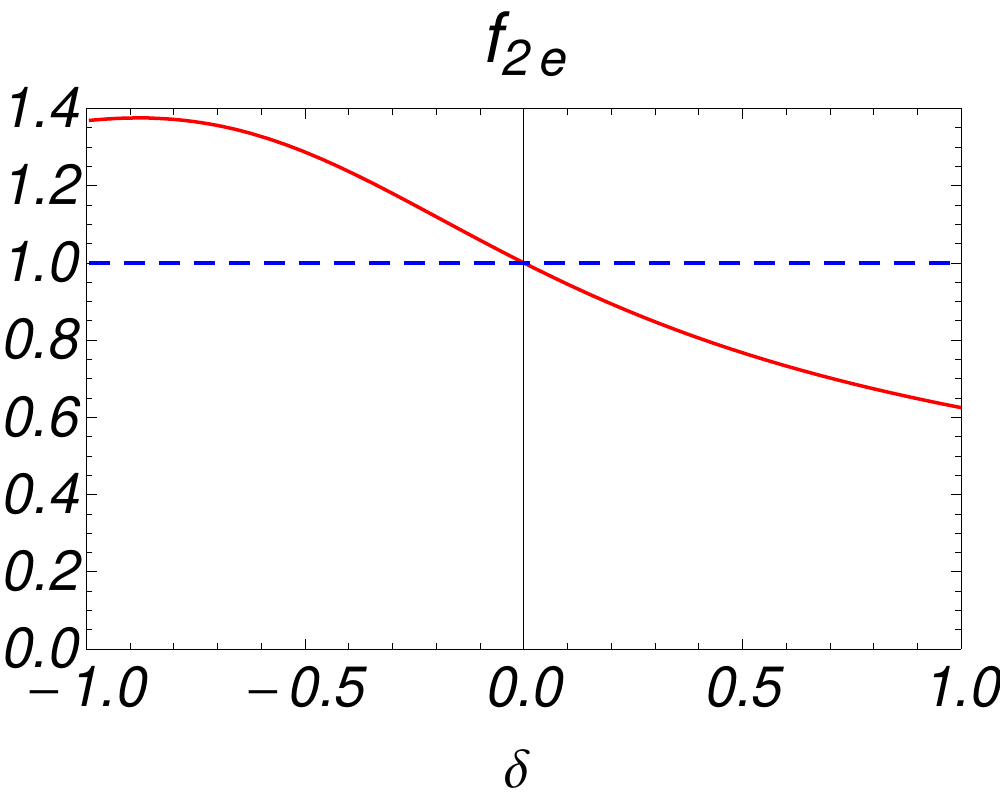}
\includegraphics[width=0.245\textwidth]{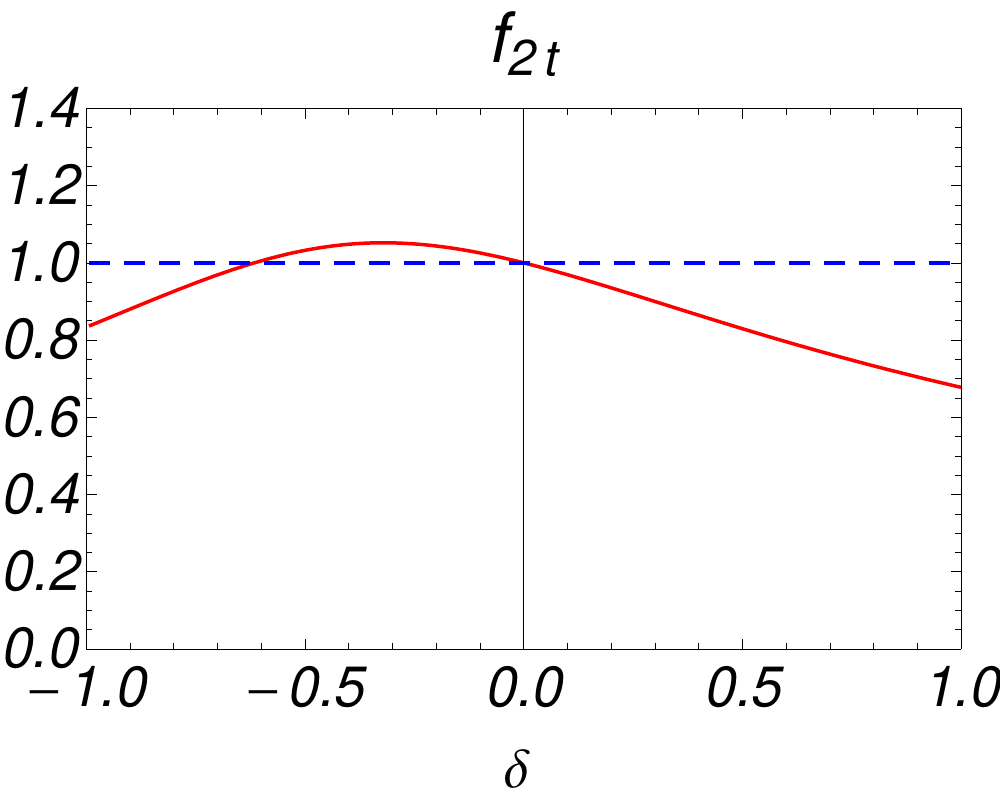}\\
\includegraphics[width=0.245\textwidth]{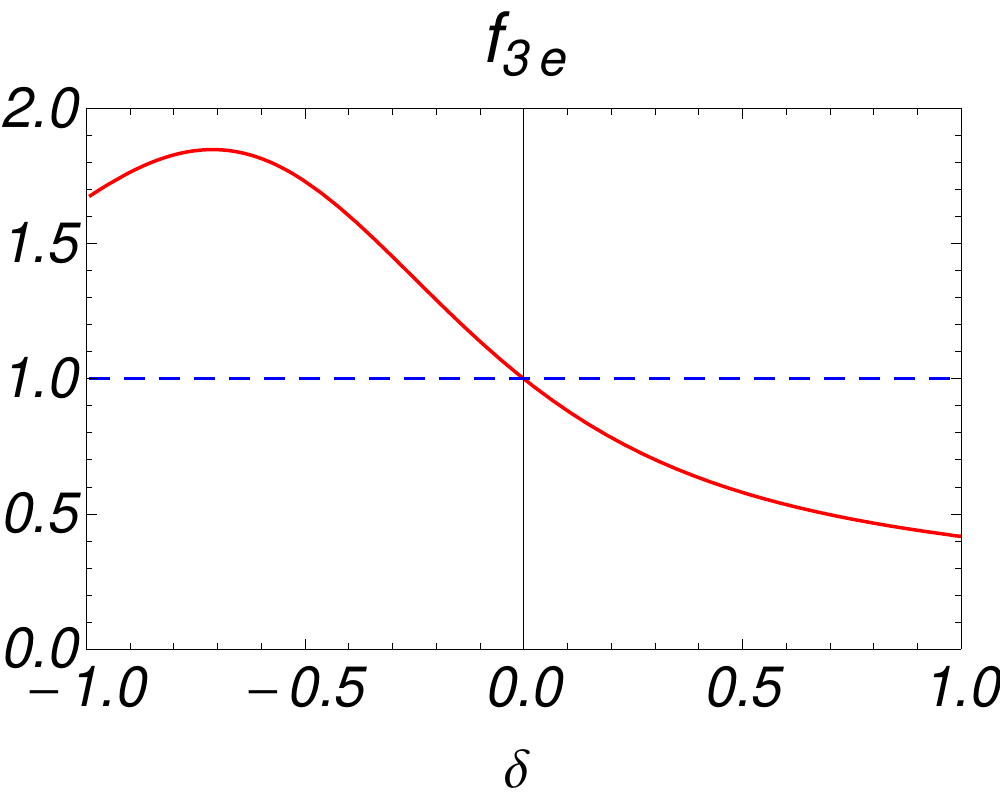}
\includegraphics[width=0.245\textwidth]{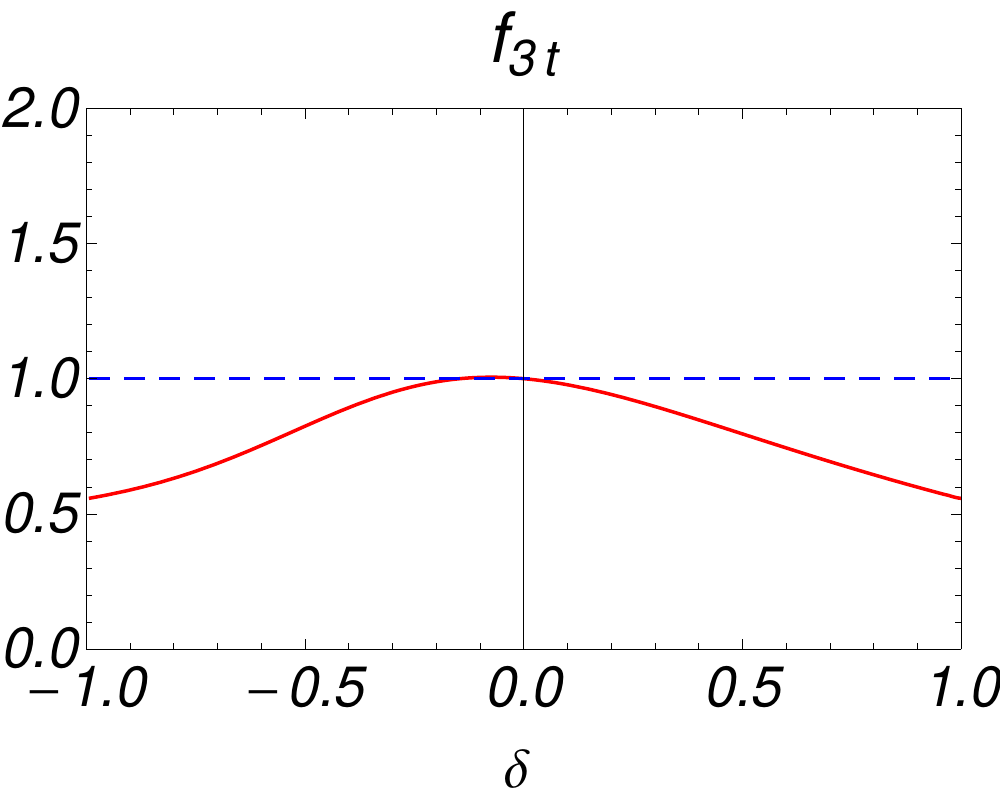}
\includegraphics[width=0.245\textwidth]{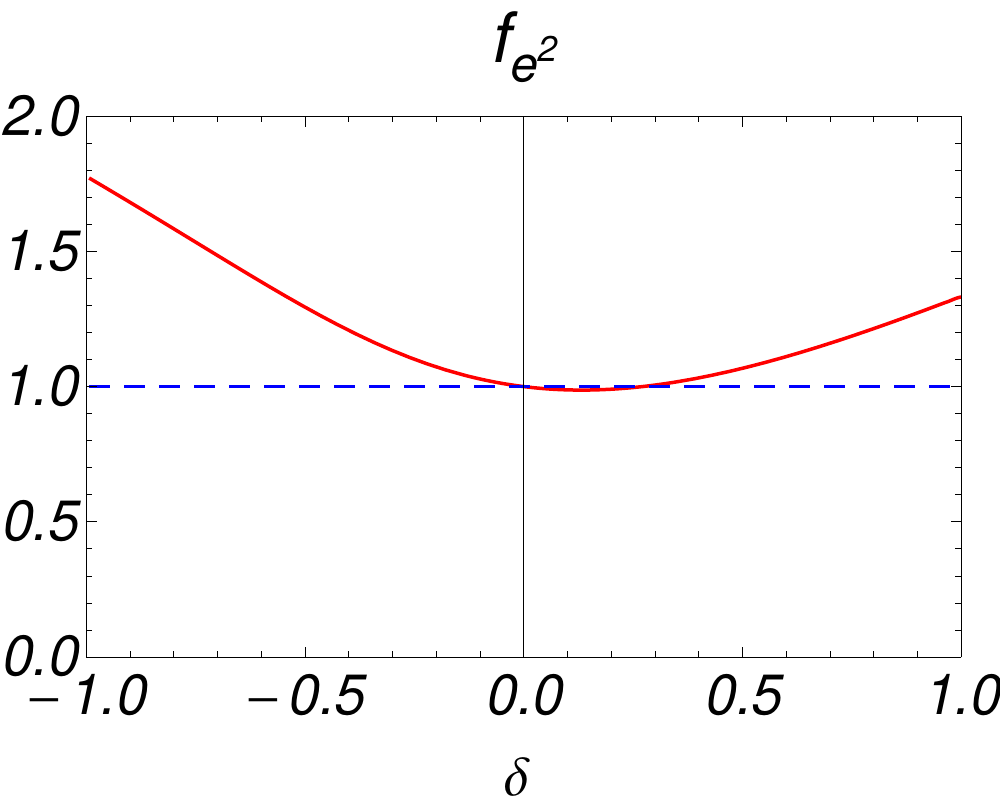}
\includegraphics[width=0.245\textwidth]{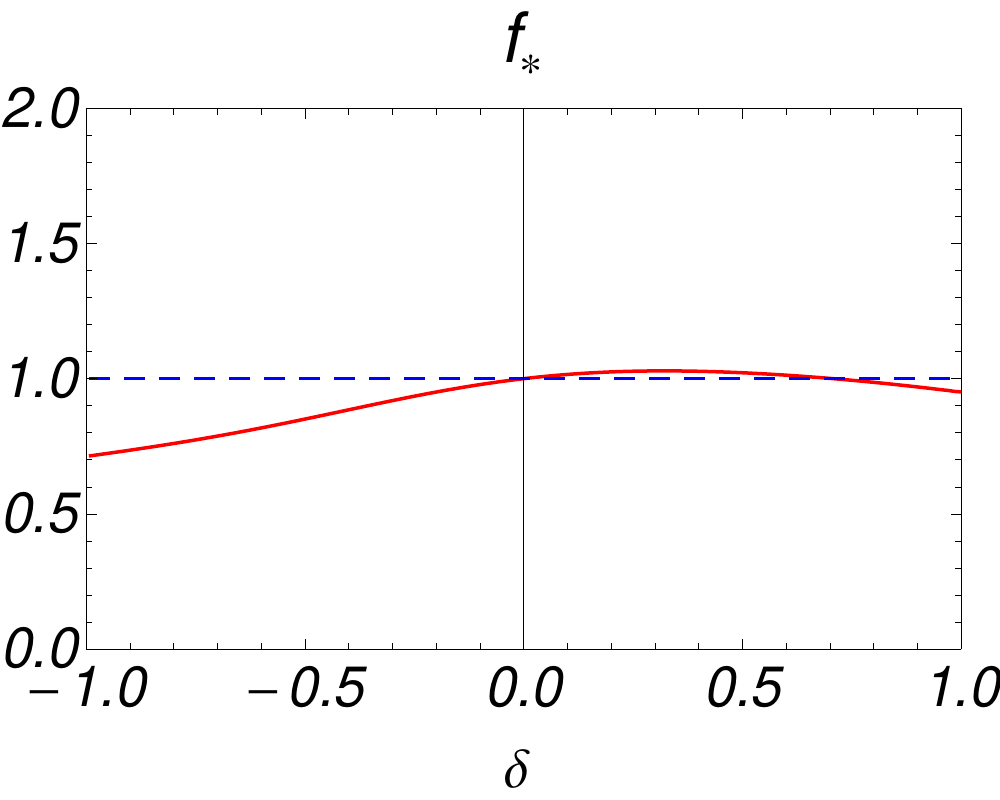}
\caption{The functions $f_i(\delta)$ defined in Eqs. (\ref{cub1b})-(\ref{cub8}) and used at various places in the analysis. The solid red lines display $f_i(\delta)$, whereas the dashed blue lines constitute unity as a guide to the eye. We also show $f_{e^2}(\delta)$ and $f_{\star}(\delta)$ characterizing the charge fixed point. As essential features the functions $f_i$ satisfy $f_i(0)=1$ and are positive of order unity for all $\delta$. Consequently they barely influence the qualitative RG flow, although, of course, they induce quantitative changes.}
\label{Figfi}
\end{minipage}
\end{figure*}

We parametrize momentum integrations by means of
\begin{align}
 \label{cub1} \int_{\textbf{q}} (\dots) = \int_0^{2\pi}\mbox{d}\phi \int_0^\pi \mbox{d}\theta \int_0^\infty \mbox{d}q\ q^2\sin \theta\ (\dots).
\end{align}
We write $X=\sum_{a=1}^5 (1+\delta s_a)^2d_a^2$ and $d_a\equiv d_a(\textbf{q})$. Further let $\chi(q^2)$ be any function of $q^2$ with compact support (such as in the  momentum shell $\Lambda/b\leq q\leq \Lambda$). The functions $f_i(\delta)$ are then defined for $|\delta|<1$ according to
\begin{align}
 \label{cub1b} \int_{\textbf{q}} \frac{1}{X^{1/2}}\ \chi(q^2) = f_1(\delta) \int_{\textbf{q}} \frac{1}{q^2}\ \chi(q^2),\\
  \label{cub1c} \int_{\textbf{q}} \frac{1}{X^{3/2}}\ \chi(q^2) = \frac{f_2(\delta)}{(1-\delta^2)} \int_{\textbf{q}} \frac{1}{q^6}\ \chi(q^2),\\
  \label{cub1d} \int_{\textbf{q}} \frac{1}{X^{5/2}}\ \chi(q^2) = \frac{f_3(\delta)}{(1-\delta^2)^3} \int_{\textbf{q}} \frac{1}{q^{10}}\ \chi(q^2),
\end{align}
and
\begin{align}
 \nonumber &a=1,2:\\
 \label{cub2} &\int_{\textbf{q}} \frac{d_a^2}{X^{1/2}}\ \chi(q^2) = \frac{1}{5} f_{1\rm e}(\delta) \int_{\textbf{q}} q^2\ \chi(q^2),\\
 \label{cub3} &\int_{\textbf{q}} \frac{d_a^2}{X^{3/2}}\ \chi(q^2)  = \frac{1}{5} \frac{f_{2\rm e}(\delta)}{(1-\delta)} \int_{\textbf{q}} \frac{1}{q^2}\ \chi(q^2) ,\\
 \label{cub4} &\int_{\textbf{q}} \frac{d_a^2}{X^{5/2}}\ \chi(q^2)  = \frac{1}{5} \frac{f_{3\rm e}(\delta)}{(1+\delta)(1-\delta)^3} \int_{\textbf{q}} \frac{1}{q^6}\ \chi(q^2) ,
\end{align}
and
\begin{align}
 \nonumber &a=3,4,5:\\
 \label{cub5} &\int_{\textbf{q}} \frac{d_a^2}{X^{1/2}}\ \chi(q^2)  = \frac{1}{5} f_{1\rm t}(\delta) \int_{\textbf{q}} q^2\ \chi(q^2) ,\\
 \label{cub6} &\int_{\textbf{q}} \frac{d_a^2}{X^{3/2}}\ \chi(q^2)  = \frac{1}{5} \frac{f_{2\rm t}(\delta)}{(1+\delta)} \int_{\textbf{q}} \frac{1}{q^2}\ \chi(q^2) ,\\
 \label{cub7} &\int_{\textbf{q}} \frac{d_a^2}{X^{5/2}}\ \chi(q^2)  = \frac{1}{5} \frac{f_{3\rm t}(\delta)}{(1-\delta)(1+\delta)^3} \int_{\textbf{q}} \frac{1}{q^6}\ \chi(q^2) .
\end{align}
We also need
\begin{align}
 \label{cub8} \int_{\textbf{q}} \frac{d_3d_4d_5}{X^{5/2}}\ \chi(q^2)  &= \frac{\sqrt{3}}{35} \frac{f_{345}(\delta)}{(1+\delta)^3} \int_{\textbf{q}} \frac{1}{q^4}\ \chi(q^2) .
\end{align}
The integrals on the left hand sides of these equations  generically appear in the derivation of RG equations carried out in this work. We see that the difference to the isotropic case ($\delta=0$) appears through scaling factors $(1\pm \delta)^{-1}$, which may qualitatively change the flow for $|\delta|\to 1$, and rather unimportant prefactors $f_i(\delta)$ representing the fine-print. The functions $f_i(\delta)$ are shown in Fig. \ref{Figfi}. 

The definitions (\ref{cub1b})-(\ref{cub8}) imply some immediate relations between the $f_i(\delta)$. Employing $\sum_a d_a^2=q^4$ we find
\begin{align}
 \label{cub8b} 5 f_1(\delta) &= 2f_{1\rm e}(\delta) + 3f_{1\rm t}(\delta),\\
 \label{cub8c} 5 f_2(\delta) &= 2(1+\delta)f_{2\rm e}(\delta)+3(1-\delta)f_{2\rm t}(\delta),\\
 \label{cub8d} 5 f_3(\delta) &= 2(1+\delta)^2f_{3\rm e}(\delta) + 3(1-\delta)^2 f_{3\rm t}(\delta).
\end{align}
Further we have
\begin{align}
\label{cub8e} 5f_1(\delta) &= 2(1-\delta) f_{2\rm e}(\delta)+3(1+\delta)f_{2\rm t}(\delta),\\
 \label{cub8f} 5f_2(\delta) &= 2f_{3\rm e}(\delta)+3f_{3\rm t}(\delta).
\end{align}
In particular, Eq. (\ref{cub8e}) implies that
\begin{align}
 \label{cub8g} 2F_-(\delta) +3F_+(\delta) =5
\end{align}
for the functions $F_\pm$ defined in Eqs. (\ref{flow21}) and (\ref{flow22}). Accordingly,
\begin{align}
 \label{cub32} F_-(-1) &= \frac{5}{2},\\
 \label{cub33} F_+(1) &= \frac{5}{3}.
\end{align}
The latter two results are relevant for the analysis of short-range interactions in the limit of strong anisotropy.

The angular $\phi$-integration can be performed analytically with the help of the complete elliptic integrals of the first and second kind given by
\begin{align}
 \label{cub9} K(m) = \int_0^{\pi/2}\mbox{d}\alpha \frac{1}{\sqrt{1-m \sin^2\alpha}},\\
 \label{cub10} E(m) = \int_0^{\pi/2}\mbox{d}\alpha \sqrt{1-m \sin^2\alpha}.
\end{align}
These functions are implemented in common computer algebra packages. We have $K(0)=E(0)=\frac{\pi}{2}$ and
\begin{align}
 \label{cub11} 2m K'(m) &= - K(m) + \frac{1}{1-m}E(m),\\
 \label{cub12} 2m E'(m)  &= E(m) -K(m).
\end{align}
Upon substituting $\alpha=2\phi$ we have
\begin{align}
 \nonumber \tilde{K}_n(A) &:=\int_0^{2\pi} \mbox{d}\phi\frac{1}{(1+A\cos^2\phi\sin^2\phi)^{n/2}} \\
 \nonumber &= 8 \int_0^{\pi/4} \mbox{d}\phi \frac{1}{(1+A\cos^2\phi\sin^2\phi)^{n/2}} \\
 \label{cub13} &= 4\int_0^{\pi/2}\mbox{d}\alpha \frac{1}{(1+\frac{A}{4}\sin^2\alpha)^{n/2}}.
\end{align}
We then find
\begin{align}
 \label{cub14} \tilde{K}_1(A) =\ & 4K\Bigl(-\frac{A}{4}\Bigr),\\
 \label{cub15} \tilde{K}_3(A) =\ & \frac{4}{1+\frac{A}{4}}E\Bigl(-\frac{A}{4}\Bigr),\\
 \nonumber \tilde{K}_5(A) =\ & \frac{4}{3(1+\frac{A}{4})^2}\Bigl[2\Bigl(2+\frac{A}{4}\Bigr)E\Bigl(-\frac{A}{4}\Bigr)\\
 \label{cub16} &-\Bigl(1+\frac{A}{4}\Bigr)K\Bigl(-\frac{A}{4}\Bigr)\Bigr].
\end{align}
For this note that an $m$-derivative of Eq. (\ref{cub9}) yields
\begin{align}
 \label{cub17} 2m K'(m) &= \int_0^{\pi/2} \mbox{d}\alpha\frac{m\sin^2\alpha-1+1}{(1-m\sin^2\alpha)^{3/2}}\\
 \nonumber &= -K(m)+ \int_0^{\pi/2}\mbox{d}\alpha \frac{1}{(1-m\sin^2\alpha)^{3/2}}.
\end{align}
Thus $\int_0^{\pi/2}\mbox{d}\alpha \frac{1}{(1-m\sin^2\alpha)^{3/2}}=2mK'(m)+K(m)=\frac{1}{1-m}E(m)$ due to Eq. (\ref{cub11}), which proves Eq. (\ref{cub15}). Taking another $m$-derivative of the just obtained relation we find $\int_0^{\pi/2}\mbox{d}\alpha \frac{1}{(1-m\sin^2\alpha)^{5/2}}=\frac{1}{1-m}+\frac{d}{dm}(\frac{1}{1-m}E(m))=\frac{1}{3(1-m)^2}[2(2-m)E(m)-(1-m)K(m)]$ due to Eq. (\ref{cub12}), and thus formula (\ref{cub16}). We further have
\begin{align}
 \nonumber \tilde{K}_{345}(A) &:= \int_0^{2\pi} \frac{\cos^2\phi\sin^2\phi}{(1+A\cos^2\phi\sin^2\phi)^{5/2}} = -\frac{2}{3}\frac{\partial}{\partial A} \tilde{K}_3(A)\\
 \label{cub18} &=\frac{2}{3}\Bigl[\frac{1}{(1+\frac{A}{4})^2}E\Bigl(-\frac{A}{4}\Bigr)+\frac{1}{1+\frac{A}{4}}E'\Bigl(-\frac{A}{4}\Bigr)\Bigr].
\end{align}
The functions $\tilde{K}_i(A)$ cover all $\phi$-integrations needed for the present analysis.

The remaining $\theta$-integrations are sufficiently simple to be evaluated numerically. Note first that in the usual spherical coordinates we have
\begin{align}
 \label{cub19} X &= (1-\delta)^2 q^4 +12\delta \sum_{i<j} q_i^2q_j^2=q^4B(1+A\cos^2\phi\sin^2\phi),
\end{align}
with
\begin{align}
 \label{cub20} A &= \frac{12\delta \sin^4\theta}{(1-\delta)^2+12\delta \sin^2\theta\cos^2\theta},\\
 \label{cub21} B &=(1-\delta)^2+12\delta\sin^2\theta\cos^2\theta.
\end{align}
Further recall $\int_{\textbf{q}} \chi(q^2) = 4\pi\int_{\Lambda/b}^\Lambda \mbox{d}q q^2\chi(q^2)$. We thus arrive at
\begin{align}
 \label{cub22} f_1(\delta) &= \frac{1}{4\pi} \int_0^\pi\mbox{d}\theta \frac{\sin\theta}{B^{1/2}} \tilde{K}_1(A),\\
 \label{cub23} f_2(\delta) &= \frac{(1-\delta^2)}{4\pi} \int_0^\pi\mbox{d}\theta \frac{\sin\theta}{B^{3/2}} \tilde{K}_3(A),\\
 \label{cub24} f_3(\delta) &= \frac{(1-\delta^2)^3}{4\pi} \int_0^\pi\mbox{d}\theta \frac{\sin\theta}{B^{5/2}} \tilde{K}_5(A).
\end{align}
Using the same derivative techniques as described in the previous paragraph one can show recursion relations such as
\begin{align}
 \label{cub24b} f_2(\delta) = f_1(\delta) + 2\delta f_1'(\delta).
\end{align}
However, for all practical purposes we found it more convenient to directly evaluate the individual integrals for numerical accuracy.

In order to evaluate the $f_i$-functions with $d_a^2$-appearances in the numerator we first observe that in order not to spoil the analytic $\phi$-integration it is convenient to exploit cubic invariance and consider
\begin{align}
  \label{cub25} a=1,2:\ d_a^2 &\to d_2^2 = \frac{1}{4}q^4 (2\cos^2\theta-\sin^2\theta)^2,\\
 \label{cub26} a=3,4,5:\ d_a^2 &\to \frac{1}{2}(d_3^2+d_4^2) =  \frac{3}{2}q^4\cos^2\theta\sin^2\theta.
\end{align}
We then arrive at
\begin{align}
 \nonumber f_{1\rm e}(\delta) &= \frac{5}{4\pi} \int_0^\pi\mbox{d}\theta \frac{\frac{1}{4}\sin\theta(2\cos^2\theta-\sin^2\theta)^2}{B^{1/2}} \tilde{K}_1(A),\\
 \nonumber  f_{2\rm e}(\delta) &= \frac{5(1-\delta)}{4\pi} \int_0^\pi\mbox{d}\theta \frac{\frac{1}{4}\sin\theta(2\cos^2\theta-\sin^2\theta)^2}{B^{3/2}} \tilde{K}_3(A),\\
 \nonumber f_{3\rm e}(\delta) &= \frac{5(1+\delta)(1-\delta)^3}{4\pi} \\
 \label{cub27} &\times \int_0^\pi\mbox{d}\theta \frac{\frac{1}{4}\sin\theta(2\cos^2\theta-\sin^2\theta)^2}{B^{5/2}} \tilde{K}_5(A).
\end{align}
and
\begin{align}
 \nonumber f_{1\rm t}(\delta) &= \frac{5}{4\pi} \int_0^\pi\mbox{d}\theta \frac{\frac{3}{2}\sin^3\theta\cos^2\theta}{B^{1/2}} \tilde{K}_1(A),\\
 \nonumber f_{2\rm t}(\delta) &= \frac{5(1+\delta)}{4\pi} \int_0^\pi\mbox{d}\theta \frac{\frac{3}{2}\sin^3\theta\cos^2\theta}{B^{3/2}} \tilde{K}_3(A),\\
 \label{cub28} f_{3\rm t}(\delta) &= \frac{5(1-\delta)(1+\delta)^3}{4\pi} \int_0^\pi\mbox{d}\theta \frac{\frac{3}{2}\sin^3\theta\cos^2\theta}{B^{5/2}} \tilde{K}_5(A).
\end{align}
To obtain the function $f_{345}$ note that $d_3d_4d_5=3\sqrt{3}q_x^2q_y^2q_z^2=3\sqrt{3}\cos^2\phi\sin^2\phi\cos^2\theta\sin^4\theta$ and, therefore,
\begin{align}
 \label{cub29} f_{345}(\delta) = \frac{105(1+\delta)^3}{4\pi} \int_0^\pi \mbox{d}\theta \frac{\sin^5\theta\cos^2\theta}{B^{5/2}}\tilde{K}_{345}(A).
\end{align}

\end{appendix}

\bibliographystyle{apsrev4-1}
\bibliography{refs_nfl}

\end{document}